\documentclass[twocolumn,aps,prd,superscriptaddress,nofootinbib,amsmath,amssymb,floats,floatfix,showkeys,notitlepage,longbibliography,author-year]{revtex4-2}
\usepackage[dvips]{graphicx,color}
\usepackage{subfigure}
\usepackage{multirow}
\usepackage{times}
\usepackage{float}
\usepackage{tabularx}
\usepackage{makecell}
\usepackage{booktabs,subcaption,amsfonts,dcolumn}
\newcolumntype{d}[1]{D..{#1}}
\usepackage{array}
\usepackage{mathrsfs}  
\usepackage{bm}
\usepackage{xcolor}
\usepackage[%
  colorlinks=true,
  urlcolor=blue,
  linkcolor=red,
  citecolor=blue
]{hyperref}
\usepackage{orcidlink}

\begin{document}

\title{Observational Signatures of Static and Rotating Wormholes Embedded in Dark Matter}

\author{Zinnat Hassan\orcidlink{0000-0002-6608-2075}}
\email{zinnathassan980@gmail.com}
\affiliation{Department of Mathematics, School of Arts, Sciences, Humanities and Education, SASTRA Deemed University, Trichy-Tanjore Road, Thirumalaisamudram, Thanjavur, Tamil Nadu 613401, India}

\author{Paras Balani}
\email{f20230738@hyderabad.bits-pilani.ac.in}
\affiliation{Department of Mathematics and Department of Computer Science, Birla Institute of Technology and Science, Pilani, Hyderabad Campus, Jawahar Nagar, Kapra Mandal, Medchal District, Telangana 500078, India}

\author{P.K. Sahoo\orcidlink{0000-0003-2130-8832}}
\email{pksahoo@hyderabad.bits-pilani.ac.in}
\affiliation{Department of Mathematics, Birla Institute of Technology and Science, Pilani, Hyderabad Campus, Jawahar Nagar, Kapra Mandal, Medchal District, Telangana 500078, India.}


\date{\today}
\begin{abstract}
We examine how static and slowly rotating traversable wormholes behave observationally when embedded in two contrasting dark matter environments, the cuspy Navarro--Frenk--White (NFW) halo and the cored solitonic wave dark matter ($\psi$DM) profile, using parameters constrained to the rotation curve of the dwarf galaxy NGC\,2366. For each profile, we solve the Einstein field equations within the Morris--Thorne framework to obtain the corresponding shape and redshift functions, then extend the analysis to slow rotation through the Teo metric with a Lense--Thirring frame-dragging term. Across the four resulting spacetimes and six throat radii, we trace null geodesics in detail, compute specific intensity profiles and polar shadow maps, and construct accretion disk images that account for the full relativistic Doppler effect. We notice that these two dark matter models are found to behave quite differently at the level of photon dynamics. The NFW potential proves too shallow to support a detached photon sphere, so its critical impact parameter remains close to the throat radius throughout. 
In contrast, the soliton core is sufficiently focused to host a truly unstable photon orbit that increases the critical impact parameter to approximately $1.18-1.26$ times the wormhole throat radius under otherwise identical conditions. This discrepancy in photon-sphere configuration propagates via almost every observable considered, including photon-ring sharpness,  shadow size, accretion-disk analysis, and, in the rotating case, the degree of shadow asymmetry produced by frame dragging. A noteworthy result is that the soliton shadow size follows the mass of the underlying ultralight boson through the connection $\rho_c\propto m_b^{-2}$, with a transition between total capture and primarily deflecting lensing regimes occurring about $m_b\sim10^{-18}$ \,eV, a signature missing in the NFW case that offers a direct probe of dark matter microphysics. Overall, these results demonstrate that the photon dynamics of wormhole solutions are shaped by the surrounding dark matter profile, proposing a potential observational path to explore the nature of dark matter itself.
\end{abstract}

\maketitle
\section{Introduction}\label{sec1}
Observations at horizon scales have unlocked a new route for testing gravity in the strong field regime, and the imaging of the supermassive compact object at the center of our galaxy, Sagittarius $A^*$ (Sgr $A^*$), by the Event Horizon Telescope (EHT) stands as one of the most evident examples of this progress \cite{Event}. The EHT data confirm that an enormous mass is packed into an incredibly small volume, yet whether this object truly retains an event horizon is still an open question \cite{Event}. Should Sgr $A^*$ turn out to lack a horizon entirely, the possibility of a physical surface would appear, unlocking the door to alternative explanations such as naked singularities, boson stars, and other horizonless compact objects, including wormholes, among them \cite{Cardoso_2019}.
In General Relativity (GR), wormholes are solutions with nontrivial spacetime topology \cite{Visser:1995cc}, and their ability to mimic the observational signatures of black holes has made them a natural candidate for such alternative descriptions \cite{PhysRevD.76.024016,PhysRevD.78.024040}. A wormhole is basically a bridge joining two distant regions of spacetime, or even two separate universes. Einstein and Rosen \cite{PhysRev.48.73} were the first to propose such a thought in 1935, through what became known as the Einstein-Rosen bridge, a concept later developed by Fuller and Wheeler \cite{PhysRev.128.919}. In 1988, Morris and Thorne \cite{10.1119/1.15620} laid the foundation for traversable wormhole solutions in GR. Building on this foundation, researchers have since put forward numerous variations of the wormhole geometry, among them Wheeler-type constructions \cite{GARFINKLE1991146,doi:10.1142/S0217732391003109}, Kerr-like rotating wormholes \cite{PhysRevD.97.024040,JOUR2019}, and wormholes built on Schwarzschild geometry \cite{10.1119/1.3672848,Cataldo2017}.
Two key elements represent the essential structure of a wormhole. First is the throat, the narrowest surface that joins the two mouths, whose geometry holds both the stability of the structure and its overall physical viability. The other is the nature of the matter required to keep the throat open. In classical theories, this typically demands the use of exotic matter, matter that fails to meet the null energy condition, one of the descriptive principles of classical GR. This condition arises from the flaring-out condition, the geometric restriction that the throat must meet in order for the wormhole to remain open and traversable.
In GR, an energy condition violation is considered a sign of something physically doubtful, which has motivated a wide range of studies aimed at finding matter descriptions that bypass this difficulty. Some of this work has centered on wormhole geometries supported by unconventional energy momentum tensors \cite{PhysRevD.46.603,PhysRevD.52.7318,Chianese2017}, while other studies have turned to phantom and quintom type energy as a means of holding the throat open \cite{PhysRevD.71.124022,Kuhfittig_2006,PhysRevD.87.084030}. A further investigation has considered configurations in which dark matter and dark energy interact non-minimally to sustain the geometry \cite{PhysRevD.89.064002}.
Quantum field theory complicates this concept further, since several of its consequences give rise to localized violations of the classical energy conditions on their own, such as Hawking radiation \cite{PhysRevD.53.1988} and the Casimir effect \cite{Epstein1965} are two well-known examples. Beyond these, quantum processes such as vacuum squeezing can cause negative energy densities, precisely the ingredient required to keep a traversable wormhole open \cite{HOCHBERG1991377}.\\
The Lambda Cold Dark Matter ($\Lambda$CDM) model is the dominant framework in cosmology, providing a coherent understanding of the large-scale structure and mass-energy composition of the Universe. This model expresses a spatially flat Universe that is mainly composed of two mysterious components, such as dark energy, which represents approximately $68 \pm 1\%$ of the total energy density, and dark matter, which accounts for about $27 \pm 1\%$ \cite{refId0}. Dark matter is generally hypothesized to consist of yet-undiscovered cold, non-relativistic particles that were generated in the early Universe and whose velocity dispersion has negligible influence on structure formation. In contrast, baryonic or ordinary matter represents a mere $5\%$ of the cosmic budget, with neutrinos and other light particles contributing even less. This cosmological composition, combined with small primordial density perturbations measured against fluctuations in the cosmic microwave background (CMB), successfully explains the formation and distribution of cosmic structures across various scales and epochs. The primordial power spectrum of initial perturbations is nearly scale-invariant, with a spectral index of $n_s = 0.965 \pm 0.006$, in agreement with predictions from inflationary models. Numerical simulations, particularly $N$-body simulations seeded with these initial conditions, reveal a hierarchical structure of dark matter clustering. In these simulations, dark matter aggregates into gravitationally bound entities called ``halos'', which may contain smaller substructures or ``sub-halos'' in a self-similar pattern. 
These exact simulations demonstrate that halos settle into density profiles with a cuspy inner region, where the density scales approximately as $\rho(r) \propto r^{-1}$ \cite{1991ApJ,Navarro_1997}. The $\Lambda$CDM model performs well on large scales, but difficulties arise once smaller scales are considered. High resolution simulations that contain only dark matter continue to favor this cuspy inner profile \cite{10.1111/j.1365-2966.2009.15878.x}, yet observational studies of dark matter dominated systems, including low surface brightness (LSB) galaxies \cite{deBlok_2001} and dwarf spheroidal galaxies \cite{Oh_2011,Walker_2011}, point rather toward flatter, core like profiles closer to $\rho(r) \propto r^{-0.2}$. This mismatch is usually referred to as the ``cusp-core problem''. Further, the CDM raises additional concerns at galactic scales beyond this issue. It predicts more massive satellite galaxies around the Milky Way than are actually observed \cite{10.1111/j.1745-3933.2011.01074.x,10.1111/j.1365-2966.2011.20181.x}, struggles to produce the spatial structure and motion of satellites seen in both the Milky Way and Andromeda \cite{10.1111/j.1365-2966.2012.20937.x,Ibata2014}, and delivers no clear justification for the inner dynamics observed in tidal dwarf galaxies \cite{refId0,Kroupa_2012}.\\
Against this background, the Navarro--Frenk--White (NFW) profile is considered as one of the most noteworthy descriptions of dark matter structure to arise from the $\Lambda$CDM paradigm \cite{Navarro_1996,Navarro_1997}. It produces the hierarchical clustering behavior expected of cold dark matter and has been widely applied to studies of galaxy formation, cluster dynamics, and gravitational lensing. 
Later studies have challenged its reliability across a broad range of halo masses and have proposed refinements, such as the Einasto profile, to match high-resolution simulation results better \cite{Navarro2004,Merritt2006}. Despite the success of the NFW profile at large scales, its cuspy behavior is inconsistent with observations in dwarf galaxies and low surface brightness systems. This discrepancy has directed researchers to explore alternative dark matter models that can naturally deliver cored profiles.
One effective alternative is ultralight or wave-like dark matter, known as fuzzy dark matter (FDM) or $\psi$DM, composed of incredibly light bosonic particles with masses around $m \sim 10^{-22}\,\mathrm{eV}$ \cite{PhysRevLett.85.1158}. Under this concept, quantum consequences begin to matter at kiloparsec scales, leading to a halo interior that differs fundamentally from the standard view. The behavior of such systems follows from the associated Schr\"odinger--Poisson equations, which allow stable, self-gravitating solitonic solutions to form at the halo core. Quantum pressure, as constrained by the uncertainty principle, supports these soliton cores and gives them a flat central density profile \cite{Schive2014}.
Cosmological simulations constructed about wave dark matter consistently create a core-halo structure, with a central soliton sitting inside an outer halo that closely follows the NFW form, a combination that upholds the large-scale achievements of CDM unchanged while addressing its small-scale shortcomings \cite{Schive2014,Marsh_2016,Hui_2017}. Further work has examined into how these solitonic cores behave dynamically, protecting their scaling relations, stability, and sensitivity to their surroundings, along with the interference patterns and fine consistencies that arise from the wave nature of the underlying field \cite{Mocz_2017,Levkov2018}.
The gap between the cuspy NFW profile and the cored solitonic profile makes clear just how sharp astrophysical and gravitational phenomena are to the underlying identity of dark matter. The inner structure of a halo, in particular, can shape spacetime geometry under strong-field conditions, with consequences for photon trajectories, shadow formation, and gravitational lensing. This perceptiveness is particularly significant in the study of compact objects and exotic geometries such as wormholes, where the surrounding matter distribution directly affects what can be observed. Keeping this in mind, the present work incorporates both the NFW and solitonic wave dark matter profiles into a wormhole framework to systematically examine how each model shapes null geodesics, ray-tracing behavior, and shadow characteristics. A comparison of this kind offers a way to probe the microscopic nature of dark matter through the observational signatures it leaves behind in strong gravitational fields.\\
The paper is organized as follows: In section \ref{sec2}, we provide a brief discussion of the field equations related to the Morris-Thorne static traversable wormhole. In section \ref{sec3}, we introduce the metric for rotating wormholes. Section \ref{sec4} focuses on investigating wormhole solutions based on two different dark matter profiles. We analyze ray tracing, accretion disks, and shadow maps for both static and rotating wormholes within the frameworks of NFW and soliton dark matter models. Next, in section \ref{sec5}, we present a detailed comparison of these two models. Finally, we conclude our paper in the last section.

\section{Light Deflection in a Static Traversable Wormhole}\label{sec2}
We consider a static, spherically symmetric spacetime that is also asymptotically flat, with its geometry encoded in the line element
\begin{equation}
    ds^{2} = -e^{2\Phi}\,dt^{2}
    + \frac{dr^{2}}{1 - b(r)/r}
    + r^{2}\!\left(d\theta^{2} + \sin^{2}\!\theta\,d\phi^{2}\right),
    \label{eq:metric}
\end{equation}
where $b(r)$ and $\Phi(r)$ represents the shape function and the redshift function, respectively \cite{10.1119/1.15620}. The radial coordinate $r$ ranges outward from the wormhole throat $r_0$ to an upper boundary, say $a$, at which the interior geometry is matched to an exterior vacuum solution. Also, a set of geometric and physical requirements must be satisfied for a wormhole to be traversable \cite{10.1119/1.15620}. The first is the absence of event horizons, which demands that $\Phi(r)$ remain finite and well-defined everywhere. Furthermore, the shape functions must satisfy the flare-out condition, which is necessary for guaranteeing that a wormhole remains traversable. This condition is defined as \cite{10.1119/1.15620}
\begin{equation}
\frac{b - b'r}{b^2} > 0.
\end{equation}  
At the throat, the above expression reduces to
\begin{equation}
b'(r_0) < 1.
\end{equation}
Moreover, when this condition is combined with the gravitational field equations, it directly leads to the violation of the null energy condition, a hallmark feature of wormhole spacetimes.
In addition, for any region outside the throat (\( r > r_0 \)), the inequality  
\begin{equation}
1 - \frac{b(r)}{r} > 0
\end{equation}  
must retain to keep an open wormhole configuration. Besides, the wormhole should demonstrate asymptotic flatness, i.e.,  
\begin{equation}
\frac{b(r)}{r} \to 0 \quad \text{as} \quad r \to \infty.
\end{equation}
Applying the Einstein field equation $G_{\mu\nu} = T_{\mu\nu}$ ($\kappa^2 \equiv 8\pi$ with $c = G = 1$), the non-vanishing components of the stress-energy tensor take the form
\begin{equation}
\rho(r)=\frac{1}{\kappa^2} \;\frac{b'}{r^2},  \label{rhoWH}
\end{equation}
\begin{equation}
p_r(r)=\frac{1}{\kappa^2} \left[2 \left(1-\frac{b}{r}
\right) \frac{\Phi'}{r} -\frac{b}{r^3}\right]  \label{prWH},
\end{equation}
\begin{multline}
p_t(r)=\frac{1}{\kappa^2} \left(1-\frac{b}{r}\right)\Bigg[\Phi
''+ (\Phi')^2+\frac{\Phi'}{r}- \frac{b'r-b}{2r(r-b)}\Phi'\\
-\frac{b'r-b}{2r^2(r-b)} \Bigg]
\label{ptWH},
\end{multline}
where $\rho(r)$ is the energy density, $p_r(r)$ is the radial pressure, $p_t(r)$ is the tangential pressure measured orthogonal to the radial direction.\\
The propagation of light in a wormhole background is governed by the null geodesic equations.
Applying the Euler--Lagrange formalism to the Lagrangian
\begin{multline}
    \mathcal{L}
    = \frac{1}{2}\,g_{\mu\nu}\dot{x}^{\mu}\dot{x}^{\nu}
    = \frac{1}{2}\!\left(
        -e^{2\Phi}\dot{t}^{2}
        + \frac{\dot{r}^{2}}{1 - b/r}
        + r^{2}\dot{\theta}^{2} \right.\\\left.
        + r^{2}\sin^{2}\!\theta\,\dot{\phi}^{2}
    \right),
    \label{eq:lagrangian}
\end{multline}
and restricting to the equatorial plane ($\theta = \pi/2$), one obtains two conserved quantities: the energy $E$ and the angular momentum $L$, giving
\begin{equation}
    \dot{t} = \frac{E}{e^{2\Phi}}, \qquad
   \dot{\phi} = \frac{L}{r^{2}\sin^2\theta}.
    \label{eq:conserved}
\end{equation}
The null condition $\mathcal{L} = 0$ then yields, after rescaling the affine parameter
$\lambda \to \lambda/|L|$,
\begin{equation}
    T + V = \frac{1}{\tilde{b}^{2}},
    \label{eq:null}
\end{equation}
where
\begin{equation}
    T \equiv \frac{e^{2\Phi}}{1 - b/r}\,\dot{r}^{2}, \qquad
    V \equiv \frac{e^{2\Phi}}{r^{2}},
\end{equation}
and $\tilde{b} \equiv |L/E|$ is the impact parameter (taking $|L|/|E|$ to ensure $\tilde{b}>0$ in the static, spherically symmetric case). The photon sphere, the unstable circular orbit that demarcates captured from deflected photon trajectories, is located at radius $r_{\rm ph}$ satisfying the two simultaneous conditions $\dot{r} = 0$ and $\ddot{r}=0$, which together require
\begin{equation}
    \left. \frac{e^{2\Phi}}{r^2}\right|_{r=r_{\rm ph}} = \frac{1}{\tilde{b}_{\rm ph}^2}, \qquad
    \left.\frac{d}{dr}\!\left(\frac{e^{2\Phi}}{r^2}\right)\right|_{r=r_{\rm ph}} = 0,
    \label{eq:photon_sphere_full}
\end{equation}
the second condition reducing to
\begin{equation}
    \left. e^{2\Phi}\,\tilde{b}^{2} \right|_{r = r_{\rm ph}} = r_{\rm ph}^{2},
    \label{eq:photon_sphere}
\end{equation}
equivalently $r_{\rm ph}\,\Phi'(r_{\rm ph}) = 1$ after differentiation. The corresponding critical impact parameter $b_{\rm ph} = r_{\rm ph}\,e^{-\Phi(r_{\rm ph})}$ determines the boundary between photon capture and scattering. The null geodesic trajectory is governed by
\begin{equation}
    \left(\frac{du}{d\phi}\right)^{2}
    = \bigl(1 - b(u^{-1})\,u\bigr)\!\left(\frac{1}{\tilde{b}^{2}\,e^{2\Phi(u^{-1})}} - u^{2}\right) \equiv \mathcal{F}(u),
    \label{eq:orbit_eq}
\end{equation}
where $u \equiv 1/r$.  Equation~\eqref{eq:orbit_eq} is solved numerically to obtain the
photon trajectories presented in this work.  Light rays with $\tilde{b} > b_{\rm ph}$ are
deflected and escape to infinity, rays with $\tilde{b} = b_{\rm ph}$ asymptotically
approach the photon sphere (forming the photon ring), and rays with
$\tilde{b} < b_{\rm ph}$ are captured by the wormhole.

\section{Rotating Wormhole Metric}
\label{sec3}
We examine the spacetime geometry of a rotating wormhole configuration. Following
Teo~\cite{Teo1998}, such a metric is stationary and axisymmetric, admitting a timelike
Killing vector $\zeta^{a} \equiv (\partial/\partial t)^{a}$ associated with time-translation
invariance and a spacelike Killing vector $\psi^{a} \equiv (\partial/\partial\phi)^{a}$
associated with azimuthal rotation invariance. The most general stationary, axisymmetric line element may be written as~\cite{Carter1968a,Carter1968b}
\begin{equation}
    ds^{2} = g_{tt}\,dt^{2} + 2g_{t\phi}\,dt\,d\phi + g_{\phi\phi}\,d\phi^{2}
             + g_{ij}\,dx^{i}\,dx^{j},
    \label{eq:general_stationary}
\end{equation}
where $i,j$ denote the remaining spatial coordinates.
We adopt the Teo-type rotating wormhole metric in spherical polar coordinates~\cite{Teo1998}
\begin{multline}
    ds^{2} = -e^{2\Phi(r)}\,dt^{2}
    + \frac{dr^{2}}{1 - b(r)/r}\\
    + r^{2}K(r)^{2}\!\left[d\theta^{2}
    + \sin^{2}\!\theta\,\bigl(d\phi - \omega(r)\,dt\bigr)^{2}\right],
    \label{eq:teo_metric}
\end{multline}
where $-\infty < t < \infty$, and $r_0 \leq r < \infty$, $0 \leq \theta \leq \pi$, and $0 \leq \phi \leq 2\pi$ are the spherical coordinates. The functions $N$, $b$, $K$, and $\omega$ depend only on $r$ and $\theta$, and are assumed to be regular on the symmetry axis $\theta = 0,\pi$ \cite{Teo1998}. The spacetime describes two identical asymptotically flat regions connected through a throat located at $r = r_0 = b > 0$. The above metric represents a rotating generalization of the static Morris--Thorne wormhole \cite{10.1119/1.15620}.
We also require that the metric \eqref{eq:teo_metric} be asymptotically flat, in which case
\begin{equation}
    \Phi(r) \to 0, \qquad K(r) \to 1, \qquad \omega(r) \to 0
    \qquad \text{as} \quad r \to \infty.
    \label{eq:asymptotic_flatness}
\end{equation}
In particular, if
\begin{equation}
    \omega(r) = \frac{2J}{r^{3}} + \mathcal{O}(r^{-4}),
    \label{eq:omega_decay}
\end{equation}
then, by changing to Cartesian coordinates, it can be checked that $J$ is the total angular momentum of the wormhole. Its mass and charge, if any, can also be deduced in the usual manner \cite{Visser:1995cc}.

\section{Dark matter density profiles}\label{sec4}
It has been shown that DM at the galactic halo could be consistent with the wormhole's structure \cite{Rahaman_2014_GalacticHalo} as well as at the central region of the halo \cite{Rahaman_2014_Central}. In the present study, we use Solition+NFW DM density profiles to derive the corresponding shape functions, where the soliton core model is applied at the central region of the DM halo, and the outer halo follows the NFW density profile, which characterizes the CDM. This scenario is motivated by the famous core-cusp problem, where N-body CDM simulations predict a cuspy density profile as $\rho(r)\propto 1/r$ at small radii, while rotation curves of dwarf galaxies, where DM is dominating, point out a flat density profile at their cores. Another puzzle is known as the missing satellites problem, where CDM simulations predict more low-mass galaxies in the local group than have already been observed. In this sense, the central masses are too low compared to the most massive (sub)halos predicted in $\Lambda$CDM. Recent simulations of soliton+NFW DM scenario show that the galactic halos surround a dense core of dwarf spheroidal galaxies with a transition between the soliton core--characterized by a flat density profile--and the CDM halo at a radius of $\simeq 1.0$ kpc \cite{4}; and also dwarf galaxies \cite{5}. Therefore, the soliton+NFW model provides a good candidate to solve small radii problems related to $\Lambda$CDM scenario. We use the numerical values of the model parameters as obtained for the dwarf galaxy NGC 2366, using fuzzy DM (soliton+NFW) simulation, by the recent analysis \cite{5}, based on the rotation curves of the LITTLE THINGS in 3D catalog \cite{6}. In the following, we shall study wormhole solutions under NFW and solition dark matter models with astrophysical observational values.

\subsection{The NFW dark matter model}\label{subsec1}

Dark matter halo modeling emerged from the need to reconcile luminous matter distributions with dynamical and gravitational evidence on galactic and cluster scales. The approximately flat rotation curves of spiral galaxies demonstrated that the enclosed gravitating mass continues to increase well beyond the optical disk, in tension with the Keplerian decline expected from the observed baryonic component alone \cite{Rubin_1980}. Later gravitational-lensing observations, most notably in merging clusters such as 1E0657-558, strengthened the case for a collisionless, non-luminous matter component whose gravitational potential can be spatially displaced from the dominant baryonic plasma \cite{Clowe_2006}. Within the standard cold dark matter paradigm, this motivated the construction of physically interpretable density profiles for virialized halos formed by hierarchical clustering. The Navarro-Frenk-White profile arose from high-resolution cold dark matter N-body simulations by Navarro, Frenk, and White \cite{Navarro_1996,Navarro_1997}, who found that halos over a wide mass range could be described by a nearly universal spherically averaged density law and its standard form can be read as
\begin{equation}\label{eq:NFWdensprof}
\rho_\text{NFW}(r) = \frac{\rho_s}{(r/r_s)(1+r/r_s)^2}\,.
\end{equation}
Here, $r_s$ stands for the characteristic scale radius and $\rho_s$ is the corresponding scale density. 
The NFW halo density acts as $\rho_{\text{NFW}} \propto 1/r$ in the inner region of the halo, whereas at large distances it reduces as $\rho_{\text{NFW}} \propto 1/r^{3}$. This behavior naturally arises from the hierarchical formation of dark matter halos through collisionless gravitational collapse. The steep central density arises from the rapid collapse of matter during the early stages of halo formation and the subsequent continuous accretion of surrounding material. In contrast, the $r^{-3}$ decrease at larger radii reflects the gradual transition from the virialized region of the halo to the surrounding cosmological background. Accordingly, the NFW profile expresses better than an empirical fitting formula. It demonstrates a direct connection between the internal structure of a halo and its construction history, concentration, and the initial conditions of the Universe \cite{Navarro_1997}. In analyses of galaxies, the NFW profile is widely used to model dark matter, accounting for the impacts of the stellar disk and bulge to reproduce observed rotation curves. On larger scales, it performs as a traditional model for determining cluster masses from gravitational lensing and X-ray observations. It has also become relevant in relativistic astrophysics, where extended dark matter halos are assumed in analyses of compact objects and the surrounding spacetime geometry. Regardless, the NFW profile remains at the center of the well-known cusp-core problem. Observations of many dwarf and low surface brightness galaxies usually demonstrate nearly constant-density central regions rather than the steep $r^{-1}$ cusp predicted by collisionless cold dark matter simulations \cite{deBlok_2002,deBlok_2010}. Several justifications have been offered, including the consequence of baryonic matter, non-circular motions, observational uncertainties, alternative dark matter scenarios such as self-interacting or warm dark matter, and the boundaries of simulations that overlook baryonic physics. Hence, although the NFW profile continues to act as the standard reference in the $\Lambda$CDM framework, it is more suitable to consider it as a physically motivated model rather than a universally accurate description of every galactic halo.
In wormhole physics, the NFW dark matter model has attracted considerable attention because realistic halo density distributions may influence, or even sustain, wormhole geometries. In GR, the existence of a static Morris-Thorne wormhole requires a throat that satisfies the flare-out condition and for which the NEC must be violated \cite{10.1119/1.15620}. Different techniques have been focused on reducing or constraining the amount of exotic matter needed to maintain such configurations \cite{Visser_2003}. For example, the NFW model density is incorporated into Einstein's field equations, and the corresponding shape function is obtained directly from the relationship between matter density and spacetime geometry. From this perspective, the dark matter halo is treated as an integral part of the gravitational configuration rather than as a small perturbation. Following this methodology, Rahaman \textit{et al.} \cite{Rahaman_2014_GalacticHalo} presented that galactic halo regions expressed by the NFW profile together with flat rotation curves may provide requirements compatible with traversable wormholes. Later, Kuhfittig \cite{Kuhfittig_2014} investigated the gravitational lensing effects of such halo-supported wormholes and noted that the deflection angle may become unbounded near the throat, indicating that strong gravitational lensing could offer a possible observational signature. Further in \cite{Rahaman_2014_Central}, the authors considered the central regions of galaxies with alternative dark matter distributions, noting that although the NFW profile successfully describes the outer halo, cored density models may offer a better representation of the inner galactic region.
Furthermore, in Ref. \cite{Rahaman_2016}, NFW and universal rotation curve (URC) models were used to examine the possibility of wormhole formation, showing that halo properties can impose meaningful constraints on the redshift function and shape function, as well as on associated energy condition violations. In addition, in \cite{Xu_2020}, traversable wormhole solutions are obtained from several dark matter distributions, including the NFW, Thomas-Fermi, and pseudo-isothermal profiles, by assuming isotropic pressure and demonstrating that suitable parameter choices allow the flare-out condition to be satisfied; however, the weak and null energy conditions are violated near the wormhole throat. Further, the NFW-supported wormholes have also been explored in different alternate theories of gravity, such as $f(R)$ gravity \cite{MUNIZ2022169129,https://doi.org/10.1002/andp.202300178}, $f(R,T)$ gravity \cite{Mustafa_2022}, $f(Q)$ gravity \cite{Mustafa_2024}, etc. Although these extended frameworks provide valuable understandings, investigations in GR remain particularly significant because they reveal the fundamental physical issue in its simplest form. A positive density NFW halo does not, by itself, eliminate the need for exotic matter. Rather, it establishes the large-scale spacetime geometry in which the conditions required for the existence of a wormhole throat must be carefully examined. Overall, the NFW profile provides a natural connection between cosmological dark matter structures and relativistic spacetime geometries, making it an important framework for exploring the interplay between galactic halos and traversable wormholes.\\
We assume that the CDM distribution in the halo obeys the NFW profile, which is extensively used in $N$-body simulations, introduced in Eq.~\eqref{eq:NFWdensprof}.
Let us now put our notice to the point that the rotation-curve that fits to NGC~2366 using the free-concentration NFW model are poorly constrained, returning an unphysically small best-fit concentration $c\lesssim0.1$; when $c$ is instead fixed to the $\Lambda$CDM-motivated value $c=9$, the resulting best-fit halo has a scale radius of several to tens of kpc, well beyond the sub-galactic throat radii $r_0\in[0.2,2.5]$~kpc considered in this work, and is therefore incompatible with the weak-field, kiloparsec-throat regime adopted here \cite{Oh_2011}. We therefore adopt representative NFW parameters at the physical scale characteristic of a dwarf-irregular dark matter halo,
\begin{equation}
r_s = 1.447~\text{kpc}, \qquad \rho_s = 3.11\times10^{-3}~\text{M}_\odot/\text{pc}^3,
\end{equation}
consistent in order of magnitude with the core radius and central density reported for the NGC~2366 rotation curve, $R_C\approx1.47$~kpc and $\rho_0\approx4.5\times10^{-2}~\text{M}_\odot/\text{pc}^3$ \cite{Oh_2011}, and with typical dwarf-galaxy halo scales more broadly \cite{deBlok_2010}. As shown below, these values keep the NFW redshift function safely in the weak-field regime, $|\Phi_\text{NFW}(r)|\ll1$, and satisfy the flare-out condition $b'(r_0)<1$ across the full range of throat radii studied.
\\
Let us first calculate the shape function by equating Eq.~\eqref{rhoWH} with Eq.~\eqref{eq:NFWdensprof}, gives 
\begin{equation}\label{eq:NFWshape}
b(r)=r_0+8\pi\rho_s r_s^3 \left[\ln\!\left(\frac{r+r_s}{r_0+r_s}\right) + \frac{r_s}{r+r_s} - \frac{r_s}{r_0+r_s}\right],
\end{equation}
where we applied the throat condition $b(r_0)=r_0$. The derivative of the shape function is
\begin{equation}
    b'(r) = \frac{8\pi\rho_s r_s^3\, r}{(r+r_s)^2}.
    \label{eq:b_prime}
\end{equation}
At the throat $r=r_0$, this gives $b'(r_0)=8\pi\rho_s r_s^3 r_0/(r_0+r_s)^2$. For the NGC 2366 parameters and $r_0=1.0$~kpc one obtains $b'(r_0)\approx 0.022\ll 1$, confirming the flare-out condition $b'(r_0)<1$ is satisfied.
At large $r\gg r_s$, the NFW profile decays as $\rho_{\rm NFW}\propto r^{-3}$, so $b'(r)\sim 8\pi\rho_s r_s^3/r$ and the shape function grows logarithmically: $b(r)\sim 8\pi\rho_s r_s^3\ln(r/r_s)$. Consequently, $b(r)/r\sim (8\pi\rho_s r_s^3/r)\ln(r/r_s)\to 0$ as $r\to\infty$, confirming that the condition $b(r)/r<1$ and asymptotic flatness are effectively maintained for the galaxy-scale parameter values used here. In a physically complete treatment, the NFW profile should be matched to a Schwarzschild exterior at the virial radius $r_\text{vir}=5.5$~kpc, beyond which the halo density is negligible~\cite{Rahaman_2014_GalacticHalo}.
The mass profile of the dark condensate galactic halo can be read as
\begin{equation}\label{34}
    M(r)=4\pi \int_0^r \rho(r) r^2 dr,
\end{equation}
For the NFW profile, the mass function can be read as
\begin{equation}\label{mass2}
M(r)=4 \pi  \rho_s r_s^3 \left[\ln\!\left(1+\frac{r}{r_s}\right) - \frac{r}{r+r_s}\right].
\end{equation}
Now, we can easily calculate the tangential velocity of the test particle with the following relation
\begin{equation}\label{ab11}
v_{t}^2(r)=G\,M(r)/r.
\end{equation}
Note that our interest is to find the redshift function using the concept of tangential velocity. Let us recall that the rotational velocity of a test particle in spherically symmetric space-time, within the equatorial plane, is determined by \cite{Böhmer_2007}; this relation holds in the weak-field limit $|\Phi|\ll 1$, which is satisfied throughout the parameter range studied (see the discussion below)
\begin{equation}\label{3711}
v_{t}^2(r)=r\,\Phi^{'}(r).
\end{equation}
Using Eqs. \eqref{mass2}, \eqref{ab11} and \eqref{3711}, one can easily calculate the redshift function
\begin{equation}
\Phi (r)=-\frac{4 \pi  \rho_s r_s^3}{r}\ln\!\left(1+\frac{r}{r_s}\right)+c_1,
\end{equation}
where $c_1$ is an integration constant. Since the first term vanishes as $r\to\infty$ (because $\ln(1+r/r_s)/r\to 0$), we must set $c_1=0$, and hence giving the NFW redshift function
\begin{equation}\label{eq:NFW_redshift}
\Phi_{\rm NFW}(r)=-\frac{4 \pi  \rho_s r_s^3}{r}\ln\!\left(1+\frac{r}{r_s}\right).
\end{equation}
For the NGC 2366 parameter values and throat radii $r_0\in[0.2,2.5]$~kpc, one can verify that $|\Phi_{\rm NFW}(r_0)|\ll 1$ throughout. For example, at $r_0=1.0$~kpc, $|\Phi(r_0)|\approx 0.043$, justifying the weak-field approximation used in Eq.~\eqref{3711}.
For the NFW wormhole, the photon sphere radius $r_{\rm ph}$ is determined numerically by locating the extremum of the effective radial potential.  Equivalently, one seeks the root of
\begin{equation}
    \frac{d}{dr}\!\left[r\,e^{-\Phi_{\rm NFW}(r)}\right] = 0,
\end{equation}
which reduces to finding where the modified impact parameter $\tilde{b}(r) = r\,e^{-\Phi}$
is stationary. This condition is precisely Eq.~\eqref{eq:photon_sphere_full}, which after substituting $\Phi=\Phi_{\rm NFW}$ and differentiating gives $r_{\rm ph}\,\Phi'_{\rm NFW}(r_{\rm ph})=1$, namely $4\pi\rho_s r_s^3[\ln(1+r_{\rm ph}/r_s)/(r_{\rm ph}) - r_s/((r_{\rm ph}+r_s)r_{\rm ph})]=1$, solved numerically.  Once $r_{\rm ph}$ is determined, the critical impact parameter follows as
\begin{equation}
    b_{\rm ph} = r_{\rm ph}\,e^{-\Phi_{\rm NFW}(r_{\rm ph})}.
\end{equation}

\subsubsection{Ray Tracing Analysis of Static NFW Wormhole}
\label{sec:analysis}

\begin{figure*}
    \centering
    \includegraphics[width=17cm,height=10cm]{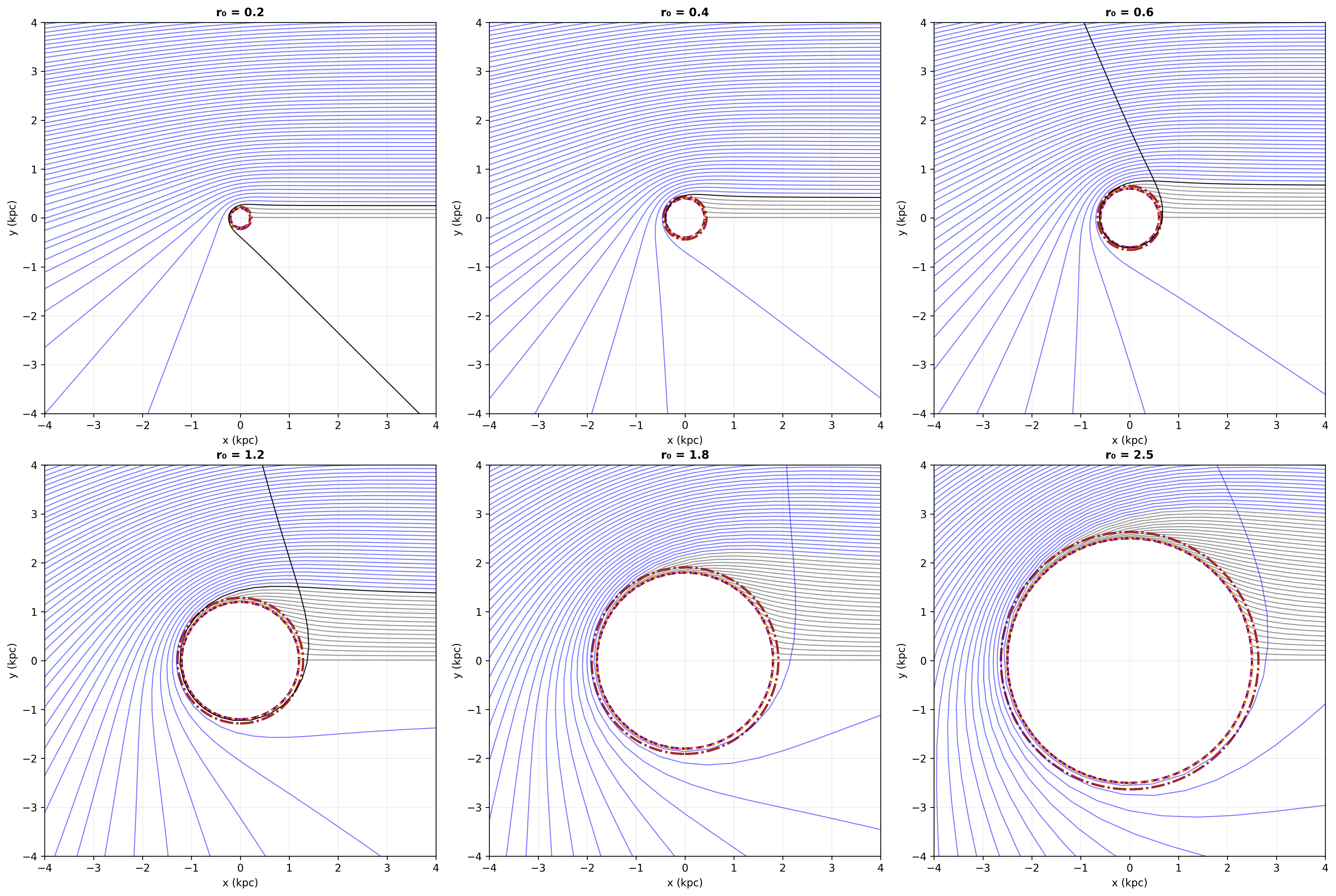}
    \caption{Null geodesics around the static NFW wormhole for six throat radii, $r_0 \in \{0.2, 0.4, 0.6, 1.2, 1.8, 2.5\}$~kpc, computed for the NGC~2366 halo parameters ($r_s = 1.447$~kpc, $\rho_s = 3.11\times10^{-3}~\text{M}_\odot/\text{pc}^3$). For each panel, $N=100$ rays are launched from $x_\text{start}=15$~kpc with impact parameters $\tilde{b}\in[0.01,8.0]$~kpc. Blue curves denote deflected rays ($\tilde{b}>b_\text{ph}$), black curves denote captured rays ($\tilde{b}\le b_\text{ph}$); the dashed purple, dotted orange, and dash-dotted brown circles mark the throat $r_0$, photon orbit radius $r_\text{ph}$, and critical impact parameter $b_\text{ph}$, respectively.}
    \label{fig:raytracing}
\end{figure*}

\begin{figure}[!ht]
    \centering
    \subfigure[Normalised intensity profiles $I(\tilde{b})$]{
    \includegraphics[width=0.42\textwidth]{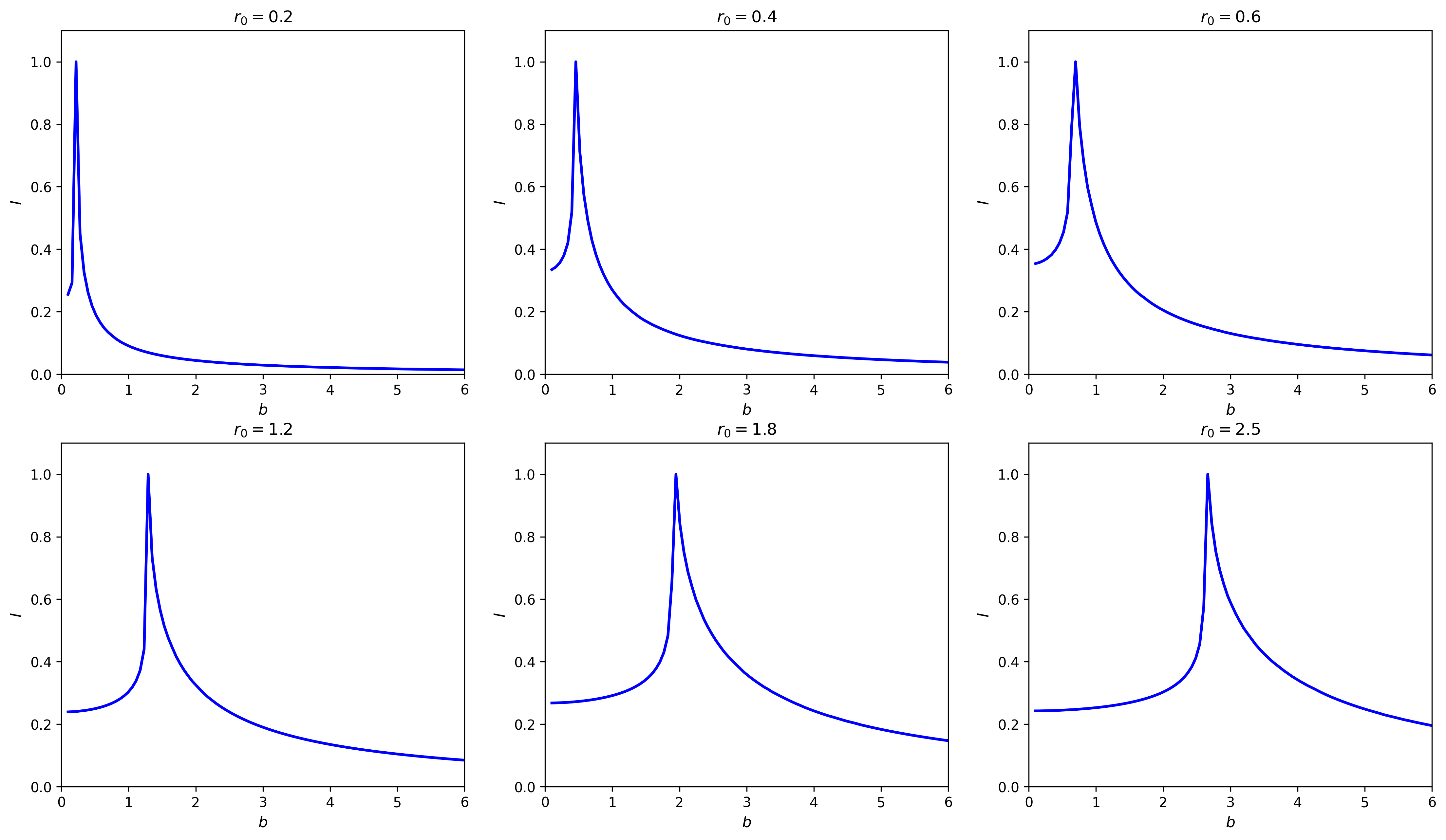}
    \label{fig:intensity}
}
\hfill
\subfigure[Polar shadow maps]{
    \includegraphics[width=0.45\textwidth]{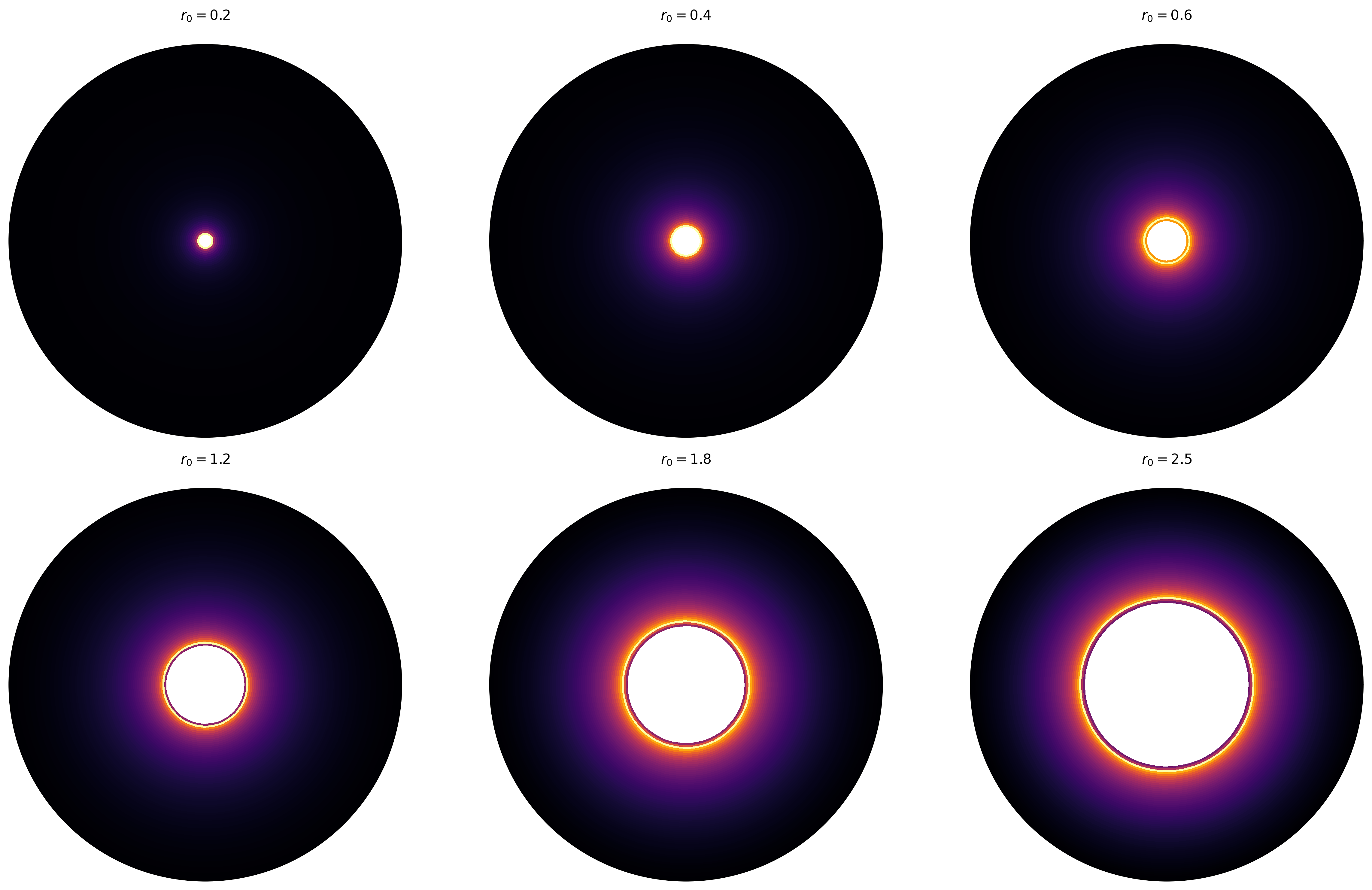}
    \label{fig:shadowmap}
}
\caption{Normalised intensity profiles and polar shadow maps of the static NFW wormhole across six throat radii, $r_0 \in \{0.2, 0.4, 0.6, 1.2, 1.8, 2.5\}$~kpc, using the NGC~2366 halo parameters ($r_s = 1.447$~kpc, $\rho_s = 3.11\times10^{-3}~\text{M}_\odot/\text{pc}^3$). The critical impact parameter scales as $b_\text{ph}\approx1.05\,r_0$ over this range, reaching $b_\text{ph}=2.63$~kpc at $r_0=2.5$~kpc.}
\label{fig:nfw_combined}
\end{figure}

Now we will analyze gravitational lensing as an observational signature of the NFW-supported wormhole, obtained by ray-tracing null geodesics in the metric~\eqref{eq:metric} with the shape
function~\eqref{eq:NFWshape} and redshift function~\eqref{eq:NFW_redshift}. The trajectories follow from the orbit equation~\eqref{eq:orbit_eq}, integrated numerically~\cite{10.1119/1.15620}. To examine how the throat size controls the deflection of light, we integrate the null geodesics for six representative throat radii spanning a wide dynamic range, $r_{0}\in\{0.2,\,0.4,\,0.6,\,1.2,\,1.8,\,2.5\}$~kpc. For each case, $N=100$ rays are launched from $x_{\rm start}=15$~kpc with impact parameters sampled uniformly on $\tilde{b}\in[0.01,8.0]$~kpc. A ray is integrated inward until it reaches a radial turning point, defined by $\dot r=0$, equivalently $r\,e^{-\Phi(r)}=\tilde b$
(Eq.~\eqref{eq:photon_sphere_full}); rays that reach no turning point before crossing the throat $r=r_{0}$ are classified as captured. Each ray is colored as deflected (blue,
$\tilde{b}>b_{\rm ph}$) or captured (black, $\tilde{b}\le b_{\rm ph}$). 
For a deflected ray, the outgoing branch is reconstructed by reflecting the trajectory about its turning point by the symmetry of the orbit equation under $\phi\to2\phi_{\rm turn}-\phi$. For the entire parameter range, the NFW redshift function in \eqref{eq:NFW_redshift} remains small, $|\Phi(r)|\ll1$, with $|\Phi(r_{0})|\lesssim0.08$ at the smallest throat and falling further outward \cite{Böhmer_2007}. Due to this, the condition for an unstable circular photon orbit given by $r\,\Phi'(r)=1$, has no solution anywhere in the exterior region as $r\,\Phi'(r)$ remains of order $|\Phi|\ll1$ throughout the spacetime. Consequently, the geometry has no photon sphere separated from the throat, and hence the throat itself that keeps the boundary between capture and escape \cite{Kuhfittig_2014}. This allows the critical impact parameter to be reduced to
\begin{equation}
b_{\rm ph}=r_{0}\,e^{-\Phi_{\rm NFW}(r_{0})},
\label{eq:bph_throat}
\end{equation}
which yields $b_{\rm ph}\approx1.05\,r_{0}$ across the range studied, the small excess over $r_{0}$ arising entirely from the redshift factor $e^{-\Phi(r_{0})}>1$ and shrinking from $\approx1.08\,r_{0}$ at $r_{0}=0.2$~kpc to $\approx1.05\,r_{0}$ at $r_{0}=2.5$~kpc, where $b_{\rm ph}=2.63$~kpc. Since the parameter $b_{\rm ph}$ determines the angular size of the shadow perceived by a distant observer, the fact that $b_{\rm ph}$ increases in proportion to $r_{0}$, indicating that larger NFW wormholes produce larger, and correspondingly more observable, shadows~\cite{Kuhfittig_2014}.\\
The figure~\ref{fig:raytracing} shows the integrated geodesics for all six throat radii, and the panels convey a common shape that scales with $r_{0}$. Rays with $\tilde{b}\gg b_{\rm ph}$ pass through with merely a tiny deflection from the weak halo potential. As $\tilde{b}$ approaches $b_{\rm ph}^{+}$, we notice that the deflection increases sharply. The rays bend around the throat and then escape towards the positive $x$ direction. This makes a densely packed band before the throat, which is clearly visible from the outside.
Rays with $\tilde{b}<b_{\rm ph}$ end at the throat are captured. It was observed that as we increased the value of $r_{0}$, the throat and its surrounding capture circle began to occupy a larger portion of the visible area, which spans $\pm4$~kpc. Also, at $r_{0}=2.5$~kpc the capture boundary nearly coincides with the throat boundary and fills the plotting window. As a ray grazes the throat of the wormhole, we observe that the deflection starts to diverge as $\tilde{b}\to b_{\rm ph}$. This behavior reflects a logarithmic divergence that has been noted previously in studies of halo-supported wormholes \cite{Kuhfittig_2014}. In the figure, the dashed purple circle indicates the position of the wormhole throat at $r = r_0$, the dotted orange circle representing the photon sphere at $r = r_{\rm ph}$, and the dash-dotted brown circle marking the critical impact parameter $b_{\rm ph}$.
To represent the observable signatures of the NFW wormhole further, we compute the specific intensity $I(\tilde{b})$ as a function of impact parameter for each throat radius using the path-length weighted transfer integral
\begin{equation}
    I(\tilde{b}) = \int e^{3\Phi(r)}\sqrt{\frac{1}{1 - b(r)/r} + \frac{u^2}{(u')^2}}\,
    |u'|\, d\phi,
    \label{eq:intensity}
\end{equation}
where $u = 1/r$ and $u' = du/d\phi$ are considered along the null geodesic. The factor $e^{3\Phi}$ is used to transform the emitted intensity into the observed specific intensity.  This transformation adheres to the Liouville invariant, which states that $I_\nu/\nu^3 = \mathrm{const}$, combined with the static redshift factor given by $g = e^{\Phi}$. The other terms reduce to $dl/r^2$, where
\begin{equation}
    dl = \frac{1}{u^2}\sqrt{\frac{(u')^2}{1 - b(r)/r} + u^2}\,d\phi
\end{equation}
represents the proper path-length element along the geodesic in the equatorial plane. Thus, we can evaluate the integral $\int e^{3\Phi}\,r^{-2}\,dl$, which corresponds to an optically thin emissivity characterized by $j \propto 1/r^2$ per unit proper length, which is the typical radial profile found in spherical accretion models \cite{Kuhfittig_2014}. When considering rays that reach a turning point and head toward the observer, we count them with a factor of two to account for both the inward and outward journeys of the trajectory. Finally, each profile is normalized to its own peak value.
The intensity profiles shown in Fig.~\ref{fig:intensity} exhibit a common structure, such that they peak at $\tilde{b} = b_{\rm ph}$, followed by a smooth tail that gradually declines. As the value of $r_0$ increases, the position of the peak shifts to larger $\tilde{b}$, which aligns with the scaling of $b_{\rm ph}$. This indicates that the angular area of the brightening near the throat expands alongside the throat itself. Further, across all six throat radii, the intensity falls below $20\%$ of its peak within roughly $\Delta\tilde{b} \sim 1$--$2$ kpc of $b_{\rm ph}$. This shows that the brightness enhancement is largely confined to the region surrounding the capture boundary. The shadow maps in Fig.~\ref{fig:shadowmap} give an azimuthally symmetric view of these dynamics. The central disk, which represents the geometric shadow cast by the wormhole, expands alongside $r_0$ in accordance with $b_{\rm ph}$. As $r_0$ increases, the bright annulus around this shadow becomes more defined against the backdrop, since the capture boundary remains at a consistent fractional offset of $e^{-\Phi_{\rm NFW}(r_0)}$ above the throat. Consequently, the radius of the shadow provides a straightforward method for determining the size of the wormhole throat within an NFW dark matter halo.
The shadow radius is the critical impact parameter
\begin{equation}\label{eq:shadow_radius}
    R_s \equiv b_{\rm ph} = r_0\,e^{-\Phi_{\rm NFW}(r_0)},
\end{equation}
which sets the physical radius of the circular shadow for this static, spherically symmetric spacetime. For an observer at a distance $d$, the angular diameter is
\begin{multline}\label{eq:shadow_angle}
    \theta_{\rm sh} = \frac{2 R_s}{d} \quad [\text{rad}]
    = \frac{2 R_s}{d}\,\frac{180\times 3600\times10^{6}}{\pi}~\mu{\rm as}\\
    = 2.0626\times10^{11}\,\frac{R_s}{d}~\mu{\rm as}.
\end{multline}
For the NGC 2366 parameters with $r_0 = 1.0$ kpc, the redshift function gives
$\Phi_{\rm NFW}(r_0) = -0.062$, so $R_s = 1.06$ kpc. At the NGC 2366 distance $d = 3.3$ Mpc, this subtends
\begin{multline}
    \theta_{\rm sh} = \frac{2\times1.06~\text{kpc}}{3.3\times10^{3}~\text{kpc}}
    = 6.4\times10^{-4}~\text{rad} \approx 2.2~\text{arcmin}\\
    \approx 1.3\times10^{8}~\mu{\rm as}.
\end{multline}
At a Galactic-centre distance $d = 8.2$ kpc the same $R_s$ subtends $\theta_{\rm sh} \approx 0.26$ rad $\approx 15^\circ$. These angular sizes are
macroscopic, exceeding the EHT resolution scale $\sim 20~\mu$as by roughly seven orders of magnitude at Mpc distances. The kiloparsec throat scale is set by the dark matter halo rather than by a horizon scale, so the resulting shadow does not correspond to a compact horizon-scale image, and the EHT/VLBI shadow framework developed for black holes does not transfer to these halo-scale geometries. The shadow angular size constrains the astrophysical interpretation: an NFW-supported wormhole with a kiloparsec throat would imprint a brightness deficit on arcminute to degree scales rather than a microarcsecond ring.

\subsubsection{Accretion Disk Analysis of Static NFW Wormhole}
\label{sec:accretion_disk}
\begin{figure*}
    \centering
    \includegraphics[width=16cm,height=9cm]{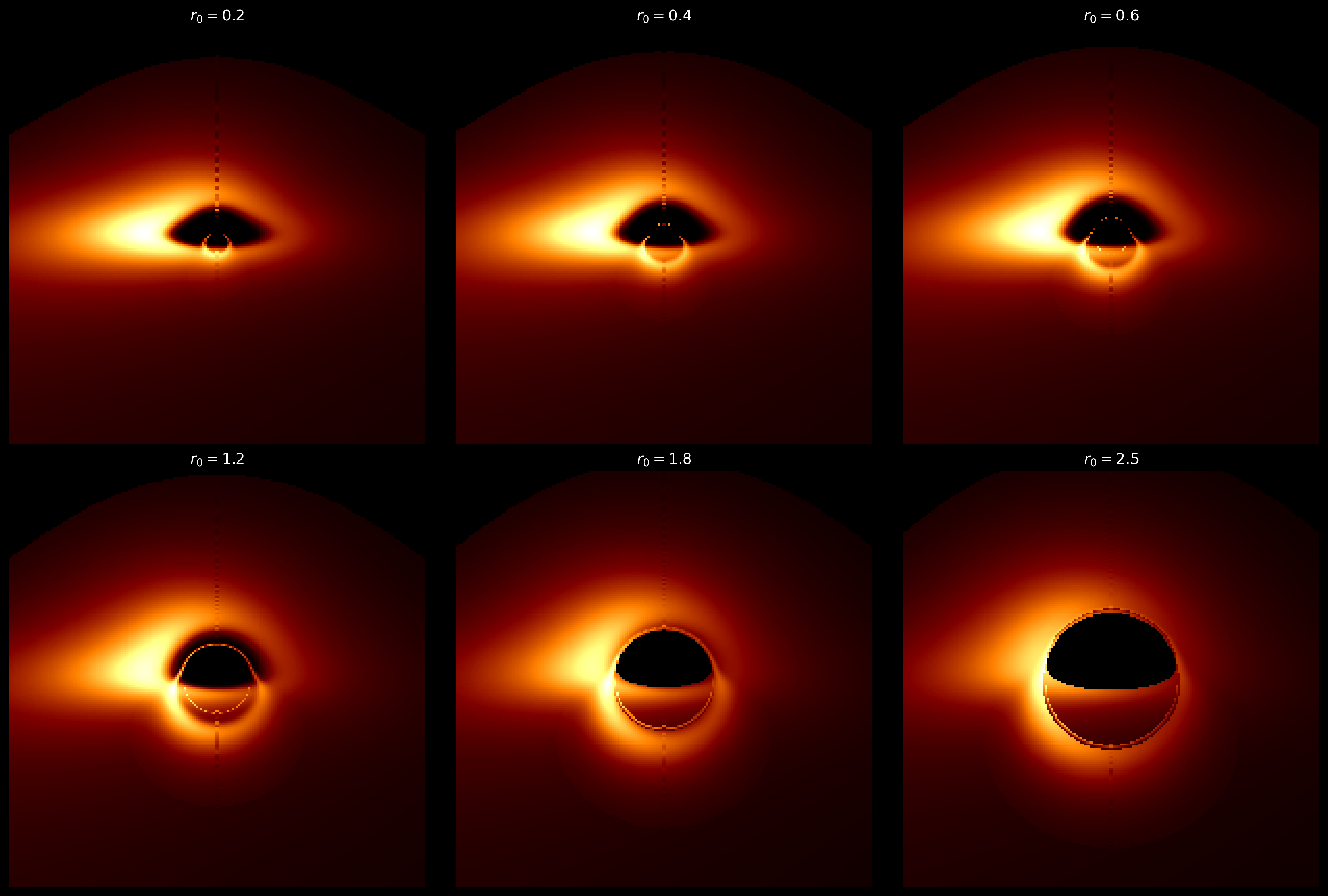}
    \caption{Accretion disc images of the static NFW wormhole for six throat radii, $r_0 \in \{0.2, 0.4, 0.6, 1.2, 1.8, 2.5\}$~kpc, using the NGC~2366 halo parameters ($r_s = 1.447$~kpc, $\rho_s = 3.11\times10^{-3}~\text{M}_\odot/\text{pc}^3$). The observer is placed at $r_\text{obs}=30$~kpc, inclination $\theta_\text{obs}=75^\circ$, with a $260\times260$ pixel screen spanning $\pm12$~kpc. The emissivity profile uses an inner cutoff $r_\text{cut}=2.5$~kpc, and each panel is normalized to its own peak intensity via a square-root transfer function $\sqrt{I/I_\text{max}}$.}
    \label{fig:accretion_disk}
\end{figure*}

Having established the ray-tracing framework and optical properties of the NFW wormhole, we now turn to the physically richer problem of imaging a geometrically thin, optically thin accretion disk surrounding the wormhole throat. This requires solving the full three-dimensional null geodesic equations in the Morris--Thorne spacetime endowed with the NFW metric functions, and accumulating the specific intensity along each photon trajectory as it crosses the equatorial plane. The metric is implemented in the general stationary rotating form with frame-dragging function $w(r)=2J/r^3$; for the static disk studied here the spin is set to $J=0$, so all $w$- and $dw$-dependent terms vanish and the spacetime reduces to the static Morris--Thorne NFW geometry. For null geodesics the Hamiltonian constraint $g^{\mu\nu}p_\mu p_\nu = 0$ gives
\begin{equation}\label{eq:hamiltonian}
    \mathcal{H} = -e^{-2\Phi}\,E^2 + \frac{L^2}{r^2\sin^2\theta} + \frac{p_\theta^2}{r^2} + \left(1 - \frac{b}{r}\right)p_r^2 = 0,
\end{equation}
where $p_r$ and $p_\theta$ are the radial and polar momenta, respectively. Hamilton's equations yield the geodesic system
\begin{align}
    \frac{dr}{d\lambda} &= \left(1 - \frac{b}{r}\right)p_r, \label{eq:dr}\\[4pt]
    \frac{d\theta}{d\lambda} &= \frac{p_\theta}{r^2}, \label{eq:dtheta}\\[4pt]
    \frac{dp_r}{d\lambda} &= -\frac{1}{2}\left[\frac{\partial g^{tt}}{\partial r}\,E^2 + \frac{\partial g^{\varphi\varphi}}{\partial r}\,L^2 + \frac{\partial g^{rr}}{\partial r}\,p_r^2 + \frac{\partial g^{\theta\theta}}{\partial r}\,p_\theta^2\right], \label{eq:dpr}\\[4pt]
    \frac{dp_\theta}{d\lambda} &= -\frac{1}{2}\,\frac{\partial g^{\varphi\varphi}}{\partial \theta}\,L^2 = \frac{L^2\cos\theta}{r^2\sin^3\theta}\,, \label{eq:dptheta}
\end{align}
where the relevant inverse metric derivatives appearing in Eq.~\eqref{eq:dpr} are
\begin{multline}
\frac{\partial g^{tt}}{\partial r} = 2\Phi'(r)\,e^{-2\Phi},\quad  \frac{\partial g^{rr}}{\partial r} = \frac{b(r) - r\,b'(r)}{r^2},\\
    \frac{\partial g^{\theta\theta}}{\partial r} = -\frac{2}{r^3}, \quad 
    \frac{\partial g^{\varphi\varphi}}{\partial r} = -\frac{2}{r^3\sin^2\theta}\,.
\end{multline}
We place a distant observer at radial coordinate $r_{\rm obs} = 30$~kpc, at an inclination angle $\theta_{\rm obs} = 75^\circ$ with respect to the polar axis. The observer's image plane is parameterized by Cartesian screen coordinates $(\alpha, \beta)$, related to the conserved quantities by
\begin{equation}
    L = -\alpha\,\sin\theta_{\rm obs}\,,\qquad p_\theta\big|_{\rm obs} = -\beta\,.
\end{equation}
Each pixel on the screen defines a photon trajectory with energy normalized to $E = 1$. The initial radial momentum $p_r$ is determined from the Hamiltonian constraint~\eqref{eq:hamiltonian}, choosing the ingoing branch ($p_r < 0$) so that rays propagate backward from the observer toward the wormhole
\begin{equation}
    p_r\big|_{\rm obs} = -\sqrt{\frac{-\mathcal{H}_0}{1 - b(r_{\rm obs})/r_{\rm obs}}}\,,
\end{equation}
where $\mathcal{H}_0 \equiv -e^{-2\Phi}E^2 + L^2/(r^2\sin^2\theta) + p_\theta^2/r^2$ collects all non-radial contributions evaluated at the observer's position, so that the constraint $\mathcal{H}=0$ gives $(1-b/r)p_r^2 = -\mathcal{H}_0>0$ for physical rays. We sample a field of view of $\pm12$~kpc on a $260 \times 260$ grid, yielding $6.76\times10^4$ rays.
For each integration step, the derivatives are computed at four intermediate points
\begin{align}
\mathbf{k}_1 &= \mathbf{f}(\mathbf{y}_n) \\
\mathbf{k}_2 &= \mathbf{f}\left(\mathbf{y}_n + \frac{\Delta \lambda}{2}\mathbf{k}_1\right) \\
\mathbf{k}_3 &= \mathbf{f}\left(\mathbf{y}_n + \frac{\Delta \lambda}{2}\mathbf{k}_2\right) \\
\mathbf{k}_4 &= \mathbf{f}\left(\mathbf{y}_n + \Delta \lambda \mathbf{k}_3\right)
\end{align}
The state vector is then updated as
\begin{equation}
\mathbf{y}_{n+1} = \mathbf{y}_n + \frac{\Delta \lambda}{6}\left(\mathbf{k}_1 + 2\mathbf{k}_2 + 2\mathbf{k}_3 + \mathbf{k}_4\right)
\end{equation}
with affine step size $\Delta\lambda = 0.05$ up to a maximum of 1500 steps. Rays are considered terminated when they either come within $1.02\,r_0$ of the throat or move beyond $35$ kpc. Along each trajectory, we identify when they cross the equatorial plane by observing sign changes in $\theta(\lambda) - \pi/2$ between consecutive integration steps. When such a crossing is located, the radial coordinate at that point $r_\times$ is obtained through linear interpolation,
\begin{equation}
    r_\times = r_{\rm old} + \frac{\pi/2 - \theta_{\rm old}}{\theta_{\rm new} - \theta_{\rm old}}\left(r_{\rm new} - r_{\rm old}\right).
\end{equation}
Only crossings that fall in the validity window $1.1\,r_0 < r_\times < 20$~kpc are allowed to contribute emission, while any crossing outside this range is assigned zero intensity. The emissivity profile of the disk is supposed to follow a power-law envelope with an inner cutoff
\begin{equation}\label{eq:emissivity}
    j_\nu^{\rm em}(r_\times) = \frac{1}{r_\times^3}\,\exp\!\left[-\left(\frac{r_{\rm cut}}{r_\times}\right)^{\!4}\right],
\end{equation}
where the factor of $r_\times^{-3}$ replicates the radial decline predicted by standard thin-disk theory,~\cite{Frolov1998}. Additionally, the exponential suppression removes emission from radii close to or inside the throat, ensuring numerical regularity.\\
The inner cutoff scale $r_{\rm cut}=2.5$~kpc is a constant that applies uniformly across all throat radii during the sweep. This effectively determines the inner boundary of the emitting region. Because the weak NFW potential ($|\Phi_{\rm NFW}|\lesssim 0.05$) admits no innermost stable circular orbit and no photon sphere, the cutoff is not tied to a marginally stable orbit; a self-consistent disk model with a dynamically determined inner edge is left for future work.\\
To account for the relativistic Doppler effect, we compute the angular velocity of the disk material from the NFW metric functions. The code evaluates the general circular-orbit frequency for the stationary rotating metric,
\begin{equation}\label{eq:Omega_disk}
    \Omega(r_\times) = w + \tfrac{1}{2}r_\times w' + \sqrt{\tfrac{1}{4}r_\times^2 w'^2 + \frac{e^{2\Phi(r_\times)}\,\Phi'(r_\times)}{r_\times}}\,.
\end{equation}
With $J=0$ the frame-dragging terms $w=w'=0$ vanish and this reduces to the static Keplerian form $\Omega(r_\times) = e^{\Phi(r_\times)}\sqrt{\Phi'(r_\times)/r_\times}$. In the weak-field limit $|\Phi|\ll 1$ adopted here ($|\Phi_{\rm NFW}|\lesssim 0.05$), the factor $e^\Phi\approx 1$ and $\Omega\approx\sqrt{\Phi'/r}$; the full form is nevertheless retained for exactness.\\
The complete relativistic Doppler factor, encoding both gravitational redshift and kinematic frequency shift between the co-rotating emitter frame and the distant observer, is computed from the emitter four-velocity normalization $u^t = 1/\sqrt{e^{2\Phi}-r^2(\Omega-w)^2}$ as
\begin{align}\label{eq:doppler_factor}
g &= \frac{1}{u^t\,\bigl(1 - \Omega(r_\times)\, L\bigr)} \\
u^t &= \frac{1}{\sqrt{e^{2\Phi(r_\times)} - r_\times^2\,(\Omega(r_\times)-w)^2}}
\end{align}
which for $w=0$ reduces to $g = \sqrt{e^{2\Phi(r_\times)} - r_\times^2\,\Omega^2(r_\times)}\,/\,(1 - \Omega(r_\times)L)$, where the numerator encodes the gravitational redshift and transverse Doppler effect and the denominator $1-\Omega L$ encodes the classical first-order Doppler shift due to the orbital velocity~\cite{Luminet1979}. The quantity $g$ is kept clamped to the range $[0.1,\,3.0]$ for numerical stability. Each equatorial crossing contributes an intensity increment
\begin{equation}\label{eq:intensity_accum}
    \Delta I = j_\nu^{\rm em}(r_\times)\,g^4
\end{equation}
to the pixel, where the $g^4$ factor arises from the bolometric transformation of specific intensity: the specific intensity transforms as $I_{\nu,\rm obs} = g^3 I_{\nu,\rm em}$, and integrating over frequency gives an additional factor $\nu_{\rm obs}/\nu_{\rm em} = g$, yielding the total (bolometric) intensity scaling $I_{\rm obs} = g^4 I_{\rm em}$. The total observed intensity at each pixel is based on the sum over all equatorial crossings along the corresponding geodesic. This means that if photons orbit multiple times near the photon sphere, they can produce higher-order images of the disk.\\
Figure~\ref{fig:accretion_disk} shows the accretion disk for the NFW wormhole across six throat radii, $r_0 \in \{0.2,\,0.4,\,0.6,\,1.2,\,1.8,\,2.5\}$~kpc expressed as a $2\times3$ panel. In each panel, the image intensity is depicted using a square-root transfer function, specifically $\sqrt{I/I_{\max}}$, which normalizes each panel to its peak intensity. This approach helps us to enhance the visibility of faint features. There are several characteristic signatures that are evident. A bright and narrow arc appears just outside the throat, which is formed by photons from the far side of the disk whose trajectories are strongly bent near the throat and loop around before reaching the observer. Since the weak NFW potential has no photon sphere, this arc is the gravitationally lensed secondary image of the disk rather than a photon-sphere photon ring. Secondly, the disk shows a clear left-to-right brightness difference from the relativistic Doppler effect. Also, the side moving toward the observer that appears brighter from blueshift ($g > 1$), the receding side looks dimmer from redshift ($g < 1$), and the steep $g^4$ scaling makes this gap even sharper. Thirdly, the interior appears dark because no emission comes from inside $r_{\rm cut}$, and rays with small impact parameters are captured at the throat. The size of this dark region depends on whichever is larger, the capture boundary $b_{\rm ph}$ or the lensed image of the inner emission edge. This alignment is consistent with the scaling of $b_{\rm ph}$ as discussed in Sec .~\ref{sec:analysis}. Finally, a faint secondary image appears as a thin band within the main one, formed by photons that cross the equatorial plane again after passing close to the throat. This is significant because the NFW wormhole derives its gravity from a galactic dark matter halo, yet it still exhibits strong-field lensing features. Because the NFW potential remains weak, i.e., ($|\Phi| \ll 1$), it keeps the dark-region boundary close to the geometric throat as shown in Sec.~\ref{sec:analysis}. In contrast, the Doppler asymmetry provides an additional handle sensitive to the disk kinematics and the observer inclination. 

\subsubsection{Ray Tracing Analysis of Rotating NFW Wormhole}
\label{sec:rot_nfw}
\begin{figure*}
    \centering
    \includegraphics[width=16cm,height=8cm]{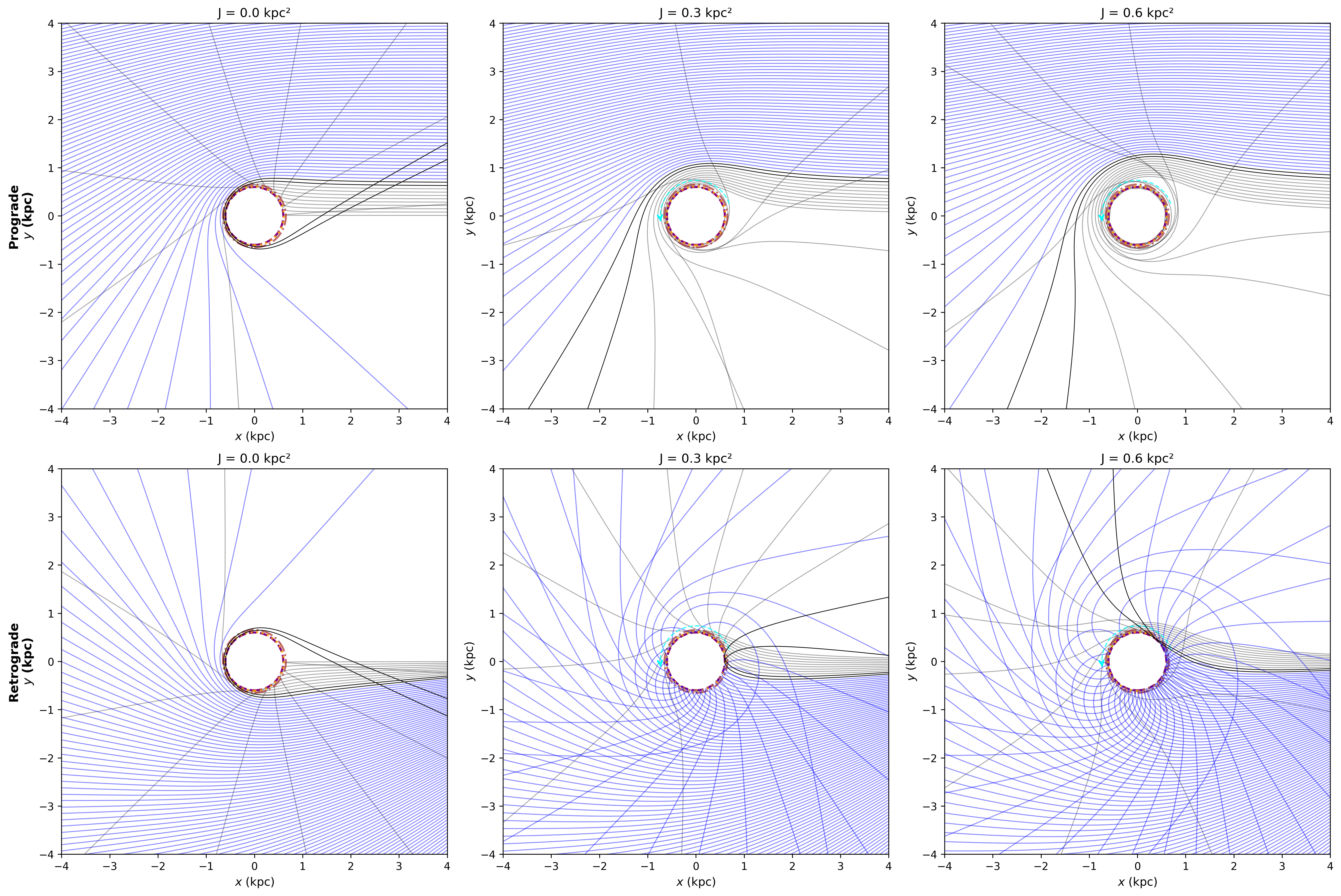}
    \caption{Ray-traced equatorial null geodesics of the slowly rotating NFW wormhole with throat radius $r_0=0.6$~kpc, for angular momenta $J\in\{0,\,0.3,\,0.6\}$~kpc$^2$ (columns) and prograde/retrograde impact parameters $\beta>0$/$\beta<0$ (rows), using the NGC~2366 halo parameters ($r_s=1.447$~kpc, $\rho_s=3.11\times10^{-3}~\text{M}_\odot/\text{pc}^3$). Rays are launched from $x_\text{start}=15$~kpc with $\beta\in[0.01,8.0]$~kpc, $N=120$ rays per panel. Dashed circles mark the throat $r_0$, the formal photon orbit radius $r_\text{ph}$, and $b_\text{ph}^{(0)}$.}
    \label{fig:rot_rt_r06}
\end{figure*}

\begin{figure*}
    \centering
    \includegraphics[width=16cm,height=8cm]{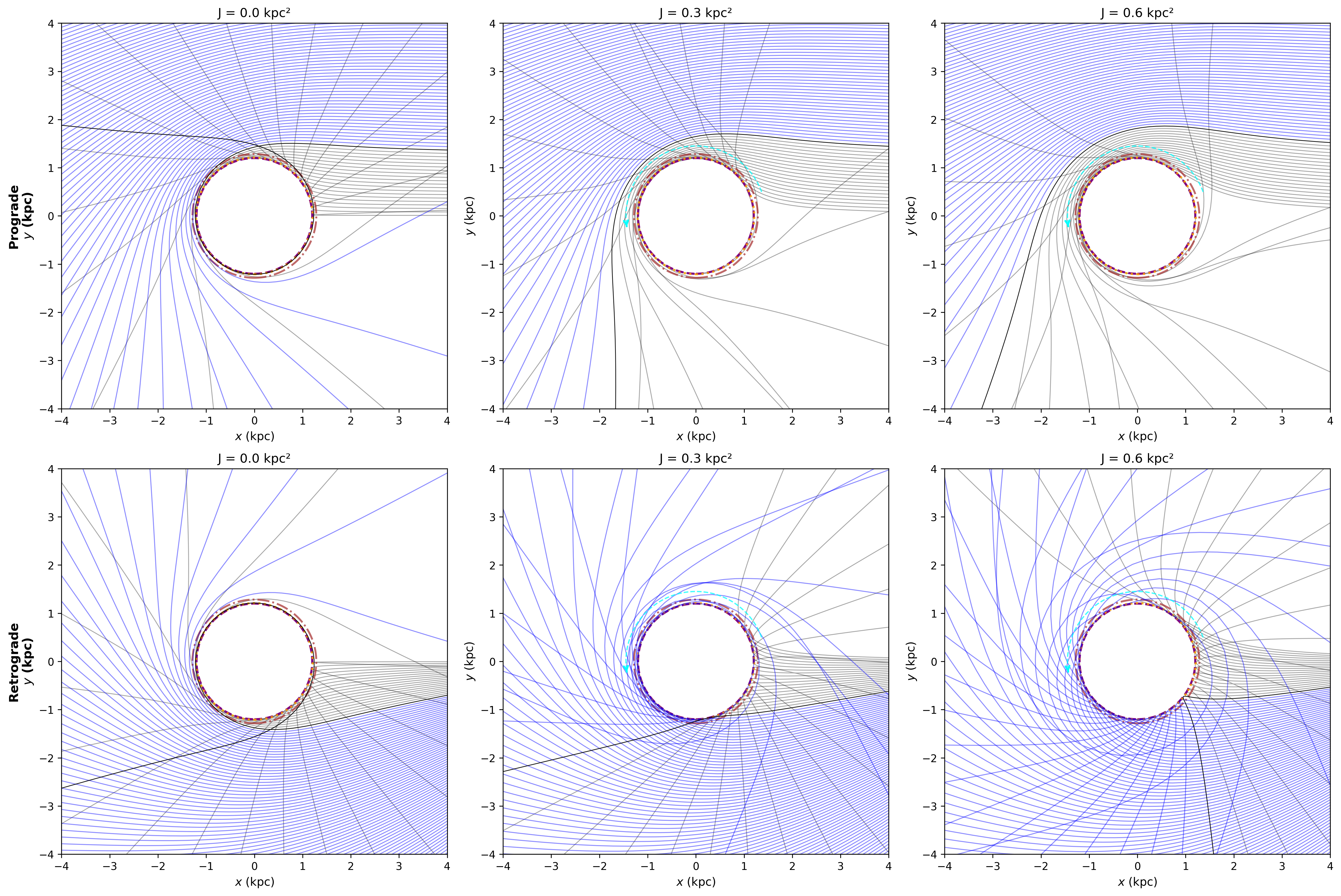}
    \caption{Ray-traced equatorial null geodesics of the slowly rotating NFW wormhole with throat radius $r_0=1.2$~kpc, for angular momenta $J\in\{0,\,0.3,\,0.6\}$~kpc$^2$ (columns) and prograde/retrograde impact parameters $\beta>0$/$\beta<0$ (rows), using the NGC~2366 halo parameters ($r_s=1.447$~kpc, $\rho_s=3.11\times10^{-3}~\text{M}_\odot/\text{pc}^3$). Rays are launched from $x_\text{start}=15$~kpc with $\beta\in[0.01,8.0]$~kpc, $N=120$ rays per panel. Dashed circles mark the throat $r_0$, the formal photon orbit radius $r_\text{ph}$, and $b_\text{ph}^{(0)}$.}
    \label{fig:rot_rt_r12}
\end{figure*}

The slow-rotation extension is obtained by inserting $\omega(r) \approx2J/r^{3}$~\cite{Teo1998,Hartle1967,HartleThorne1968} into the Teo metric (\ref{eq:teo_metric}) with the NFW shape function $b_{\rm NFW}(r)$ derived in Eq. \eqref{eq:NFWshape}. Slow rotation affects frame-dragging and Lense--Thirring precession near the throat~\cite{CiufoliniPavlis2004,Hartle1967}, while the radial structure remains governed by $b_{\rm NFW}(r)$. The equatorial Lense--Thirring precession frequency, to first order in $\omega$~\cite{Hartle1967,HartleThorne1968,CiufoliniPavlis2004}, is
\begin{equation}
    \Omega_{\rm LT}(r) = -\frac{1}{2}\,\frac{d\omega}{dr}
    = \frac{3J}{r^{4}},
    \label{eq:lt_freq}
\end{equation}
which grows steeply toward small radii, so the cuspy NFW geometry produces stronger near-throat frame-dragging than cored profiles at comparable throat radii~\cite{NedkovaTinchevYazadjiev2013,Shaikh2018}.
In the equatorial plane ($\theta = \pi/2$) with $K=1$, the metric~\eqref{eq:teo_metric} reduces to~\cite{Teo1998,Carter1968a}
\begin{equation}
    ds^{2} = -e^{2\Phi(r)}\,dt^{2}
    + \frac{dr^{2}}{1 - b(r)/r}
    + r^{2}\,\bigl(d\phi - \omega(r)\,dt\bigr)^{2}.
    \label{eq:nfw_rotating_metric}
\end{equation}
The two Killing symmetries~\cite{Carter1968a,Carter1968b,Wald1984} yield the conserved energy $E$ and angular momentum $L$, which to first order in $\omega$ read
\begin{align}
    E &= e^{2\Phi(r)}\,\dot{t} + r^{2}\,\omega(r)\,\dot{\phi}, \label{eq:conserved_E} \\
    L &= r^{2}\!\left(\dot{\phi} - \omega(r)\,\dot{t}\right), \label{eq:conserved_L}
\end{align}
giving
\begin{equation}
    \dot{t} = \frac{E - \omega(r)\,L}{e^{2\Phi(r)}}, \qquad
    \dot{\phi} = \frac{L}{r^{2}} + \omega(r)\,\frac{E}{e^{2\Phi(r)}}.
    \label{eq:tdot_phidot}
\end{equation}
The null condition $ds^{2}=0$ gives the exact radial equation
\begin{equation}
    \dot{r}^{2} = \left(1 - \frac{b(r)}{r}\right)
    \!\left[\frac{\bigl(E - \omega(r)\,L\bigr)^{2}}{e^{2\Phi(r)}}
    - \frac{L^{2}}{r^{2}}\right],
    \label{eq:radial_eq}
\end{equation}
defining $V_{\rm eff}(r) = -\dot{r}^{2}$, so that $\dot{r}^{2} + V_{\rm eff}(r) = 0$ and local maxima of $V_{\rm eff}$ correspond to unstable circular photon orbits where they exist~\cite{Synge1966,ClaudelVirbhadraEllis2001,Perlick2004}. Introducing the impact parameter $\beta = L/E$ and linearising in $\omega$, the orbit equation becomes~\cite{Teo1998,Bozza2002}
\begin{equation}
    \frac{d\phi}{dr}
    = \frac{\dfrac{\beta}{r^{2}} + \dfrac{\omega(r)}{e^{2\Phi(r)}}}
      {\sqrt{\left(1 - \dfrac{b(r)}{r}\right)
      \!\left(e^{-2\Phi(r)} - \dfrac{\beta^{2}}{r^{2}}\right)}},
    \label{eq:orbit_eq_rotating}
\end{equation}
and the total deflection angle is
\begin{equation}
    \hat{\alpha} = 2\int_{r_{\rm min}}^{\infty}\frac{d\phi}{dr}\,dr - \pi,
    \label{eq:deflection_angle}
\end{equation}
with $r_{\rm min}$ the radial turning point~\cite{VirbhadraEllis2000,Bozza2002,BozzaTsukamoto2009}. Co-rotating photons ($\beta>0$) experience slightly reduced deflection and counter-rotating photons ($\beta<0$) enhanced deflection relative to the static case~\cite{NedkovaTinchevYazadjiev2013,Shaikh2018,Jusufi2018}.
A circular null orbit satisfies $\dot{r}=0$ and $\ddot{r}=0$, giving
\begin{equation}
    \frac{\bigl(E - \omega L\bigr)^{2}}{e^{2\Phi}} - \frac{L^{2}}{r^{2}} = 0,
    \qquad
    \frac{d}{dr}\!\left[\frac{\bigl(E - \omega L\bigr)^{2}}{e^{2\Phi}}
    - \frac{L^{2}}{r^{2}}\right] = 0.
    \label{eq:photon_sphere_cond}
\end{equation}
A real solution $r_{\rm ph}$ exists only when the static condition $r\,\Phi'(r)=1$ admits a root~\cite{Synge1966,ClaudelVirbhadraEllis2001,Perlick2004}. For the weak NFW redshift function \eqref{eq:NFW_redshift}, $|\Phi|\lesssim 0.05$ and $r\Phi'\ll 1$ throughout, so no detached photon sphere forms and the capture boundary coincides with the throat~\cite{Kuhfittig_2014,Rahaman_2014_Central}; the expressions below are exact for redshift profiles or parameter regimes in which a photon sphere is present, and reduce to throat-set quantities in the present NFW case. Expanding Eq.~\eqref{eq:photon_sphere_cond} to first order in $\omega$, the co-rotating ($-$) and counter-rotating ($+$) critical impact parameters are~\cite{Teo1998,Hioki2009,Shaikh2018}
\begin{equation}
    b_{\pm}(r_{\rm ph})
    \approx \frac{r_{\rm ph}}{e^{\Phi(r_{\rm ph})}}
    \!\left(1 \pm \frac{\omega(r_{\rm ph})\,r_{\rm ph}}{e^{\Phi(r_{\rm ph})}}\right),
    \label{eq:bpm}
\end{equation}
reducing in the static limit $\omega\to 0$ to $b_{\rm ph}^{(0)} = r_{\rm ph}\,e^{-\Phi(r_{\rm ph})}$. Defining $K \equiv b_{\rm ph}^{(0)}$, the linearised splitting reads $b_{\pm} = K(1 \pm K\omega(r_{\rm ph}))$; $b_{-}$ is the co-rotating (inner) edge and $b_{+}$ the counter-rotating (outer) edge of the shadow. The rotational shift of the photon orbit radius is
\begin{equation}
    r_{\rm ph} \approx r_{\rm ph}^{(0)} + \delta r_{\rm ph}, \qquad
    \delta r_{\rm ph}
    = \frac{2\,r_{\rm ph}^{(0)\,3}\,\omega\!\left(r_{\rm ph}^{(0)}\right)}
           {2 - r_{\rm ph}^{(0)}\,\Phi'\!\left(r_{\rm ph}^{(0)}\right)}.
    \label{eq:rph_shift}
\end{equation}
For an illustrative inverse-radius redshift model $\Phi(r)=-a/r$ (so $\Phi'=a/r^{2}$), Eq.~\eqref{eq:rph_shift} with $\omega=2J/r^{3}$ gives $\delta r_{\rm ph} \approx 4J/(2 - a/r_{\rm ph}^{(0)})$; for the NFW profile $\Phi'$ must be taken from $\Phi_{\rm NFW}$. The differential shadow radius follows from Eq.~\eqref{eq:bpm}~\cite{Hioki2009,Konoplya2018,Tsupko2017},
\begin{equation}
    \Delta b_{\rm ph} \equiv b_{+}-b_{-}
    = \frac{2\,\omega(r_{\rm ph})\,r_{\rm ph}^{2}}{e^{2\Phi(r_{\rm ph})}}
    \approx \frac{4J}{r_{\rm ph}}.
    \label{eq:delta_bph}
\end{equation}
A measurement of $\Delta b_{\rm ph}$, together with a known photon orbit radius, yields the wormhole angular momentum $J$ independently of the halo profile parameters~\cite{Hioki2009,Bambi2013}. The factor $e^{-\Phi(r_{\rm ph})}$ in Eq.~\eqref{eq:bpm} sets the overall shadow scale, and the frame-dragging term introduces the splitting $\Delta b_{\rm ph}$ that makes the rotating shadow asymmetric~\cite{Bardeen1973,FalckeMeliaAgol2000,EHT2019}.\\
We visualize the equatorial geodesics of the rotating wormhole defined by Eqs.~(\ref{eq:nfw_rotating_metric}--\ref{eq:tdot_phidot}), integrating with a fourth-order Runge--Kutta scheme. Each ray starts at $x_{\rm start}=15$~kpc with impact parameter $\beta=L/E$ and is integrated inward until it reaches the throat ($r = r_{0}+10^{-7}$) or passes a radial turning point and escapes.
Here we used the first-order system for integration in Eq.~\eqref{eq:radial_eq} and expressed it in terms of $\beta$, along with Eq.~\eqref{eq:tdot_phidot}. The angular velocity picks up an extra term $\omega\,e^{-2\Phi}$ for prograde rays ($\beta>0$) and loses this term for retrograde rays ($\beta<0$). For rays that reach a turning point, we reflect the incoming trajectory about the turning radius to determine the outgoing path. This method is exact in the static limit and accurate to first order in $\omega$ \cite{Teo1998,Bambi2013}. We assume $\beta\in[0.01,8.0]$~kpc with $N=120$ rays per panel and demonstrate trajectories for two throat radii, $r_{0}=0.6$~kpc (Fig.~\ref{fig:rot_rt_r06}) and $r_{0}=1.2$~kpc (Fig.~\ref{fig:rot_rt_r12}). Each figure contains a $2\times 3$ grid with prograde and retrograde rays depicted in separate rows, along with three angular momentum values, $J\in\{0,\,0.3,\,0.6\}$~kpc$^{2}$. In this figure, the blue curves represent the deflected rays ($\beta>b_{\rm ph}$), the faint black curves illustrate the captured rays ($\beta<b_{\rm ph}$), and rays close to the critical value, i.e., $|\beta-b_{\rm ph}|<0.05$~kpc, are depicted as thin black curves. Additionally, the dashed circles denote the throat $r_{0}$, the formal photon orbit radius $r_{\rm ph}$, and $b_{\rm ph}^{(0)}$, which nearly coincide in the weak NFW regime. Lastly, the cyan color arc indicates the sense $\omega(r)>0$.
For $J=0$ the prograde and retrograde panels are mirror images across the $x$-axis, as required by the reflection symmetry $\phi\to-\phi$ of the static spacetime; near-critical rays wind around the throat before escape or capture, reproducing the static lensing pattern of Sec. \ref{subsec1}~\cite{Kuhfittig_2014,Bambi2013}. For $J>0$ this row symmetry is broken: prograde near-critical rays wind in the same sense as $\omega(r)>0$ and execute additional angular sweep, while retrograde rays receive contributions of opposite sign in $\dot\phi$, sweep through a wider azimuthal range, and remain trapped in long windings near the throat, visible as the denser spirals in the lower half-planes of the retrograde panels~\cite{NedkovaTinchevYazadjiev2013,Shaikh2018,Jusufi2018,Tsupko2017}. The captured fractions in the two rows separate monotonically with $J$, consistent with the splitting $\Delta b_{\rm ph}\approx 4J/r_{\rm ph}$ from Eq.~\eqref{eq:delta_bph}~\cite{Hioki2009}. We do not extract $\Delta b_{\rm ph}$ numerically from these panels because the formal $r_{\rm ph}$ used in the splitting formula coincides with the throat in the absence of a true photon sphere, so the absolute critical parameter is set by the throat-grazing condition $b_{\pm}\to r_{0}\,e^{-\Phi(r_{0})}$ rather than by an unstable circular orbit~\cite{Kuhfittig_2014,Rahaman_2014_Central}. Comparing the two figures, the dashed circles dilate in proportion to $r_{0}$ and the captured/near-critical band thickens accordingly; the relative rotational asymmetry $\Delta b_{\rm ph}/b_{\rm ph}\propto J/r_{\rm ph}^{2}e^{-\Phi}$ is smaller for $r_{0}=1.2$~kpc than for $r_{0}=0.6$~kpc at fixed $J$, so the prograde--retrograde contrast is more pronounced in Fig.~\ref{fig:rot_rt_r06}, the geometric manifestation of $\Omega_{\rm LT}=3J/r^{4}$ steepening at small radii~\cite{Hartle1967,CiufoliniPavlis2004}.

\subsubsection{Intensity profiles and shadow maps}
\begin{figure}[!ht]
    \centering
    \subfigure[$r_0=1.5$ kpc]{
    \includegraphics[width=0.48\textwidth]{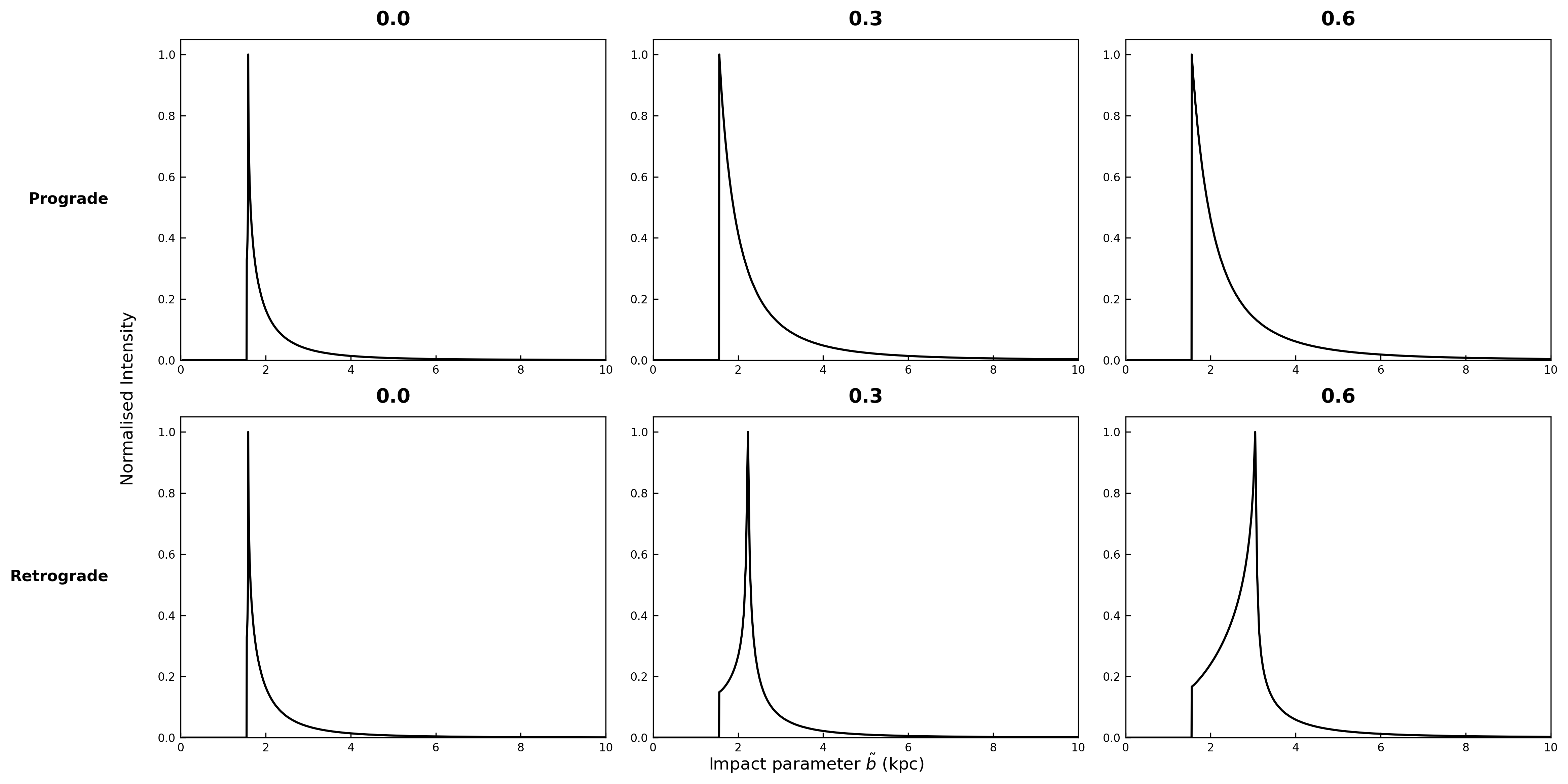}
    \label{fig:intensity_r15}
}
\hfill
\subfigure[$r_0=2.5$ kpc]{
    \includegraphics[width=0.48\textwidth]{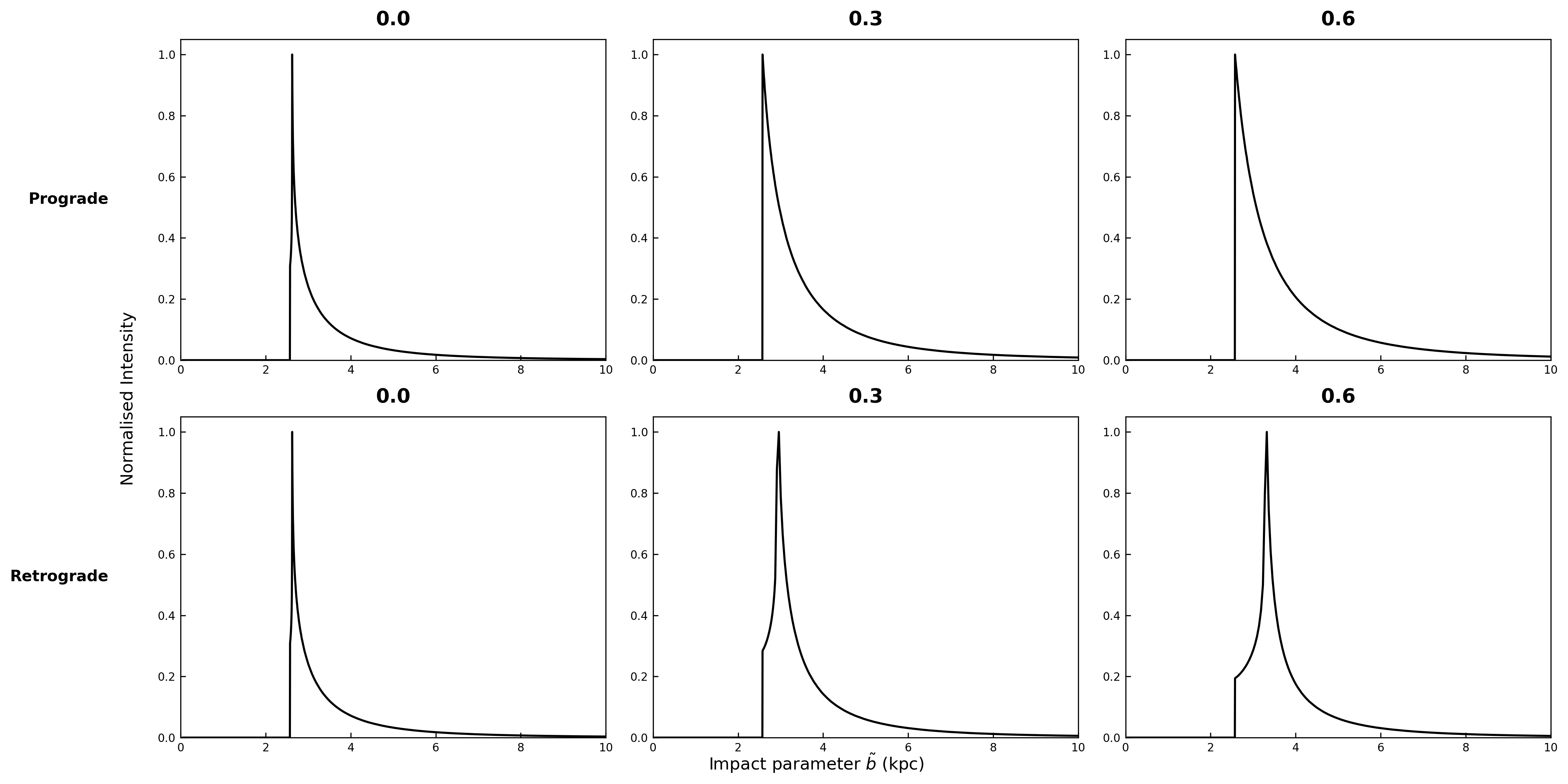}
    \label{fig:intensity_r25}
}
\caption{Normalised equatorial intensity profiles $I(\tilde{b})$ for the slowly rotating NFW wormhole at $J\in\{0,\,0.3,\,0.6\}$~kpc$^2$, for throat radii $r_0=1.5$~kpc (left, static peak $b_\text{ph}^{(0)}\approx1.6$~kpc) and $r_0=2.5$~kpc (right, static peak $b_\text{ph}^{(0)}\approx2.6$~kpc), using the NGC~2366 halo parameters ($r_s=1.447$~kpc, $\rho_s=3.11\times10^{-3}~\text{M}_\odot/\text{pc}^3$). Rays are launched from $r_\text{obs}=50$~kpc; upper rows show prograde ($\beta>0$) and lower rows retrograde ($\beta<0$) trajectories.}
\label{fig:intensity_combined}
\end{figure}

We now compute the equatorial intensity profile $I(\tilde{b})$ and the corresponding two-dimensional shadow map for two throat radii, $r_{0}=1.5$~kpc and $r_{0}=2.5$~kpc, at $J\in\{0,\,0.3,\,0.6\}\,\mathrm{kpc}^{2}$, using the first-order geodesic system of Eqs.~(\ref{eq:radial_eq}--\ref{eq:orbit_eq_rotating}), an optically thin emissivity model, and the splitting expressions Eqs.~\eqref{eq:bpm} and~\eqref{eq:delta_bph}~\cite{Teo1998,Bambi2013,Synge1966,FalckeMeliaAgol2000,Younsi2016}. Each ray is launched from the observer at $r_{\rm obs}=50$~kpc with impact parameter $\beta=L/E$ and integrated inward until either the throat or a radial turning point is reached. The observed specific intensity is accumulated along each null geodesic following the optically thin transport relation~\cite{Cunningham1973,Luminet1979,Bambi2013}
\begin{equation}
    I(\tilde{b})\;=\;\int e^{3\Phi(r)}\,\sqrt{\frac{1}{1-b_{\rm NFW}(r)/r}+\frac{u^{2}}{(u')^{2}}}\,j_{\nu}(r)\,|u'|\,d\phi,
    \label{eq:intensity_rot}
\end{equation}
with $u=1/r$ and $u'=du/d\phi$ reconstructed from the integrated trajectory. The factor $e^{3\Phi}$ encodes the Liouville invariant $I_{\nu}/\nu^{3}=\mathrm{const}$~\cite{MisnerThorneWheeler1973,Cunningham1973}, the emissivity is $j_{\nu}(r)\propto r^{-2}$ as appropriate for spherical accretion~\cite{Bambi2013,Younsi2016,FalckeMeliaAgol2000}, and rays that reach a turning point contribute with a factor of two for the inward and outward branches. In Fig.\ref{fig:intensity_r15}, upper row corresponds to prograde ($\beta>0$) while lower row corresponds to retrograde ($\beta<0$). The static peak sits at $b_{\rm ph}^{(0)}=r_{0}\,e^{-\Phi(r_{0})}\approx 1.6$~kpc. The retrograde peak shifts outward with $J$ to $b_{+}\approx 2.2$~kpc at $J=0.3$ and to $b_{+}\approx 3.0$~kpc at $J=0.6$, in quantitative agreement with Eq.~\eqref{eq:bpm}. NFW parameters: $R_{s}=1.447$~kpc, $\rho_{s}=3.11\times10^{-3}$. As in Fig.~\ref{fig:intensity_r15}, for $r_{0}=2.5$~kpc. The static peak is at $b_{\rm ph}^{(0)}\approx 2.6$~kpc; the retrograde peak shifts to $b_{+}\approx 3.0$~kpc at $J=0.3$ and to $b_{+}\approx 3.5$~kpc at $J=0.6$. The relative displacement $\Delta b_{\rm ph}/b_{\rm ph}^{(0)}\sim K\omega(r_{\rm ph})$ is smaller than in Fig.~\ref{fig:intensity_r15} because $K\omega\propto J/r_{\rm ph}^{2}$ decreases with throat size at fixed $J$.
For both throat radii, the static profiles exhibit a single sharp peak at $b_{\rm ph}^{(0)}$ followed by a monotonically decreasing tail, the standard logarithmically divergent path-length accumulation feature near a capture boundary in optically thin transport~\cite{Luminet1979,Bambi2013,GrallaHolzWald2019}. The peak position scales linearly with $r_{0}$ via $b_{\rm ph}^{(0)}=r_{0}\,e^{-\Phi(r_{0})}$, confirming the geometric character of the boundary in the weak-field NFW regime~\cite{Kuhfittig_2014,Rahaman_2014_Central}. For $J>0$ the retrograde branch displays a clearly resolved displacement of the peak to $b_{+}=K(1+K\omega)$, growing approximately linearly in $J$; at $r_{0}=1.5$~kpc and $J=0.6\,\mathrm{kpc}^{2}$ the peak reaches $b\approx 3.0$~kpc, corresponding to $K\omega\approx 0.9$, at the edge of the linearised expansion~\cite{Teo1998,Hartle1967,HartleThorne1968}. The prograde branch is constrained from below by the integration cutoff at $0.98\,b_{\rm ph}^{(0)}$, so the analytic prograde shift $b_{-}=K(1-K\omega)$ is encoded in the asymmetric shadow maps below rather than as a separate peak in $I(\tilde{b})$~\cite{NedkovaTinchevYazadjiev2013,Shaikh2018}.\\
The two-dimensional shadow maps are constructed by tiling the normalized intensity profile along the azimuthal coordinate $\theta$ and masking the interior of the dipolar capture boundary~\cite{Hioki2009,Tsupko2017}
\begin{equation}
    b_{\rm cut}(\theta) = \tfrac{1}{2}(b_{+}+b_{-}) + \tfrac{1}{2}(b_{+}-b_{-})\,\cos(\theta+\phi_{\rm or}),
    \label{eq:bcut_dipole}
\end{equation}
with $\phi_{\rm or}=0$ for the prograde-rendered panel(top row) and $\phi_{\rm or}=\pi$ for the retrograde-rendered panel(bottom row). Equation~\eqref{eq:bcut_dipole} traces a circle of radius $\tfrac{1}{2}(b_{+}+b_{-})$ displaced from the image-plane origin by $\tfrac{1}{2}\Delta b_{\rm ph}$ along the rotation axis, the slow-rotation analogue of the laterally offset Kerr shadow boundary~\cite{Bardeen1973,FalckeMeliaAgol2000,Hioki2009,EHT2019,EHT2022}.\\
\begin{figure}[!ht]
    \centering
    \subfigure[$r_0=1.5$ kpc]{
    \includegraphics[width=0.45\textwidth]{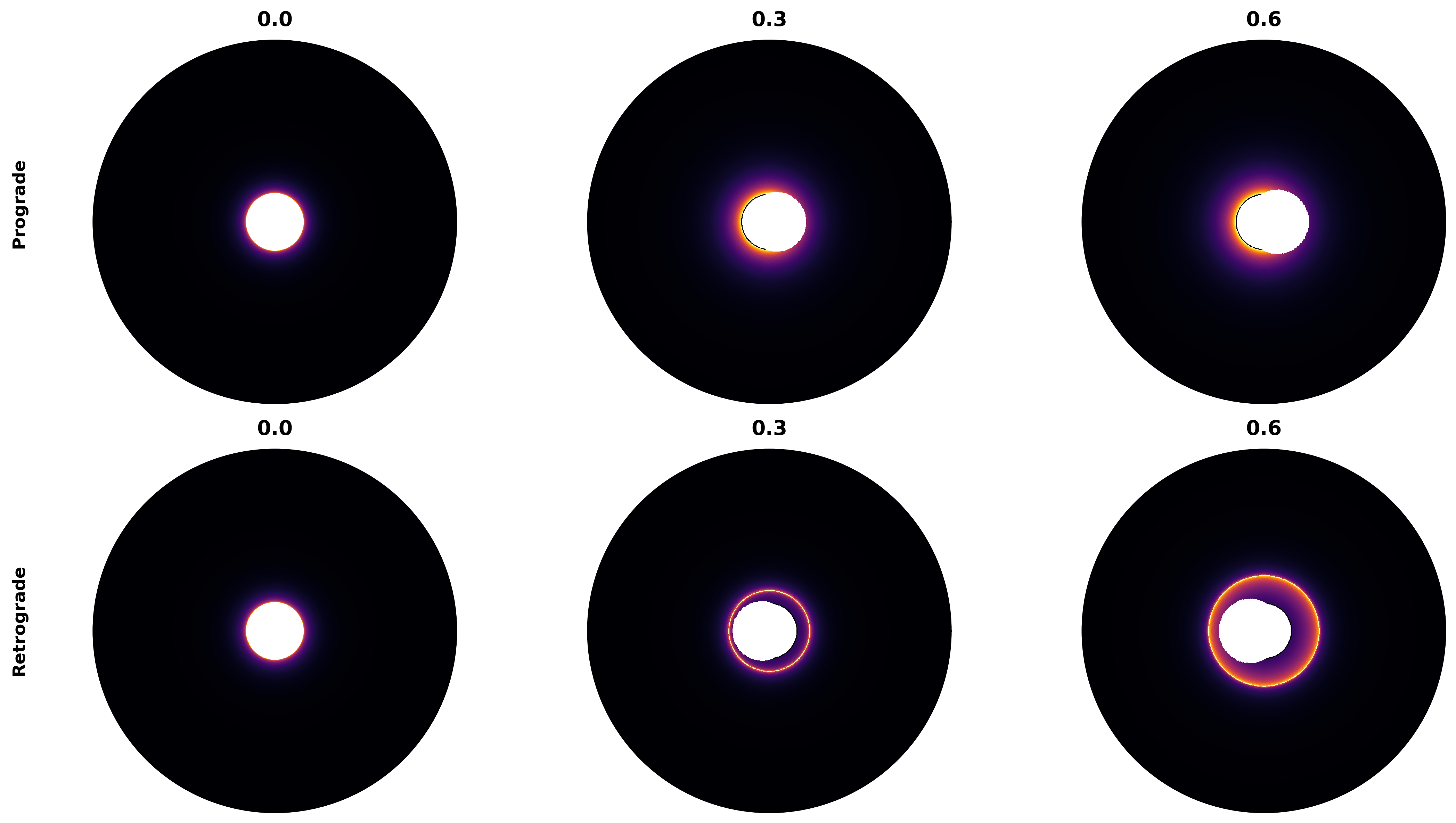}
    \label{fig:shadow_r15}
}
\hfill
\subfigure[$r_0=2.5$ kpc]{
    \includegraphics[width=0.45\textwidth]{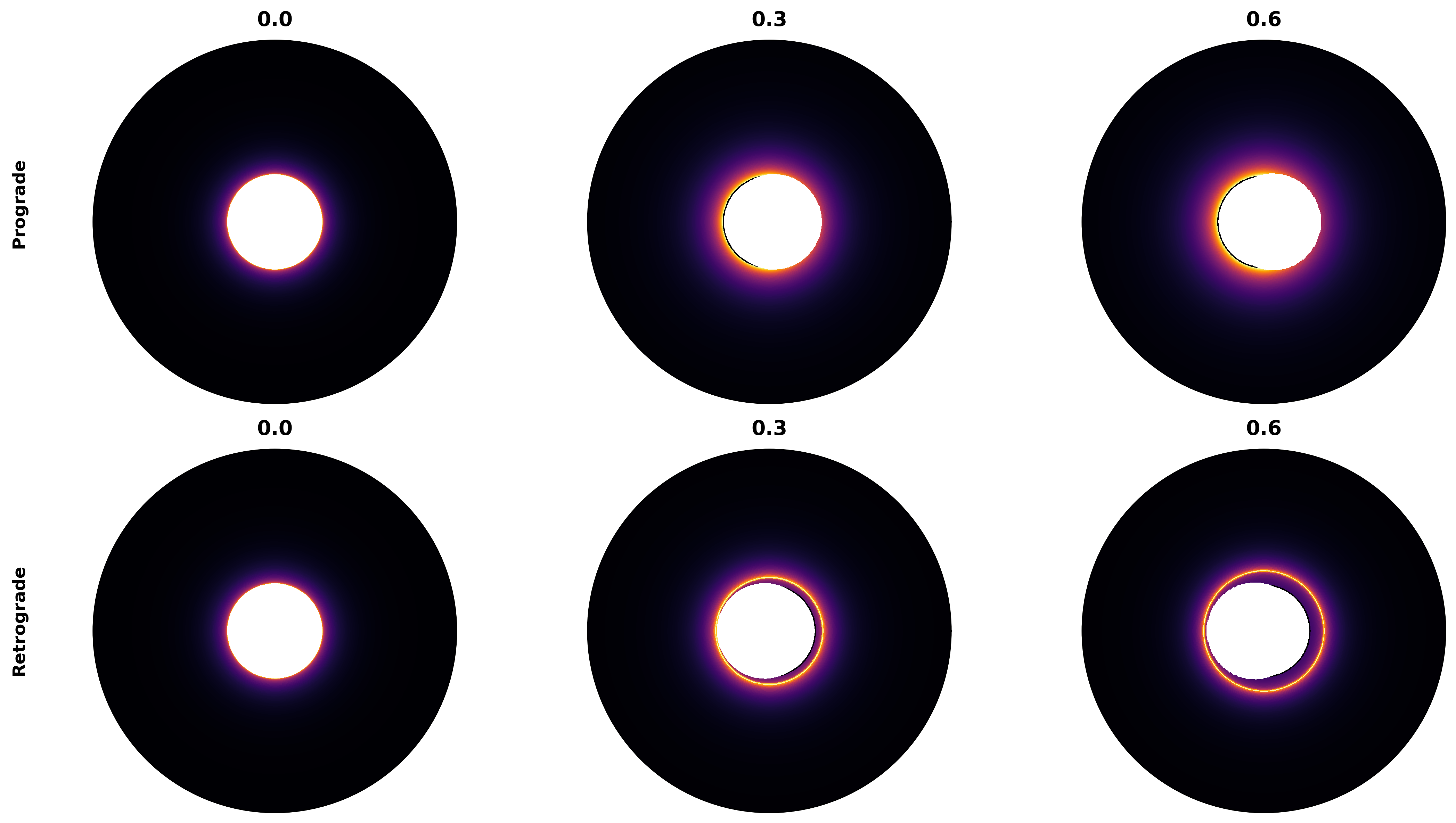}
    \label{fig:shadow_r25}
}
\caption{Polar shadow maps of the slowly rotating NFW wormhole for throat radii $r_0=1.5$~kpc and $r_0=2.5$~kpc, at angular momenta $J\in\{0,\,0.3,\,0.6\}$~kpc$^2$, using the NGC~2366 halo parameters ($r_s=1.447$~kpc, $\rho_s=3.11\times10^{-3}~\text{M}_\odot/\text{pc}^3$). Top rows show prograde-rendered panels ($\phi_\text{or}=0$) and bottom rows retrograde-rendered panels ($\phi_\text{or}=\pi$); the dipolar capture boundary follows Eq.~\eqref{eq:bcut_dipole}.}
\label{fig:shadow_combined}
\end{figure}
In the static case the shadow boundary in Figs.~\ref{fig:shadow_r15}--\ref{fig:shadow_r25} is a circle of radius $b_{\rm ph}^{(0)}$ centred at the origin, with a thin photon-ring-like brightening immediately outside~\cite{Luminet1979,Bambi2013,GrallaHolzWald2019,Tsukamoto2017}. The angular radius grows linearly with $r_{0}$, reproducing the geometric scaling of Sec.~\ref{subsec1} and contrasting with the mass-controlled Schwarzschild scale of compact-object shadows~\cite{FalckeMeliaAgol2000,EHT2019,Bambi2013}. For $J>0$ the central dark region in the prograde row is displaced toward $\theta=0$, extending out to $b_{+}$ on that side and only to $b_{-}$ on the opposite side; the retrograde panels exhibit the mirror configuration toward $\theta=\pi$. The maximum displacement reaches $\Delta b_{\rm ph}/2\approx 2J/r_{\rm ph}$, recovering Eq.~\eqref{eq:delta_bph}~\cite{NedkovaTinchevYazadjiev2013,Jusufi2018,Konoplya2018,Tsupko2017}. The brightening ring tracks the displaced boundary and is most pronounced where $b_{\rm cut}\to b_{+}$~\cite{Hioki2009,EHT2019,GrallaHolzWald2019}.
At the largest angular momentum studied ($J=0.6\,\mathrm{kpc}^{2}$, $r_{0}=1.5$~kpc) the dimensionless rotation parameter $K\omega(r_{\rm ph})\approx 0.9$ approaches unity and the linearized splitting~\eqref{eq:bpm} loses quantitative accuracy~\cite{Teo1998,Hartle1967,HartleThorne1968}; the corresponding panels in Fig.~\ref{fig:shadow_r15} therefore illustrate qualitative behaviour at the edge of the slow-rotation regime, with a quantitatively reliable description requiring either an exact rotating wormhole solution or higher-order Hartle--Thorne expansion~\cite{Hartle1967,HartleThorne1968,Bronnikov2017}
The asymmetry developed in Figs.~\ref{fig:shadow_r15}--\ref{fig:shadow_r25} suggests an observational diagnostic: a measurement of the shadow centroid offset, combined with the shadow size from $b_{\rm ph}^{(0)}$, directly yields the wormhole angular momentum via $J\approx r_{\rm ph}\,\Delta b_{\rm ph}/4$, with no further dependence on the halo profile parameters~\cite{Hioki2009,Konoplya2018,EHT2019}. For galactic-halo throat scales the predicted angular size of $\Delta b_{\rm ph}$ at the NGC~2366 distance $d=3.3$~Mpc is macroscopic (arcminute scale), well above current VLBI resolution but unrelated to horizon-scale interferometry, which probes microarcsecond structure of compact objects rather than kiloparsec-scale halo geometries~\cite{EHT2019,EHT2022,Falcke2013,Bambi2013}.

\subsubsection{Accretion Disc Analysis of Rotating NFW wormhole} \label{sec:rot_nfw_disc}
The static accretion disc image of Sec.~\ref{sec:accretion_disk} is now extended to the slowly rotating NFW wormhole by integrating the full three-dimensional null geodesic
equations in the Teo metric of Sec.~\ref{sec:rotating_metric}, with the frame-dragging term $\omega(r) = 2J/r^{3}$ retained throughout the Hamiltonian and disc-emission
model~\cite{Teo1998,Hartle1967,HartleThorne1968}. The line element
takes the form of Eq.~\eqref{eq:nfw_rotating_metric} away from the equatorial plane, so the inverse metric components carrying the rotation are
\begin{multline}
    g^{tt} = -e^{-2\Phi}, \quad g^{t\varphi} = -\omega\,e^{-2\Phi}, \quad g^{\theta\theta} = \frac{1}{r^{2}}, \\ g^{\varphi\varphi} = \frac{1}{r^{2}\sin^{2}\theta} - \omega^{2}\,e^{-2\Phi},
    \quad g^{rr} = 1 - \frac{b}{r}.
    \label{eq:inv_metric_rot}
\end{multline}
The super-Hamiltonian $\mathcal{H} = \tfrac{1}{2}g^{\mu\nu}p_{\mu}p_{\nu}=0$ for null
geodesics gives, with conserved $p_{t}=-E$ and $p_{\varphi}=L$~\cite{Carter1968a,Carter1968b,Wald1984},
\begin{equation}
    2\mathcal{H}
    = g^{tt}E^{2} - 2\,g^{t\varphi}E L + g^{\varphi\varphi}L^{2}
      + g^{rr}p_{r}^{2} + g^{\theta\theta}p_{\theta}^{2} = 0,
    \label{eq:H_rot}
\end{equation}
from which the ingoing radial momentum at the observer follows by inversion. Hamilton's equations yield the geodesic system
\begin{align}
\frac{dr}{d\lambda}
    &= \left(1-\frac{b}{r}\right)p_r, \\
\frac{d\theta}{d\lambda}
    &= \frac{p_\theta}{r^2},
\label{eq:rdot_thetadot_rot}\\[3pt]
\frac{dp_r}{d\lambda}
    &= -\frac{1}{2}\left[
    \begin{aligned}
        &\partial_r g^{tt}\,E^2
        - 2\,\partial_r g^{t\varphi}\,E L \\
        &{}+ \partial_r g^{\varphi\varphi}\,L^2
        + \partial_r g^{rr}\,p_r^2 \\
        &{}+ \partial_r g^{\theta\theta}\,p_\theta^2
    \end{aligned}
    \right],
\label{eq:pr_dot_rot}\\[3pt]
\frac{dp_\theta}{d\lambda}
    &= -\frac{1}{2}\,\partial_\theta g^{\varphi\varphi}\,L^2
     = \frac{L^2\cos\theta}{r^2\sin^3\theta}.
\label{eq:ptheta_dot_rot}
\end{align}
with the relevant inverse-metric derivatives
\begin{equation}
\begin{aligned}
    \partial_{r}g^{tt} &= 2\Phi'\,e^{-2\Phi}, \qquad
    \partial_{r}g^{t\varphi} = -\omega'\,e^{-2\Phi} + 2\omega\Phi'\,e^{-2\Phi},\\
    \partial_{r}g^{\varphi\varphi} &= -\frac{2}{r^{3}\sin^{2}\theta}
        - 2\omega\omega'\,e^{-2\Phi} + 2\omega^{2}\Phi'\,e^{-2\Phi}, \\
    \partial_{r}g^{rr} &= \frac{b-r b'}{r^{2}}, \qquad
    \partial_{r}g^{\theta\theta} = -\frac{2}{r^{3}},
\end{aligned}
\end{equation}
and $\omega(r)=2J/r^{3}$, $\omega'(r)=-6J/r^{4}$~\cite{Teo1998,Hartle1967}. The system reduces to the static Hamiltonian formulation of Sec.~\ref{sec:accretion_disk} when $J\to 0$~\cite{Carter1968a}.
An observer is placed at $r_{\rm obs}=30$~kpc and polar inclination $\theta_{\rm obs}=60^{\circ}$, with a $400\times400$ pixel screen of half-width $12$~kpc parameterised by Cartesian coordinates $(\alpha,\beta)$. The Bardeen--Cunningham camera relations are~\cite{Bardeen1973,Cunningham1973,Luminet1979,Bambi2013}
\begin{equation}
    L = -\alpha\,\sin\theta_{\rm obs}, \qquad p_{\theta}\big|_{\rm obs}=-\beta,
    \label{eq:camera_rot}
\end{equation}
with $E=1$, and $p_{r}|_{\rm obs}$ fixed to the ingoing branch from Eq.~\eqref{eq:H_rot}. Rays are integrated backward with a fixed-step fourth-order Runge--Kutta scheme ($d\lambda=0.05$, $1500$ steps), and equatorial-plane crossings $\theta(\lambda)=\pi/2$ are detected by sign changes between successive steps. The crossing radius $r_{\times}$ is obtained by linear interpolation~\cite{Cunningham1973,Younsi2016}. The disc emissivity profile is the same as in the static case~\cite{Bambi2013,Younsi2016},
\begin{equation}
    j_{\nu}^{\rm em}(r_{\times}) = \frac{1}{r_{\times}^{3}}\,
    \exp\!\left[-\left(\frac{r_{\rm cut}}{r_{\times}}\right)^{\!4}\right],
    \label{eq:emissivity_rot}
\end{equation}
with $r_{\rm cut}=2.5$~kpc. For a circular timelike geodesic in the equatorial plane of the Teo metric~\eqref{eq:nfw_rotating_metric}, the radial geodesic equation $\partial_{r}g_{tt}+2\Omega\,\partial_{r}g_{t\varphi}+\Omega^{2}\,\partial_{r}g_{\varphi\varphi}=0$ becomes the quadratic~\cite{Teo1998,Hartle1967,HartleThorne1968}
\begin{equation}
    r\,\Omega^{2} - \bigl(2r\omega + r^{2}\omega'\bigr)\Omega
    + \bigl(r\omega^{2} + r^{2}\omega\omega' - \Phi'\,e^{2\Phi}\bigr) = 0,
    \label{eq:omega_quadratic}
\end{equation}
with the prograde co-rotating solution
\begin{equation}
    \Omega(r) = \omega(r) + \tfrac{1}{2}r\,\omega'(r)
    + \sqrt{\tfrac{1}{4}r^{2}\omega'^{2}(r) + \frac{\Phi'(r)\,e^{2\Phi(r)}}{r}}.
    \label{eq:omega_circular}
\end{equation}
The retrograde branch is obtained by reversing the sign of the square root. Equation~\eqref{eq:omega_circular} collapses to $\Omega = e^{\Phi}\sqrt{\Phi'/r}$ in the static limit~\cite{Bambi2013}, recovering Eq.~(\ref{eq:Omega_disk}) of Sec.~\ref{sec:accretion_disk}. The four-velocity of the orbiting emitter is $u^{\mu}=u^{t}(1,\,0,\,0,\,\Omega)$ with normalisation $u_{\mu}u^{\mu}=-1$ giving~\cite{Cunningham1973,MisnerThorneWheeler1973}
\begin{equation}
    u^{t} = \frac{1}{\sqrt{e^{2\Phi} - r^{2}\,(\Omega-\omega)^{2}}}.
    \label{eq:ut_rot}
\end{equation}
The relativistic redshift factor between the emitter and the distant observer is then~\cite{Luminet1979,Cunningham1973,Bambi2013,Younsi2016}
\begin{equation}
    g \equiv \frac{\nu_{\rm obs}}{\nu_{\rm em}}
    = \frac{\sqrt{e^{2\Phi(r_{\times})} - r_{\times}^{2}\,(\Omega-\omega)^{2}}}
            {1 - \Omega(r_{\times})\,\beta},
    \qquad \beta = \frac{L}{E},
    \label{eq:g_doppler_rot}
\end{equation}
clamped to $g\in[0.1,3.0]$ for numerical stability~\cite{Falanga2015,Younsi2016}. The numerator, $1/u^{t}$, carries the gravitational redshift and transverse Doppler shift in the rotating spacetime; the denominator $1-\Omega\beta$ carries the longitudinal Doppler shift~\cite{Luminet1979,Bambi2013}. In the static limit $\omega\to0$, $\Omega-\omega=\Omega$ and Eq.~\eqref{eq:g_doppler_rot} reduces to Eq.~(\ref{eq:doppler_factor}). Each equatorial crossing contributes an intensity increment
\begin{equation}
    \Delta I = j_{\nu}^{\rm em}(r_{\times})\,g^{4},
    \label{eq:dI_rot}
\end{equation}
where $g^{4}$ is the bolometric transformation of specific intensity~\cite{MisnerThorneWheeler1973,Luminet1979,Cunningham1973}. The total observed intensity at each pixel is the cumulative sum over all equatorial crossings along the corresponding geodesic, so that strongly lensed rays that re-cross the equatorial plane contribute secondary images~\cite{Luminet1979,Bambi2013,GrallaHolzWald2019}.\\
\begin{figure*}
    \centering
    \includegraphics[width=16cm,height=8cm]{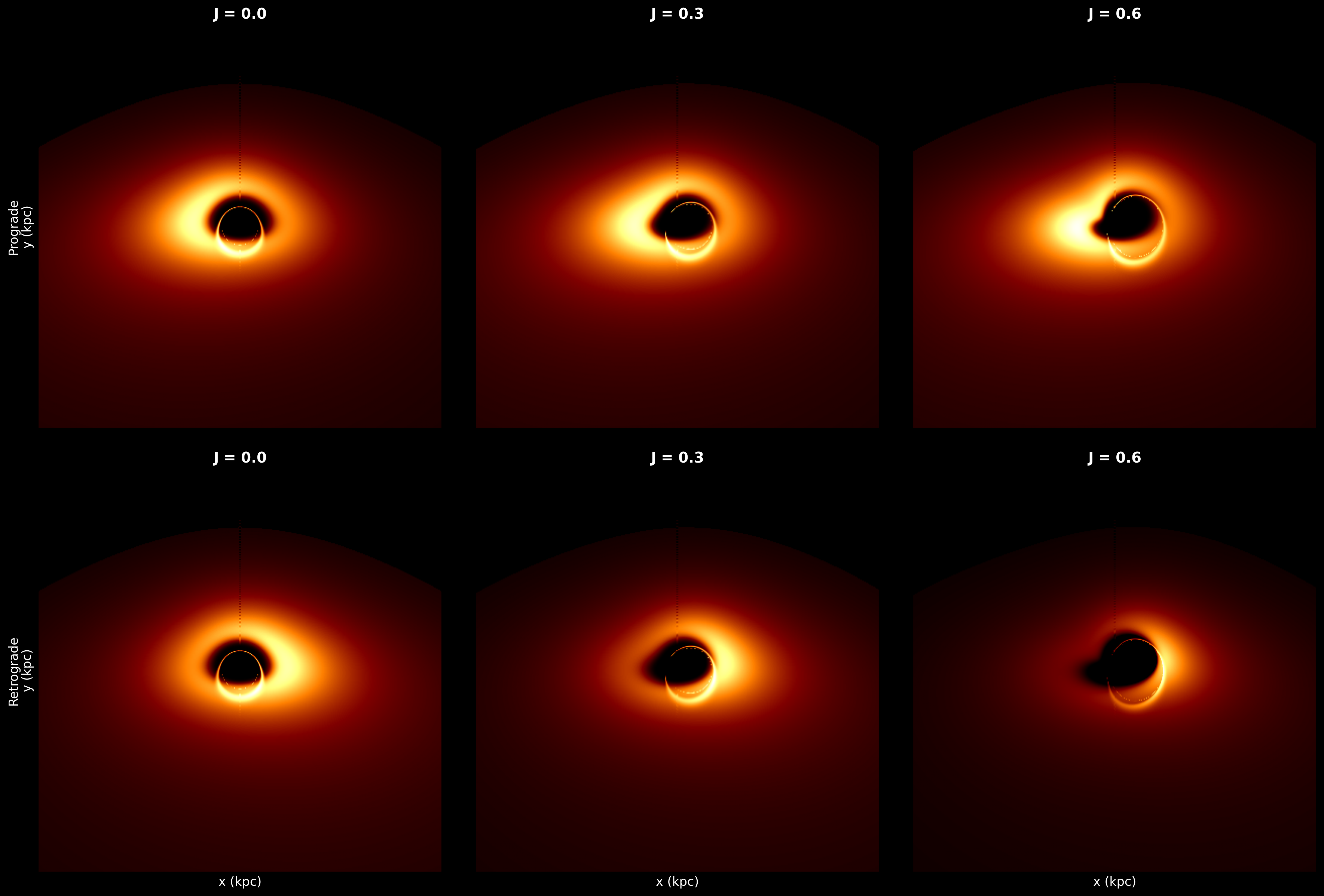}
    \caption{Accretion disc images of the slowly rotating NFW wormhole with throat radius $r_0=1.0$~kpc, for angular momenta $J\in\{0,\,0.3,\,0.6\}$~kpc$^2$, using the NGC~2366 halo parameters ($r_s=1.447$~kpc, $\rho_s=3.11\times10^{-3}~\text{M}_\odot/\text{pc}^3$). The observer is placed at $r_\text{obs}=30$~kpc, inclination $\theta_\text{obs}=60^\circ$, with a $400\times400$ pixel screen spanning $\pm12$~kpc, and an emissivity inner cutoff $r_\text{cut}=2.5$~kpc. The top and bottom rows show prograde and retrograde disc rotation, respectively.}
    \label{fig:rot_disc}
\end{figure*}
Figure~\ref{fig:rot_disc} shows the resulting images for $J\in\{0,\,0.3,\,0.6\}\,\mathrm{kpc}^{2}$. The $J=0$ panels reproduce the static Luminet-type disc image of Sec.~\ref{sec:accretion_disk}, with a lensed disc arc above the central dark region, a pronounced left--right Doppler asymmetry from the prograde Keplerian motion of the disc material, and a secondary image faintly visible immediately outside the central dark region~\cite{Luminet1979,Cunningham1973,Bambi2013,GrallaHolzWald2019}. The asymmetry is set by the disc's own rotation~\eqref{eq:omega_circular} with $\omega=0$, so the two $J=0$ panels share the morphology of the disc-driven Doppler boost~\cite{Younsi2016,FalckeMeliaAgol2000}. For $J>0$ the geometry acquires definite handedness through frame dragging. In the prograde row, where the wormhole's rotation co-rotates with the disc, the bright crescent is enhanced on the approaching side and the central dark region elongates and shifts toward the receding side, consistent with the photon-orbit splitting $b_{\pm}=K(1\pm K\omega)$ of Eq.~\eqref{eq:bpm}~\cite{NedkovaTinchevYazadjiev2013,Shaikh2018,Jusufi2018}. In the retrograde row, the dark region is offset toward the opposite side, and the asymmetry of the brightness pattern is correspondingly mirrored~\cite{Hioki2009,Konoplya2018,Tsupko2017}. The displacement of the central dark region grows monotonically with $J$ across each row, in qualitative agreement with the splitting scale $\Delta b_{\rm ph}\approx 4J/r_{\rm ph}$ of Eq.~\eqref{eq:delta_bph}~\cite{Hioki2009,Bambi2013}. Higher-order images, where present, also acquire the same handedness, providing a secondary imprint of $J$ on the image~\cite{GrallaHolzWald2019,Luminet1979}.  At the parameter choices used here the slow-rotation expansion parameter $K\omega(r_{\rm ph})$, with $K=b_{\rm ph}^{(0)}=r_{0}\,e^{-\Phi(r_{0})}\simeq 1.06$ and $\omega(r_{\rm ph})\simeq 2J$, takes values $K\omega\simeq 0.64$ at $J=0.3\,\mathrm{kpc}^{2}$ and $K\omega\simeq 1.28$ at $J=0.6\,\mathrm{kpc}^{2}$, both outside the strict slow-rotation regime $K\omega\ll 1$~\cite{Teo1998,Hartle1967,HartleThorne1968}. The corresponding panels in Fig.~\ref{fig:rot_disc} should therefore be read as qualitative illustrations of the morphological trends imposed by frame dragging; quantitatively reliable images in this regime would require either an exact rotating wormhole solution or a higher-order Hartle--Thorne expansion of the metric~\cite{Hartle1967,HartleThorne1968,Bronnikov2017}. The static and weakly rotating ($J=0.3$) panels remain within the validity of the present construction up to the leading-order corrections retained in Eqs.~\eqref{eq:omega_circular}--\eqref{eq:g_doppler_rot}. Three observational implications follow from Fig.~\ref{fig:rot_disc}. First, the centroid offset of the dark region scales linearly with $J$ at fixed $r_{0}$ in the perturbative regime, providing a direct measurement of the wormhole angular momentum via $J\simeq r_{\rm ph}\,\Delta b_{\rm ph}/4$ that is independent of the halo parameters~\cite{Hioki2009,Konoplya2018}. Second, the prograde--retrograde mirror symmetry serves as a model-independent check on the rotation sense, since the symmetry is broken only by the sign of $J$~\cite{Bardeen1973,FalckeMeliaAgol2000,EHT2019}. Third, for galactic-halo throat scales ($r_{0}\sim$~kpc) at extragalactic distances the predicted shadow and asymmetry remain macroscopic (arcminute scale at the NGC~2366 distance $d=3.3$~Mpc), well above current VLBI resolution but unrelated to the horizon-scale interferometry probing compact-object shadows at microarcsecond scales~\cite{EHT2019,EHT2022,Falcke2013,Bambi2013}. The combination of the shadow size~\eqref{eq:bcut_dipole}, the splitting~\eqref{eq:delta_bph}, and the disc-image centroid offset measured in Fig.~\ref{fig:rot_disc} thus constitutes a triple consistency test sensitive to both the halo profile and the rotation of the wormhole geometry~\cite{NedkovaTinchevYazadjiev2013,Shaikh2018,Bambi2013,EHT2019}.

\subsection{The Soliton quantum wave dark matter model} \label{subsec2}
The standard cold dark matter paradigm, embedded in the broader \(\Lambda\)CDM cosmological model, has achieved striking success in explaining the cosmic microwave background anisotropies, the large-scale distribution of galaxies, gravitational lensing statistics, and the emergence of the cosmic web from nearly scale-invariant primordial perturbations. Its central assumption is that the dark component is effectively collisionless, non-relativistic, and pressureless on astrophysical scales. Within this framework, hierarchical structure formation naturally produces virialized halos whose spherically averaged density profiles are well described by the Navarro-Frenk-White form, with a central cusp and extended outer envelope \cite{Navarro_1997}. Nevertheless, the interpretation of several small-scale observations has long motivated the exploration of alternatives to strictly cold, collisionless dark matter. The cusp-core problem, the apparent overabundance of low-mass subhalos, and the too-big-to-fail tension in the Local Group indicate that the inner structure and abundance of dwarf-scale halos may be more delicate probes of dark matter microphysics than large-scale clustering alone \cite{Klypin_1999,Moore_1999,BoylanKolchin_2012}. Although baryonic feedback, tidal stripping, reionization, and observational incompleteness may reduce these discrepancies, they do not remove the theoretical motivation for dark matter models in which the particle nature leaves a coherent dynamical imprint on galactic scales.
Soliton quantum wave dark matter, commonly discussed under the names fuzzy dark matter, wave dark matter, scalar-field dark matter, or ultra-light axion dark matter, is one of the most theoretically economical realizations of this idea. The model assumes that the dark sector is composed of bosonic particles with masses typically near \(m_\psi \sim 10^{-22}\,\mathrm{eV}\), although current constraints and phenomenological applications span a broader range \cite{PhysRevLett.85.1158,Hui_2017,Ferreira_2021}. Such an ultra-light mass implies an astrophysically large de Broglie wavelength,
\[
\lambda_{\rm dB}\sim {h\over m_\psi v},
\]which can reach kiloparsec scales for velocity dispersions characteristic of dwarf galaxies. 
The behavior of dark matter on halo scales cannot always be accurately described as a cluster of classical point particles. Instead, ultralight bosonic dark matter acts as a coherent, self-gravitating wave. Its gradient energy, often called quantum pressure, can prevent gravitational collapse below a characteristic scale. This concept is closely related to axion-like fields and moduli predicted in high-energy theories, including models inspired by string theory, such as the axiverse, where extremely light scalar fields can occur naturally \cite{Marsh_2016}. Cosmologically, these fields behave like pressureless matter once their coherent oscillation frequency exceeds the Hubble rate, although wave effects remain significant at small scales.
In the non-relativistic weak field regime, minimally coupled ultralight bosonic dark matter is described by the Schr\"odinger--Poisson system,
\begin{equation}
\begin{aligned}
 i\hbar\frac{\partial \psi}{\partial t}
&=
-\frac{\hbar^2}{2m_\psi}\nabla^2\psi
+m_\psi\Phi\psi
+g|\psi|^2\psi,\\
\nabla^2\Phi &=4\pi G m_\psi |\psi|^2 ,
\end{aligned}
\end{equation}
where \(\psi\) denotes the macroscopic wavefunction, \(\Phi\) is the Newtonian gravitational potential, and \(g\) represents a possible contact self interaction. When self-interaction is negligible, the model corresponds to the fuzzy dark matter or free scalar field limit. If self-interaction is significant, it becomes similar to the Gross-Pitaevskii-Poisson description used in Bose-Einstein condensate dark matter models \cite{Böhmer_2007,Harko_2011}.
Applying the Madelung transformation,
\begin{equation}
\psi=\sqrt{\frac{\rho}{m_\psi}}\exp\left(\frac{iS}{\hbar}\right),
\end{equation}
transforms the system into a fluid form with a velocity potential and an additional quantum potential. This term creates an effective pressure that depends on scale and resists collapse in regions with strong density gradients. The model is therefore distinct from warm dark matter and cannot be represented by a cored halo profile. Its halo structure emerges from wave interference, gravitational cooling, and the ground state solutions of the Schr\"odinger-Poisson equations.
One important prediction is the presence of a gravitationally bound solitonic core at the center of a virialized halo. High-resolution simulations show that wave dark matter halos contain a dense coherent soliton surrounded by an outer halo whose time-averaged density profile resembles an NFW envelope \cite{Schive2014,Schive_2014_PRL}. The soliton density profile is approximated by
\begin{equation}
\rho_{\rm sol}(r)=\rho_c\left[1+0.091\left(\frac{r}{r_c}\right)^2\right]^{-8},
\end{equation}
where \(r_c\) is the half density core radius and \(\rho_c\propto m_\psi^{-2}r_c^{-4}\). This relation reflects the balance between self-gravity and wave support required by the uncertainty principle. A more compact soliton must have a higher central density, while a lighter boson produces a more extended core.
The relation between the soliton and its host halo is still being investigated. Nevertheless, simulations support a consistent picture in which wave interference and gravitational cooling form a low-entropy ground-state core at the halo center. The outer halo retains density fluctuations of order unity on scales close to the local de Broglie wavelength \cite{Mocz_2017}. Such fluctuations may influence stellar heating, perturb stellar streams, affect black hole environments, and alter lensing substructure.
The suppression of the linear matter power spectrum below the wave Jeans scale gives soliton quantum wave dark matter its relevance for small-scale cosmology. Modes with wavelengths shorter than the effective de Broglie or Jeans scale cannot collapse efficiently, delaying the formation of low-mass halos and reducing the number of substructures relative to collisionless CDM \cite{PhysRevLett.85.1158}. At the same time, the central soliton replaces the divergent CDM cusp with a finite-density core, offering a possible explanation for the slowly rising rotation curves of dwarf and low-surface-brightness galaxies \cite{Marsh_2015,Robles_2012}. 
Rotation curve studies demonstrate that a universal particle mass and a fixed relation between the soliton and its host halo are difficult to reconcile with the diversity of observed galaxies. This is especially evident when baryonic effects and halo assembly histories are taken into account \cite{Bar_2018,Bar_2022}. Lyman-\(\alpha\) forest observations also provide significantly lower limits on the mass of bosons. If the boson were too light, it would restrict the formation of structures at the redshifts examined through intergalactic absorption spectra \cite{Irsic_2017}. As a result, the mass range that can produce kiloparsec-scale cores in dwarf galaxies faces increasing observational tension. Current studies examine whether self-interactions, mixed dark-matter models, environmental effects, or revised relations between solitons and host halos can maintain the small-scale advantages of the model while remaining consistent with high-redshift structure.\\
Astrophysically, wave dark matter is important because it connects galactic dynamics with the underlying properties of a quantum field. In galaxy rotation curves, the soliton contributes a central mass component whose size and amplitude depend on \(m_\psi\), whereas the outer halo produces the approximately flat part of the rotation curve. In halos, interference granules generate time-dependent fluctuations in the gravitational potential and may lead to relaxation processes that do not occur in an ideal collisionless cold dark matter halo. In gravitational lensing, the reduced abundance of low-mass halos alters the expected substructure population, while coherent density granulation may introduce small perturbations in lensing signals. Near supermassive black holes, a soliton may be compressed, disrupted, or depleted. The black hole can also modify the scalar field density profile and the surrounding dynamical friction environment \cite{Davies_2020}. Galactic nuclei, stellar streams, ultra-faint dwarf galaxies, and strongly lensed systems provide useful environments for testing the wave properties of dark matter. Generally, this model links cosmological structure formation with macroscopic quantum mechanics. Its observable signatures arise from the coherence length of a cosmic quantum field rather than from short-distance particle scattering.
These properties also make solitonic quantum wave dark matter relevant to relativistic compact objects, such as wormholes. Errehymy et al.\ \cite{Errehymy_2024} constructed wormhole solutions in \(f(R)\) gravity using cold dark matter and solitonic quantum wave halo density profiles and demonstrated that NEC is violated at the wormhole throat for the solitonic model, whereas it is satisfied for the cold dark matter model. Further, Alshammari et al. \cite{Alshammari_2026} constructed minimally deformed fuzzy dark matter wormholes surrounded by solitonic quantum wave dark matter halos by employing the minimal geometric deformation technique in the context of GR. They concluded that wormhole geometries are possible, smoothly capturing the characteristics of solitonic quantum wave dark matter while exhibiting an appropriate level of finely tuned exoticity near the throat.\\
Axions and other ultralight bosons, with masses $m_b \sim 10^{-23}-10^{-21}$ eV, are a well-known contender to address the aforementioned issues. At cosmological scales, these particles are consistent with the CDM. However, these particles behave as self-gravitating DM waves and populate galactic halos with significant occupation numbers at distances comparable to their de Broglie wavelengths, which can be on the kpc scale. The production of a flat-core ``soliton" at the center of galaxies, with a rather marked transition to a less dense outer region that follows a CDM-like distribution, is one of the repercussions of this, as it appears to exert a pressure-like effect on macroscopic scales.\\
Assuming the simple case of ultralight bosons, when self-interaction is ignored, then the boson mass is the only free parameter. If the corresponding de Broglie wavelength exceeds the mean free path set by the density of dark matter, these bosons can satisfy the ground state condition for a Bose-Einstein condensate described by the coupled Schr\"{o}dinger-Poisson equation. This can be written in comoving coordinates as
\begin{align*}
    &\left[i \frac{\partial}{\partial \tau}+\frac{1}{2}\nabla^2-a V\right]\psi=0,\\
    &\nabla^2 V=4\pi (|\psi|^2-1),
\end{align*}
where $\psi$ is the wave function, $V$ is the gravitation potential and $a$ is the cosmological scale factor. The fitting formula for the density profile of the solitonic core in a $\psi$DM halo is obtained from cosmological simulations \cite{Schive2014,Schive_2014_PRL}
\begin{equation} \label{eq:solitondensprof}
    \rho_\text{sol}(r) = \frac{\rho_c}{\left[1+\alpha \left(r/r_c\right)^2 \right]^{8}}\,,
\end{equation}
Here, $\rho_c$ and $r_c$ represent the central density and size of the soliton core \cite{Herrera-Mart2019}. The leading study thoroughly examines the distribution of the matter mentioned previously~\cite {Schive_2014_PRL}. The precise calculation for the half-density radius will be a specific radius that is determined as a constant $\alpha = \sqrt[8]{2}-1 \sim 0.09051$ \cite{Schive2014,Schive_2014_PRL}. Furthermore, the value of $\rho_c$ in Eq.~\eqref{eq:solitondensprof} is specified as \cite{Herrera-Mart2019}
\begin{equation} \label{e2}
    \rho_c = 2.4\times 10^{12} \left(\frac{m_b}{10^{-22}\text{eV}}\right)^{-2}  \left(\frac{r_c}{\text{pc}}\right)^{-4}\frac{M_\odot}{\text{pc}^{3}}.
\end{equation}
We use the numerical values of the model parameters, $r_c$ and $\rho_c$, using fuzzy DM simulation \cite{5}, based on the rotation curves of the LITTLE THINGS in 3D catalog \cite{6}. For the dwarf galaxy NGC 2366, the core radius $r_c=3$ kpc and the central density $\rho_c = 15\times 10^{-3}~\text{M}_{\odot}/\text{pc}^3$ \cite{5}.\\
Now, let us calculate the shape function of the wormhole under the soliton quantum wave dark matter. Comparing the Eq. \eqref{eq:solitondensprof} with Eq. \eqref{rhoWH} and impose to initial condition $b(r_0)=r_0$, we obtain
\begin{multline}\label{eq:soliton_shape}
b(r)=\frac{1}{215040}\left(\frac{\rho_c r r_c^4 \Lambda_1}{\alpha  \left(\alpha  r^2+r_c^2\right)^7}+\frac{3465 \rho_c r_c^3 \Lambda_3}{\alpha ^{3/2}}\right.\\\left.
+r_0 \left(215040-\frac{\rho_c r_c^4 \Lambda_2}{\alpha  \left(r_c^2+\alpha  r_0^2\right)^7}\right)\right).
\end{multline}
where 
\begin{multline}\nonumber
\Lambda_1=\left(23100 \alpha ^5 r^{10} r_c^2+65373 \alpha ^4 r^8 r_c^4 
+101376 \alpha ^3 r^6 r_c^6 \right.\\\left.
+92323 \alpha ^2 r^4 r_c^8+48580 \alpha  r^2 r_c^{10}+3465 \alpha ^6 r^{12}-3465 r_c^{12}\right)
\end{multline}
\begin{multline}\nonumber
\Lambda_2=\left(48580 \alpha  r_c^{10} r_0^2+92323 \alpha ^2 r_c^8 r_0^4+101376 \alpha ^3 r_c^6 r_0^6 \right.\\\left.
+65373 \alpha ^4 r_c^4 r_0^8+23100 \alpha ^5 r_c^2 r_0^{10}+3465 \alpha ^6 r_0^{12}-3465 r_c^{12}\right)  
\end{multline}
\begin{equation}\nonumber
\Lambda_3=\left(\tan ^{-1}\left(\frac{\sqrt{\alpha } r}{r_c}\right)-\tan ^{-1}\left(\frac{\sqrt{\alpha } r_0}{r_c}\right)\right)
\end{equation}
Also, for this model, the mass function can be obtained using Eq. \eqref{34}
\begin{multline}\label{mass2}
M(r)=\frac{\pi  \rho_c r_c^3}{53760 \alpha ^{3/2}}\left(\frac{\sqrt{\alpha } r r_c}{\left(\alpha  r^2+r_c^2\right)^7}\left(3465 \alpha ^6 r^{12} \right.\right.\\\left.\left.
+23100 \alpha ^5 r^{10} r_c^2+65373 \alpha ^4 r^8 r_c^4+101376 \alpha ^3 r^6 r_c^6 \right.\right.\\\left.\left.
+92323 \alpha ^2 r^4 r_c^8 +48580 \alpha  r^2 r_c^{10}-3465 r_c^{12}\right)+\Lambda_4\right),
\end{multline}
where $\Lambda_4=3465 \tan ^{-1}\left(\frac{\sqrt{\alpha } r}{r_c}\right)$. Now, using Eqs. \eqref{mass2}, \eqref{ab11}, and \eqref{3711}, one can easily calculate the redshift function
\begin{multline}
\Phi (r)=-C_2+\frac{\pi  \rho_c r_c^3}{53760 \alpha ^{3/2}}\left(-\frac{\sqrt{\alpha } r_c}{\left(\alpha  r^2+r_c^2\right)^6}\left(3465 \alpha ^5 r^{10} \right.\right.\\\left.\left.
+19635 \alpha ^4 r^8 r_c^2+45738 \alpha ^3 r^6 r_c^4+55638 \alpha ^2 r^4 r_c^6 \right.\right.\\\left.\left.
+36685 \alpha  r^2 r_c^8 +11895 r_c^{10}\right)-\frac{\Lambda_4}{r}\right),
\label{eq:soliton_redshift}
\end{multline}
where $C_2$ is an integrating constant. Now, following the same methodology used in the subsection- , we will investigate Ray tracing, Accretion disk, and shadow of wormhole solutions under this soliton model.

\subsubsection{Ray Tracing Analysis of Static Soliton Wormhole}
\label{sec:soliton_analysis}
\begin{figure*}
    \centering
    \includegraphics[width=16cm,height=8cm]{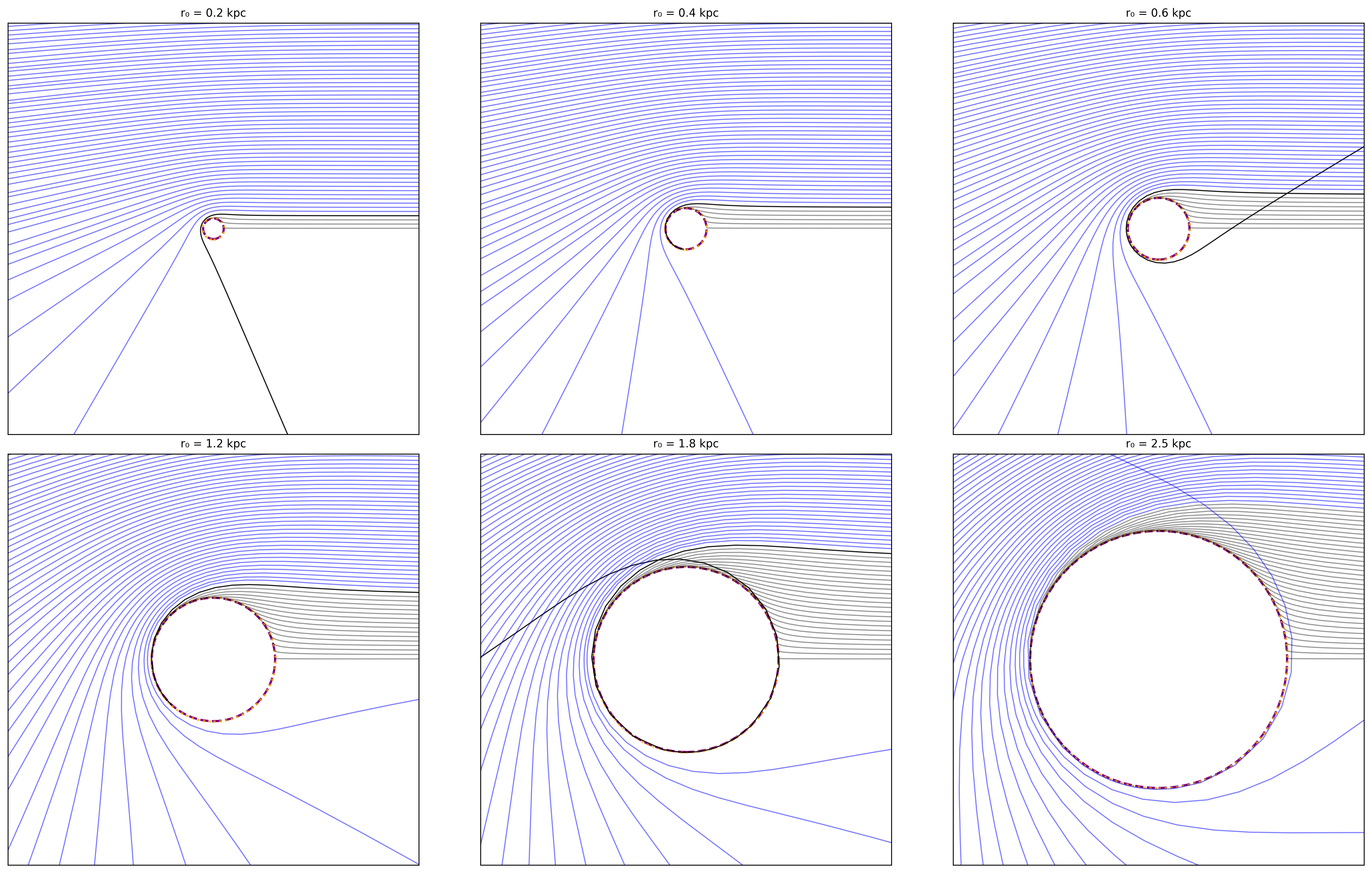}
    \caption{Null geodesics around the static soliton-supported wormhole for six throat
    radii, $r_{0}\in\{0.2,\,0.4,\,0.6,\,1.2,\,1.8,\,2.5\}$~kpc, using boson mass
    $m_{b}=1.2\times10^{-17}$~eV and core radius $r_{c}=3$~kpc, giving central density
    $\rho_{c}=2.06\times10^{-12}\,M_{\odot}\,\mathrm{pc}^{-3}$. The dashed circle marks the
    wormhole throat $r=r_{0}$; the solid black curve is the critical geodesic at the
    critical impact parameter $b_{\rm ph}$, Eq.~\eqref{eq:bph_soliton}, separating deflected
    rays from those captured by the wormhole.}
    \label{fig:soliton_raytracing}
\end{figure*}

\begin{figure}[!ht]
    \centering
    \subfigure[Normalised intensity profiles $I(\tilde{b})$]{
    \includegraphics[width=0.45\textwidth]{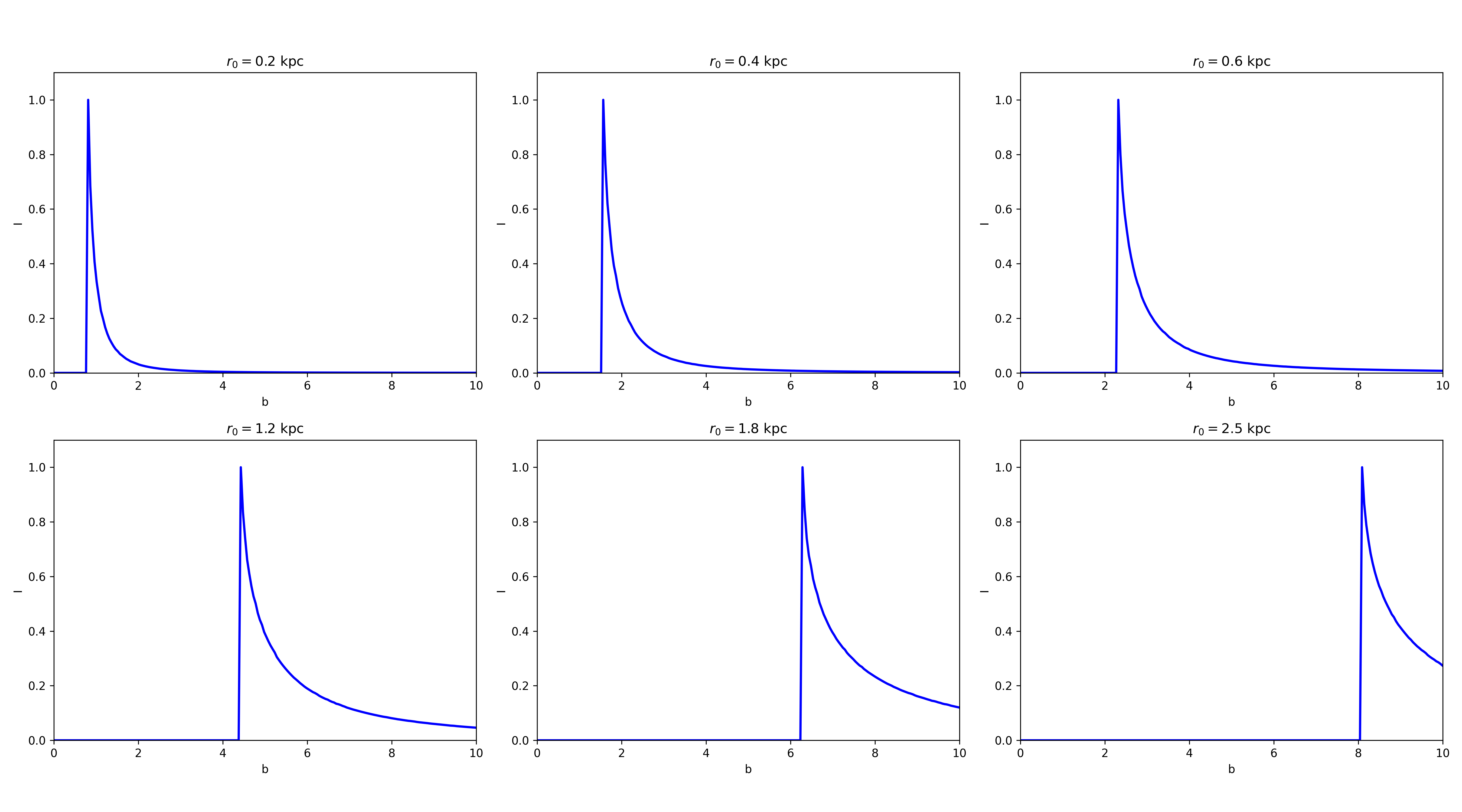}
    \label{fig:soliton_intensity}
}
\hfill
\subfigure[Polar shadow maps]{
    \includegraphics[width=0.43\textwidth]{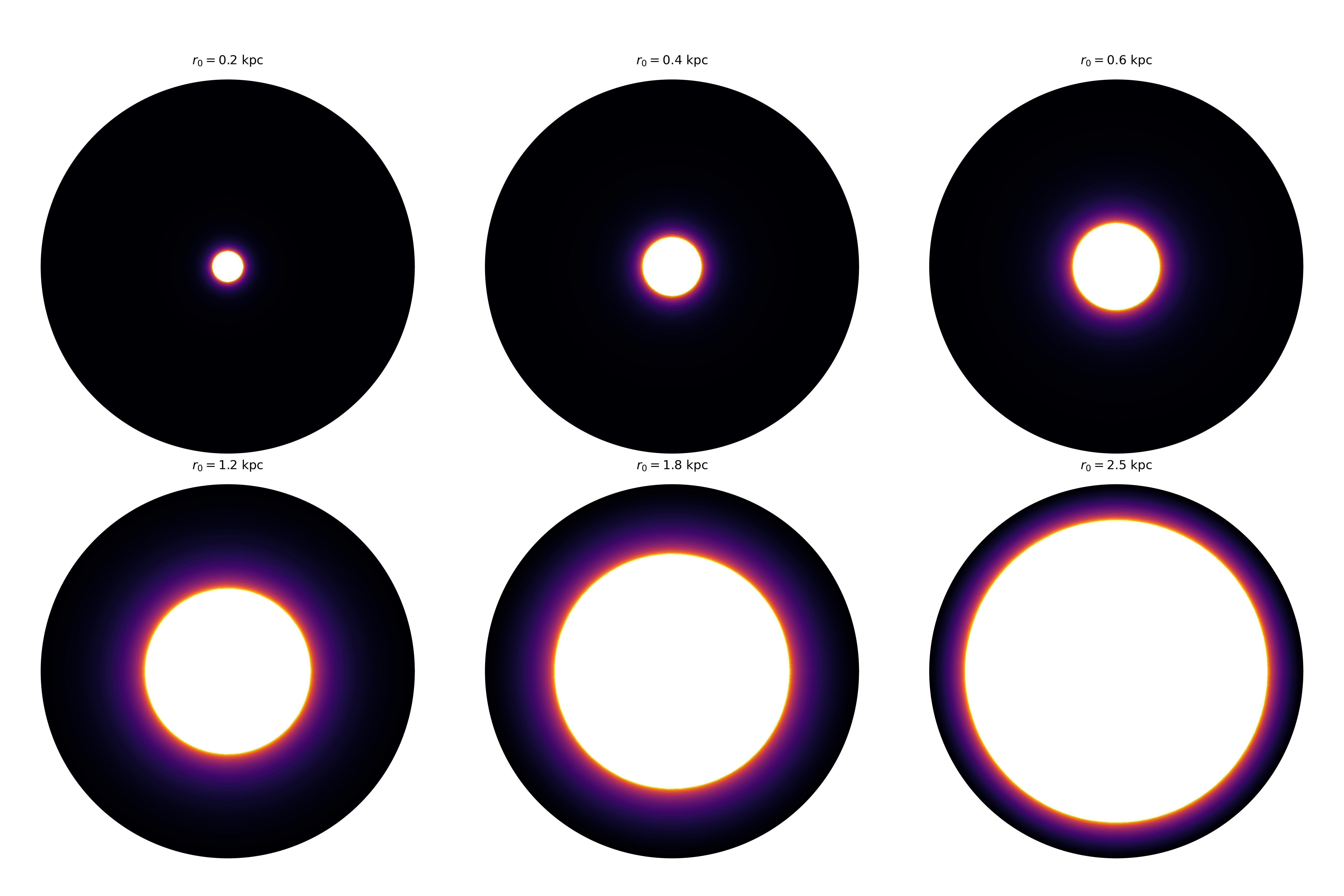}
    \label{fig:soliton_shadowmap}
}
\caption{Normalised intensity profiles and polar shadow maps of the static soliton ($\psi$DM) wormhole across six throat radii, $r_0\in\{0.2,\,0.4,\,0.6,\,1.2,\,1.8,\,2.5\}$~kpc, for boson mass $m_b=1.2\times10^{-17}$~eV, core radius $r_c=3$~kpc, and central density $\rho_c=2.06\times10^{-12}\,M_\odot\,\mathrm{pc}^{-3}$.}
\label{fig:soliton_combined}
\end{figure}

As discussed previously in the Sec.~\ref{sec:analysis}, we repeat the ray-tracing analysis of the soliton-supported wormhole using the same orbit equation~\eqref{eq:orbit_eq}, intensity
transfer integral~\eqref{eq:intensity}, and shadow definitions~\eqref{eq:shadow_radius}--\eqref{eq:shadow_angle}, with the soliton shape function~\eqref{eq:soliton_shape} and
redshift function~\eqref{eq:soliton_redshift}. The same six throat radii $r_{0}\in\{0.2,\,0.4,\,0.6,\,1.2,\,1.8,\,2.5\}$~kpc, sampling geometry ($N=100$ rays, $x_{\rm start}=15$~kpc, $\tilde{b}\in[0.01,8.0]$~kpc), and capture criterion
($\dot r=0$, $r\,e^{-\Phi(r)}=\tilde b$) are used throughout. We adopt boson mass $m_{b}=1.2\times10^{-17}$~eV and core radius $r_{c}=3$~kpc, giving central density $\rho_{c}=2.06\times10^{-12}\,M_{\odot}\,{\rm pc}^{-3}$, for the representative figures.
The soliton core concentrates its mass within $r_{c}$ (Eqs.~\eqref{eq:solitondensprof}--\eqref{mass2}), and the redshift function~\eqref{eq:soliton_redshift} is deep enough that the photon-sphere condition $r\,\Phi'(r)=1$, equivalently $v_{t}^{2}(r)=1$ from Eq.~\eqref{3711}, admits a solution at finite $r_{\rm ph}$ mildly outside the throat, $r_{\rm ph}\approx r_{0}+0.01$~kpc across the range studied. The critical impact parameter
\begin{equation}
b_{\rm ph}=r_{\rm ph}\,e^{-\Phi_{\rm sol}(r_{\rm ph})}
\label{eq:bph_soliton}
\end{equation}
therefore exceeds the throat radius by a modest but non-negligible factor. For this fiducial boson mass we find $b_{\rm ph}\approx1.18$--$1.26\,r_{0}$ across the range studied, set by the redshift factor
$e^{-\Phi_{\rm sol}(r_{0})}$ at the capture radius, with the ratio decreasing mildly toward larger throats (from $1.26$ at $r_0=0.2$~kpc to $1.18$ at $r_0=2.5$~kpc). This is a somewhat larger offset than the NFW case, where the shallow halo potential pins $b_{\rm ph}\approx 1.05\,r_{0}$ at the throat~\cite{Kuhfittig_2014}; a substantially more detached photon sphere requires a denser, lighter-boson core, as discussed below.\\
Figure~\ref{fig:soliton_raytracing} presents the integrated geodesics around the soliton wormhole for six radii of the throat. The morphology
follows that of the Soliton grid: rays with $\tilde{b}\gg b_{\rm ph}$ undergo weak deflection,
rays approaching $b_{\rm ph}^{+}$ wind around the capture region and produce the dense
near-critical band, and rays with $\tilde{b}<b_{\rm ph}$ terminate at the throat. The capture region and the near-critical band grow with $r_{0}$, occupying a larger fraction of the field as the throat enlarges. The deflection diverges logarithmically as $\tilde{b}\to b_{\rm ph}$, the standard signature of a capture boundary in halo-supported
geometries~\cite{Kuhfittig_2014,Cardoso_2019}.
As \(\tilde{b}\) approaches \(b_{\rm ph}\), the deflection angle diverges logarithmically. This behaviour is the standard signature of a capture boundary in halo supported geometries \cite{Kuhfittig_2014,Cardoso_2019}. The intensity profiles shown in Fig.~\ref{fig:soliton_intensity} and the polar shadow maps in Fig.~\ref{fig:soliton_shadowmap} depict the expected features of the soliton geometry. A sharp intensity maximum occurs at \(\tilde{b}=b_{\rm ph}\), followed by a tail that decreases monotonically. Both the peak position and the radius of the central shadow disk scale with \(r_{0}\) through \(b_{\rm ph}\). The bright annulus also becomes sharper as the throat radius increases. For the fiducial boson mass, the factor \(e^{-\Phi_{\rm sol}(r_{0})}\) sets the soliton peak in the range \(\tilde{b}\approx 1.18\,r_{0}\) to \(1.26\,r_{0}\). Consequently, at fixed \(r_{0}\), the bright annulus and the shadow disk are slightly wider than those obtained for the NFW geometry. The shadow angular size follows Eq.~\eqref{eq:shadow_angle} with $R_{s}=b_{\rm ph}$.
Unlike the NFW geometry, whose lensing is
controlled by $r_{0}$ alone, the soliton wormhole carries an additional dependence on the
boson mass through the central density $\rho_{c}\propto m_{b}^{-2}$ (Eq.~\eqref{e2}). Since
$\rho_{c}$ sets the depth of the potential~\eqref{eq:soliton_redshift}, varying $m_{b}$
moves the system between two regimes.
For lighter bosons, the core is denser and the potential deeper. At $m_{b}=1.2\times10^{-18}$~eV ($\rho_{c}=2.06\times10^{-10}\,M_{\odot}\,{\rm pc}^{-3}$) a photon sphere detaches further from the throat at $r_{\rm ph}\approx1.1$--$5.0$~kpc, $b_{\rm ph}$
grows large, and all $100$ sampled rays are captured at every throat radius: the capture region engulfs the full impact-parameter range $\tilde{b}\le8$~kpc and no deflected branch survives.
For heavier bosons, the core is dilute, and the geometry approaches the weak-field, NFW-like regime more closely. At $m_{b}=1.2\times10^{-17}$~eV
($\rho_{c}=2.06\times10^{-12}\,M_{\odot}\,{\rm pc}^{-3}$), the fiducial value used in Figs.~\ref{fig:soliton_raytracing}--\ref{fig:soliton_combined}, the photon sphere sits close to
the throat, $r_{\rm ph}\approx r_{0}$ and $b_{\rm ph}\approx1.18$--$1.26\,r_{0}$, and the
sampled rays are predominantly deflected. The captured fraction rises monotonically with throat size; the remaining rays escape after reaching a turning point. A critical boson mass near
$m_{b}\sim1.2\times10^{-18}$~eV separates the totally capturing regime from the deflecting regime, providing a lensing diagnostic of the wave-dark-matter particle mass absent in the NFW construction~\cite{Schive2014,Mocz_2017,Errehymy_2024,Alshammari_2026}.

\subsubsection{Accretion Disk Analysis of Static Soliton Wormhole}
\label{sec:soliton_accretion_disk}

\begin{figure*}
    \centering
    \includegraphics[width=15cm,height=8cm]{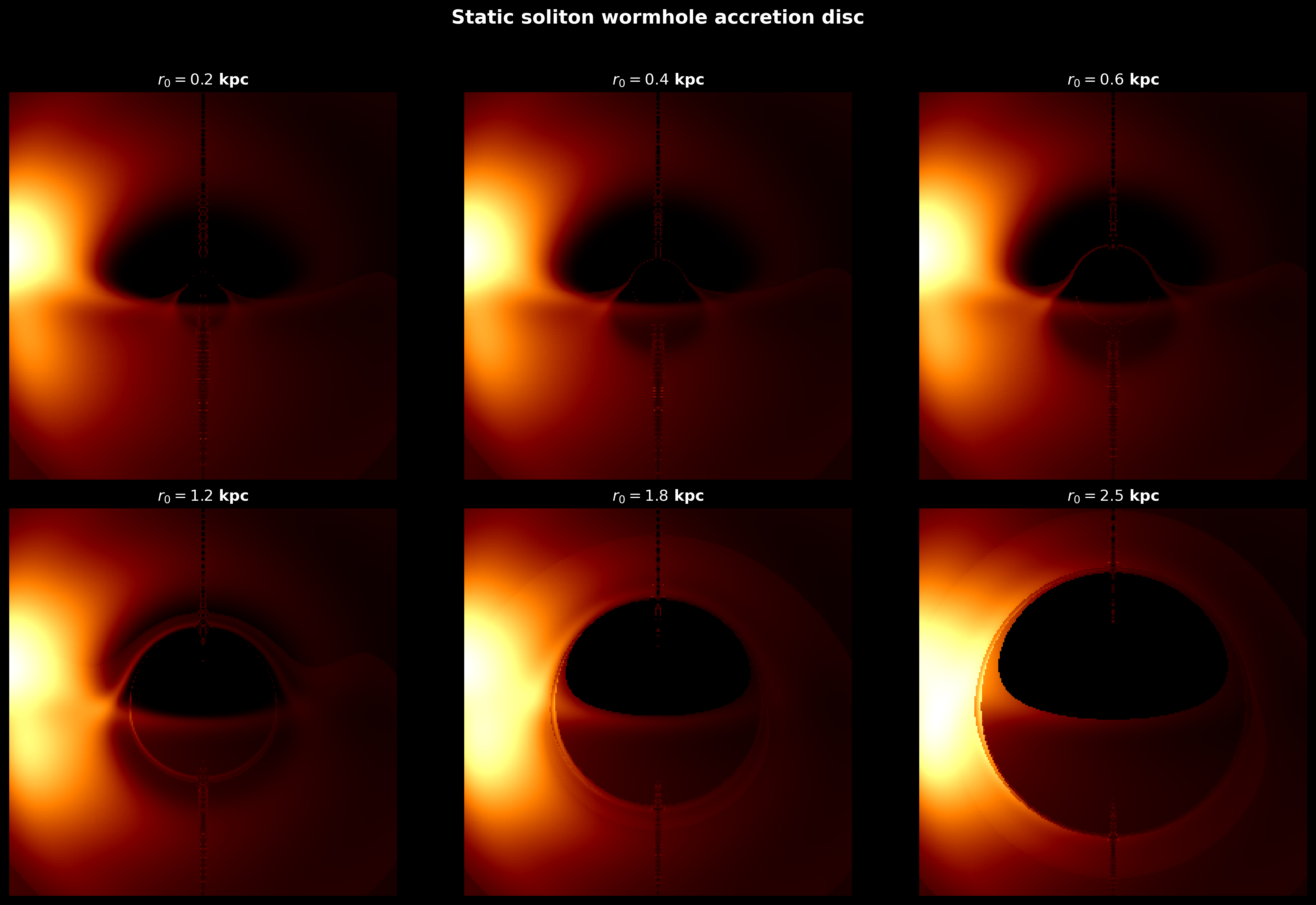}
    \caption{Accretion discs of the static soliton ($\psi$DM) wormhole for six throat radii, $r_0\in\{0.2,\,0.4,\,0.6,\,1.2,\,1.8,\,2.5\}$~kpc, using boson mass $m_b=1.2\times10^{-17}$~eV, core radius $r_c=3$~kpc, and central density $\rho_c=2.06\times10^{-12}\,M_\odot\,\mathrm{pc}^{-3}$. }
    \label{fig:soliton_accretion_disk}
\end{figure*}

Having analyzed the soliton ($\psi$DM) wormhole shadow and intensity maps, we now image a geometrically thin, optically thin accretion disk around the soliton wormhole throat. The ray-tracing framework is identical to that of the static NFW wormhole in Sec.~\ref{sec:accretion_disk}: the same null Hamiltonian constraint Eq.~\eqref{eq:hamiltonian}, the same geodesic system Eqs.~\eqref{eq:dr}--\eqref{eq:dptheta}, the same RK4 integrator with $\Delta\lambda = 0.05$ and $1500$ maximum steps, the same observer setup ($r_{\rm obs} = 30$~kpc, $\theta_{\rm obs} = 75^\circ$, $E=1$, ingoing $p_r$ branch), the same $\pm 12$~kpc field of view on a $260\times260$ grid ($6.76\times10^4$ rays), and the same termination conditions ($r < 1.02\,r_0$ or $r > 35$~kpc). The frame-dragging function $w(r)=2J/r^3$ is again set to $J=0$ for the static disk. The emissivity profile Eq.~\eqref{eq:emissivity} with $r_{\rm cut}=2.5$~kpc, the emission validity window $1.1\,r_0 < r_\times < 20$~kpc, the orbital frequency Eq.~\eqref{eq:Omega_disk}, the Doppler factor Eq.~\eqref{eq:doppler_factor} clamped to $[0.1,3.0]$, and the $g^4$ intensity accumulation Eq.~\eqref{eq:intensity_accum} are all carried over unchanged.\\
The soliton density profile arises in the Bose--Einstein condensate / fuzzy dark matter scenario, where ultralight bosons form a self-gravitating ground-state core~\cite{PhysRevLett.85.1158,Schive2014,Schive_2014_PRL,Hui_2017,Marsh_2016,Ferreira_2021}, and has been invoked as a dark-matter source for traversable wormholes in the galactic halo~\cite{Errehymy_2024,Alshammari_2026}.
The redshift function $\Phi(r)$ and shape function $b(r)$ here are the closed-form soliton expressions evaluated from the soliton density profile (Sec.~\ref{sec:soliton_theory}). In contrast to the NFW case, where $\Phi'(r)$ and $b'(r)$ are available analytically, the radial derivatives of the soliton functions are obtained by second-order central finite differences,
\begin{align*}\label{eq:soliton_fd}
\Phi'(r) \approx \frac{\Phi(r+h) - \Phi(r-h)}{2h},\\
b'(r) \approx \frac{b(r+h) - b(r-h)}{2h},
\end{align*}
with step $h = 10^{-5}$, owing to the algebraic complexity of the soliton $\Phi(r)$ and $b(r)$. These finite-difference estimates replace the analytic $\Phi'$ and $b'$ in the inverse-metric derivatives entering Eq.~\eqref{eq:dpr}. The radial coordinate is floored at $r_0 + 10^{-5}$ before each metric evaluation to prevent sampling inside the throat.\\
Figure~\ref{fig:soliton_accretion_disk} presents the resulting accretion disk images for the soliton wormhole over the same sweep of six throat radii, $r_0 \in \{0.2,\,0.4,\,0.6,\,1.2,\,1.8,\,2.5\}$~kpc, displayed as a $2\times3$ panel with the square-root transfer function $\sqrt{I/I_{\max}}$ normalised per panel. The images reproduce the same qualitative strong-lensing signatures found for the NFW wormhole~\cite{Bambi2013,NedkovaTinchevYazadjiev2013,Shaikh2018}: a bright lensed arc just outside the throat, a pronounced left--right Doppler brightness asymmetry (approaching side blueshift-brightened, receding side redshift-dimmed, amplified by the $g^4$ dependence), a central brightness deficit, and a faint nested secondary image. As with NFW, the weak soliton potential ($|\Phi|\ll1$) admits no photon sphere, so the dark-region boundary stays close to the geometric throat.
These results show that the soliton wormhole, sourced by a fuzzy/Bose--Einstein dark matter core rather than a compact object, reproduces the strong-field lensing signatures established for the NFW case. The soliton and NFW images are closely comparable, reflecting the similarly weak halo-scale potentials, while the differing radial fall-off of $\Phi(r)$ and $b(r)$ between the cored soliton and the cuspy NFW halo~\cite{deBlok_2010,deBlok_2001,Marsh_2015} provides a kinematic handle in principle accessible through the Doppler-induced brightness asymmetry.

\subsubsection{Ray Tracing Analysis of Rotating Soliton Wormhole}
\label{sec:rot_soliton}
\begin{figure*}
    \centering
    \includegraphics[width=15cm,height=8cm]{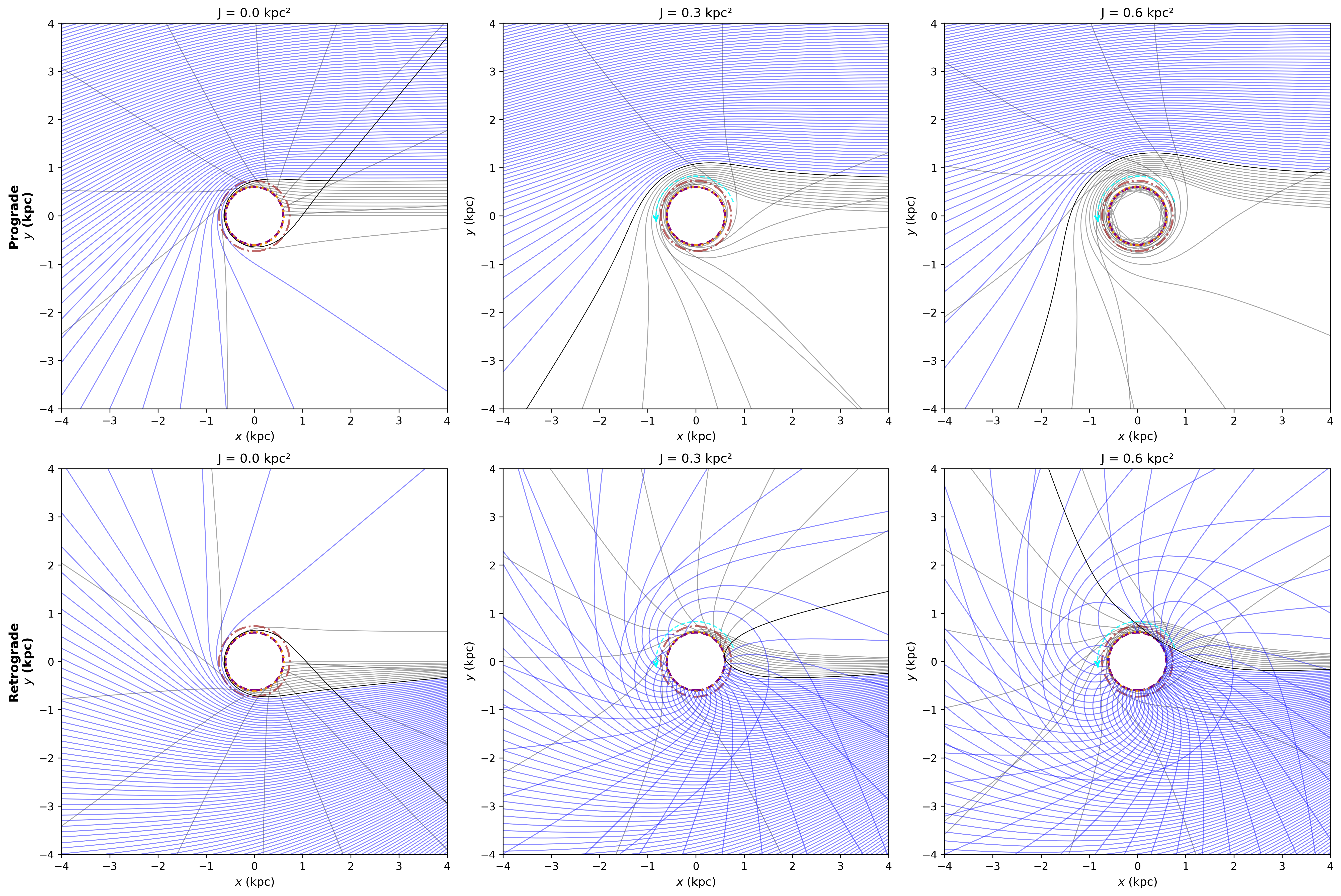}
    \caption{Ray-traced equatorial null geodesics of the slowly rotating soliton ($\psi$DM) wormhole with throat radius $r_0=0.6$~kpc, for angular momenta $J\in\{0,\,0.3,\,0.6\}$~kpc$^2$ (columns) and prograde/retrograde branches (top/bottom rows), using core radius $r_c=3.0$~kpc, boson mass $m_b=1.2\times10^{-17}$~eV, and $\alpha=2^{1/8}-1\approx0.0905$. Dashed purple, dotted orange, and dash-dotted brown circles mark the throat $r_0$, photon orbit radius $r_\text{ph}$, and $b_\text{ph}^{(0)}$, respectively.}
    \label{fig:rot_sol_rt_r06}
\end{figure*}

\begin{figure*}
    \centering
    \includegraphics[width=15cm,height=8cm]{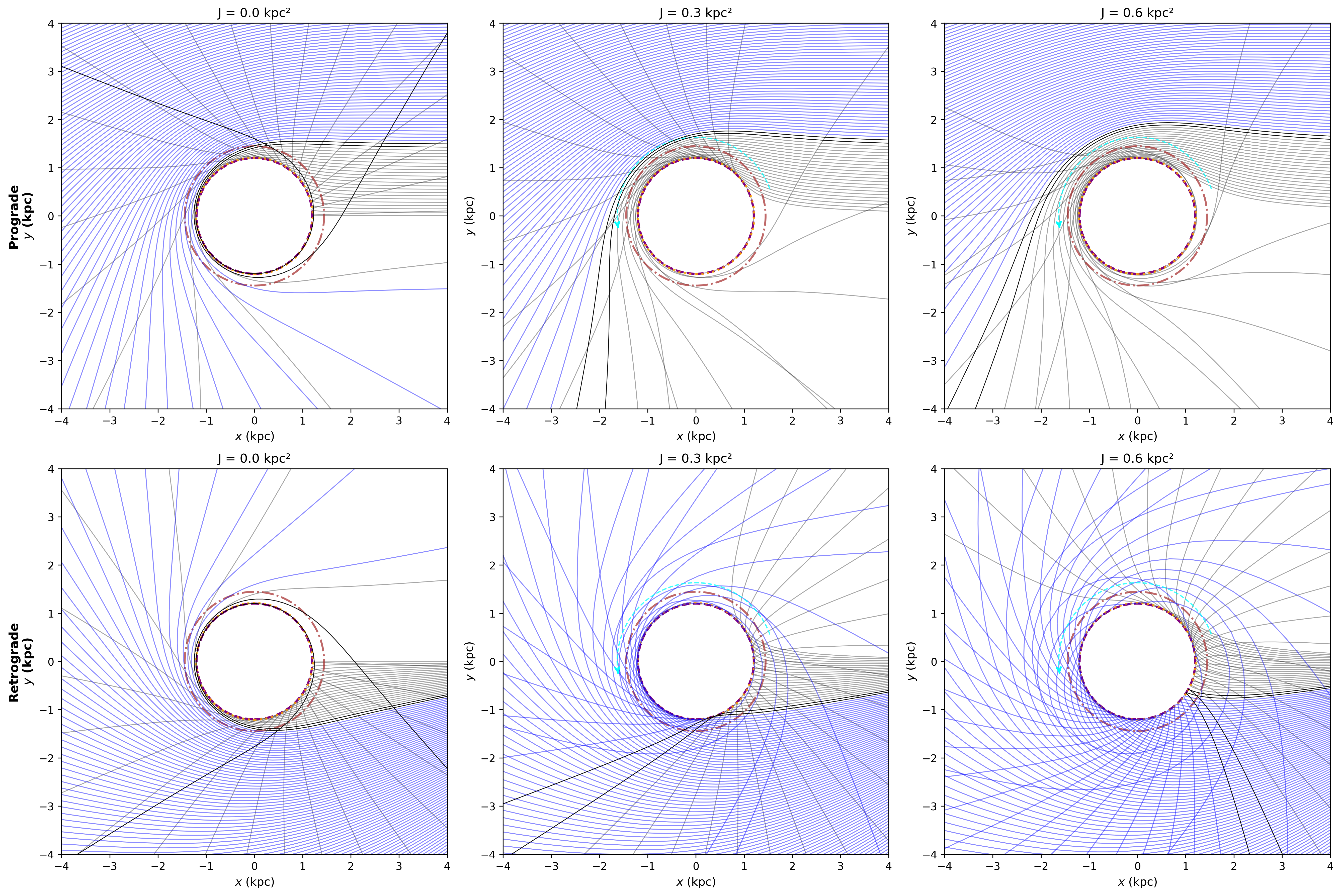}
    \caption{Ray-traced equatorial null geodesics of the slowly rotating soliton ($\psi$DM) wormhole with throat radius $r_0=1.2$~kpc, for angular momenta $J\in\{0,\,0.3,\,0.6\}$~kpc$^2$ (columns) and prograde/retrograde branches (top/bottom rows), using core radius $r_c=3.0$~kpc, boson mass $m_b=1.2\times10^{-17}$~eV, and $\alpha=2^{1/8}-1\approx0.0905$. Dashed purple, dotted orange, and dash-dotted brown circles mark the throat $r_0$, photon orbit radius $r_\text{ph}$, and $b_\text{ph}^{(0)}$, respectively.}
    \label{fig:rot_sol_rt_r12}
\end{figure*}

The slow-rotation soliton ($\psi$DM) wormhole is treated exactly as the rotating NFW case of Sec.~\ref{sec:rot_nfw}; the frame-dragging function $\omega(r)\approx2J/r^{3}$ is inserted into the equatorial Teo metric \eqref{eq:nfw_rotating_metric}, and the Lense--Thirring frequency~\eqref{eq:lt_freq}, conserved quantities, geodesic system, radial and orbit equations~\eqref{eq:conserved_E}--\eqref{eq:orbit_eq_rotating}, and splitting formulae~\eqref{eq:bpm}--\eqref{eq:delta_bph} all carry over, now with the soliton functions $\Phi_{\rm sol}(r)$, $b_{\rm sol}(r)$ of Sec.~\ref{sec:soliton_theory} replacing their NFW counterparts~\cite{Teo1998,HartleThorne1968,Carter1968a,Carter1968b,Wald1984}. We adopt soliton core parameters $r_{c}=3.0$~kpc, ultralight boson mass $m_{b}=1.2\times10^{-17}$~eV, and $\alpha=2^{1/8}-1\approx0.0905$~\cite{PhysRevLett.85.1158,Schive2014,Schive_2014_PRL,Hui_2017,Marsh_2016,Ferreira_2021}.
The essential structural difference from the NFW wormhole is that the soliton redshift function is strong enough near the throat to admit a genuinely unstable circular photon orbit. For each throat radius we locate $r_{\rm ph}$ as the extremum of $\tilde b(r)=r\,e^{-\Phi_{\rm sol}(r)}$, solving $d\tilde b/dr=0$ by central-difference bracketing bisection, giving $b_{\rm ph}^{(0)}=r_{\rm ph}\,e^{-\Phi_{\rm sol}(r_{\rm ph})}$~\cite{Synge1966,ClaudelVirbhadraEllis2001,Perlick2004,Tsukamoto2017}. In contrast to the NFW geometry, for which \(r\Phi'\ll 1\) constrains the capture boundary to the throat, the soliton geometry reveals a genuine circular null orbit at \(r_{\rm ph}>r_{0}\). Therefore, we can apply the splitting relations from Eqs.~\eqref{eq:bpm} and~\eqref{eq:delta_bph} using \(r_{\rm ph}\) rather than at \(r_{0}\). In this context, defining \(K\equiv b_{\rm ph}^{(0)}\), the corresponding impact parameters are $$b_{\pm}=K\left[1\pm K\omega(r_{\rm ph})\right],$$ as discussed in \cite{Hioki2009,Konoplya2018,Tsupko2017,Shaikh2018}.
The integration setup matches Sec.~\ref{sec:rot_nfw}: RK4 launch from $x_{\rm start}=15$~kpc, inward integration to the throat ($r=r_{0}+10^{-7}$~kpc) or a turning point, $\beta\in[0.01,8.0]$~kpc with $N=120$ rays per panel, near-critical band $|\beta-b_{\rm ph}|<0.05$~kpc, and the same blue/black ray colouring and cyan $\omega>0$ arc~\cite{Teo1998,Bambi2013}. We present two throat radii, $r_{0}=0.6$~kpc (Fig.~\ref{fig:rot_sol_rt_r06}) and $r_{0}=1.2$~kpc (Fig.~\ref{fig:rot_sol_rt_r12}), each a $2\times3$ grid over prograde (top), retrograde (bottom), and $J\in\{0,\,0.3,\,0.6\}\,\mathrm{kpc}^{2}$. A difference from the NFW figures is that the purple dashed \(r_{0}\), the orange dotted \(r_{\rm ph}\), and the brown dash-dotted \(b_{\rm ph}^{(0)}\) are visible separately. In particular, both \(r_{\rm ph}\) and \(b_{\rm ph}^{(0)}\) lie far from the throat.\\
For \(J=0\), the rows are symmetric about the \(x\)-axis. Near critical rays orbit the photon sphere several times before either escaping or being captured. This behavior produces strong lensing spirals that are absent in the NFW geometry, which does not contain a photon sphere \cite{Bozza2002,Tsukamoto2017}. For $J>0$ this symmetry breaks: prograde near-critical rays co-wind with $\omega>0$ into dense spirals against the photon ring, while retrograde rays sweep a wider azimuthal range and remain trapped in long windings, visible as the denser lower-half spiral families~\cite{NedkovaTinchevYazadjiev2013,Shaikh2018,Jusufi2018,Tsupko2017}. The captured fractions separate monotonically with $J$, consistent with $\Delta b_{\rm ph}\approx4J/r_{\rm ph}$~\cite{Hioki2009}. The relative asymmetry $\Delta b_{\rm ph}/b_{\rm ph}\propto J/r_{\rm ph}^{2}e^{-\Phi}$ is larger for $r_{0}=0.6$~kpc than for $r_{0}=1.2$~kpc at fixed $J$, so the prograde--retrograde contrast is more pronounced in Fig.~\ref{fig:rot_sol_rt_r06}, the geometric manifestation of $\Omega_{\rm LT}=3J/r^{4}$ steepening at small radii~\cite{Hartle1967,CiufoliniPavlis2004}.

\subsubsection{Intensity profiles and shadow maps}
\begin{figure*}
    \centering
    \subfigure[$r_0=1.5$ kpc]{
    \includegraphics[width=15cm,height=8cm]{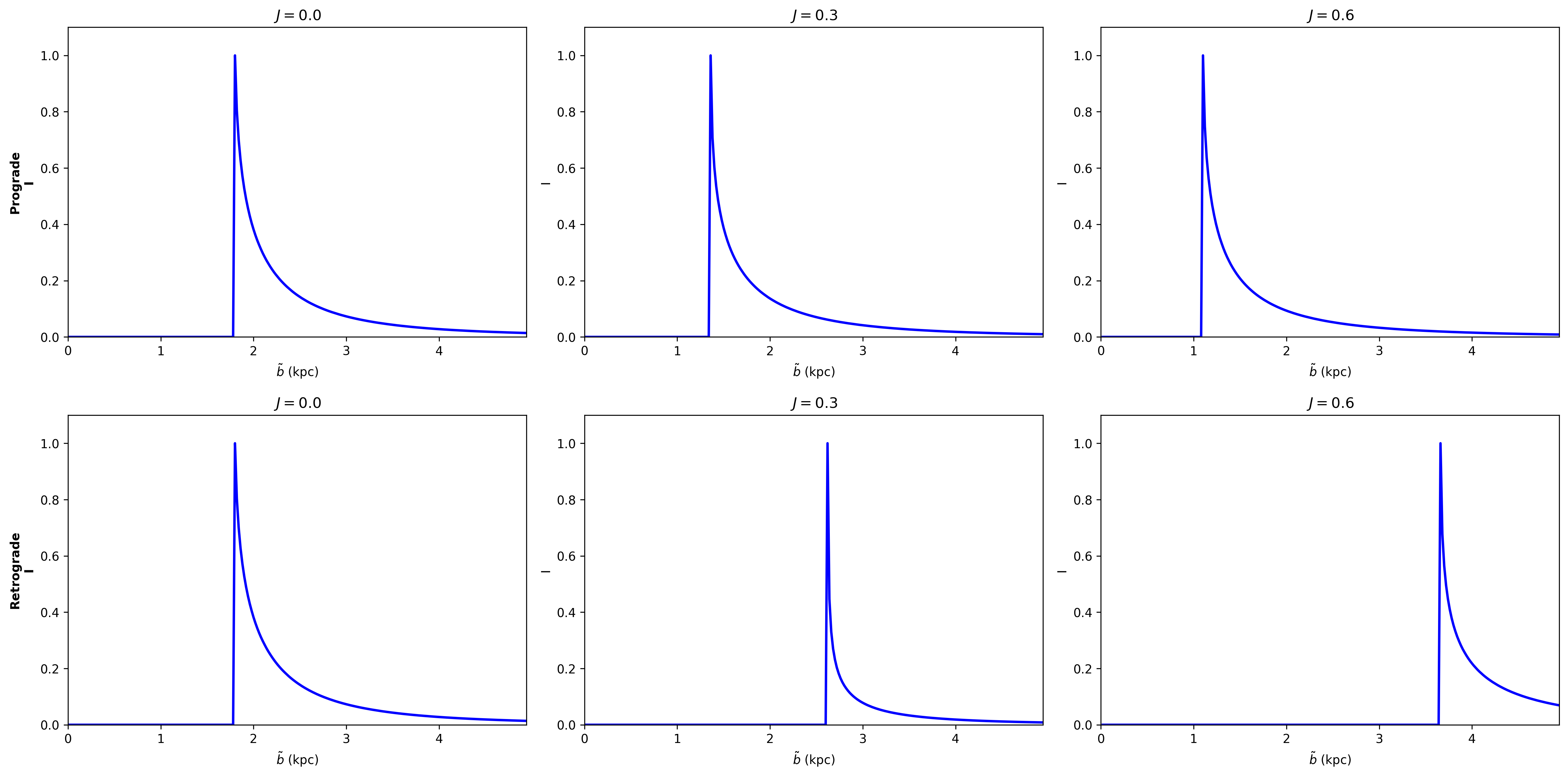}
    \label{fig:sol_intensity_r15}
}
\hfill
\subfigure[$r_0=2.5$ kpc]{
    \includegraphics[width=15cm,height=8cm]{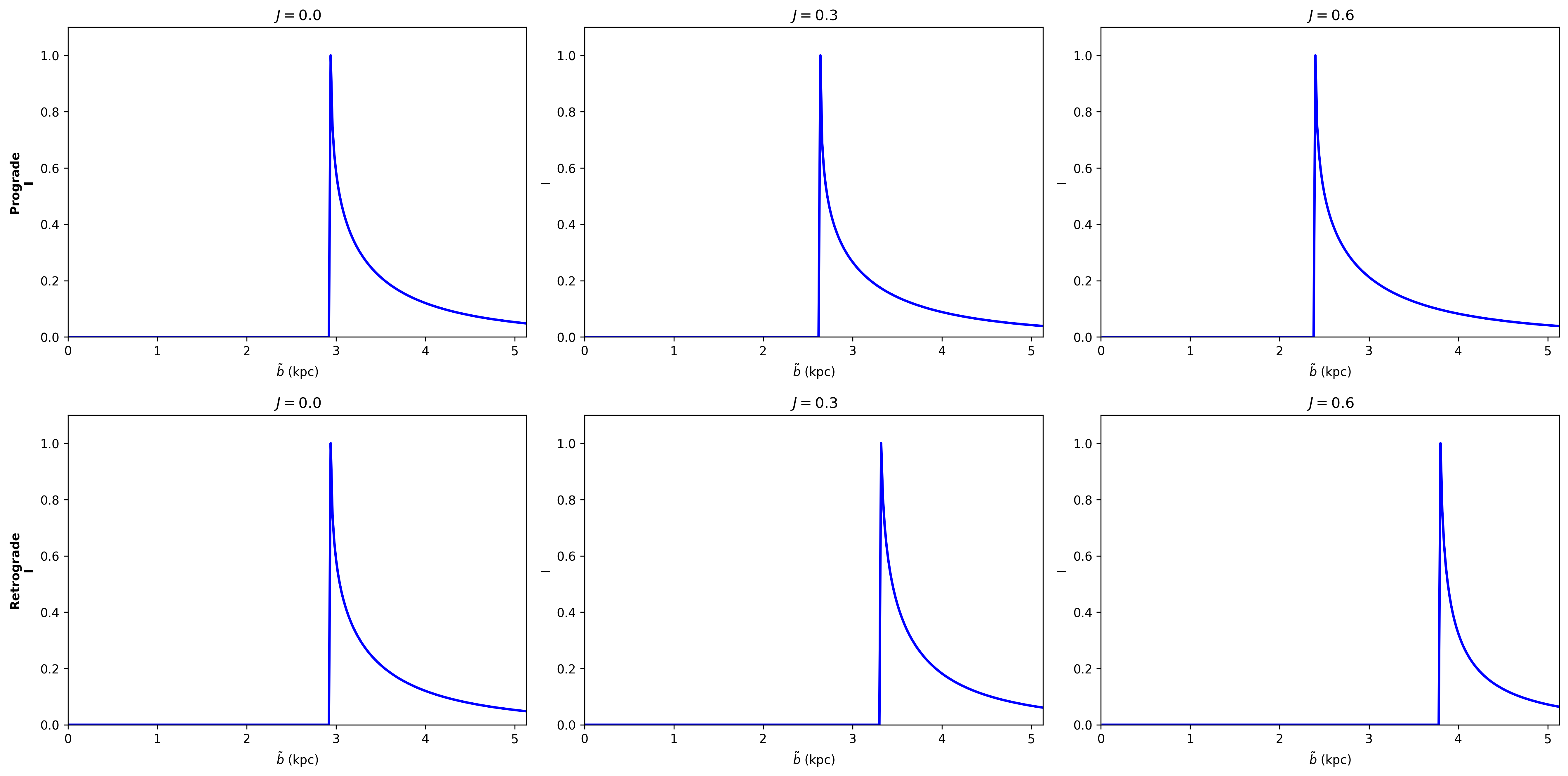}
    \label{fig:sol_intensity_r25}
}
\caption{Normalised equatorial intensity profiles $I(\tilde{b})$ of the slowly rotating soliton ($\psi$DM) wormhole for throat radii $r_0=1.5$~kpc (left, static peak $b_\text{ph}^{(0)}\approx1.75$~kpc) and $r_0=2.5$~kpc (right, static peak $b_\text{ph}^{(0)}\approx2.9$~kpc), at $J\in\{0,\,0.3,\,0.6\}$~kpc$^2$, using core radius $r_c=3.0$~kpc and boson mass $m_b=1.2\times10^{-17}$~eV. For $r_0=1.5$~kpc, the prograde peak shifts inward to $b_-\approx1.35$~kpc ($J=0.3$) and $\approx1.1$~kpc ($J=0.6$), while the retrograde peak shifts outward to $b_+\approx2.6$~kpc and $\approx3.65$~kpc. Top rows show prograde, bottom rows retrograde branches, scanned over $400$ impact parameters $\tilde{b}\in[0.02,8.0]$~kpc.}
\label{fig:sol_intensity_combined}
\end{figure*}

The intensity-transport pipeline and the dipolar shadow-boundary construction are those of Sec.~\ref{sec:rot_nfw}, Eqs.~\eqref{eq:intensity_rot} and~\eqref{eq:bcut_dipole}, with $b_{\rm NFW}\to b_{\rm sol}$ and the prograde/retrograde edges $b_{-}$, $b_{+}$ measured from ray fate; rays that plunge to the throat give zero intensity and define the shadow interior. We scan $400$ impact parameters over $\tilde b\in[0.02,8.0]$~kpc per branch for $r_{0}=1.5$~kpc and $r_{0}=2.5$~kpc at $J\in\{0,\,0.3,\,0.6\}\,\mathrm{kpc}^{2}$, normalised to unit peak per panel (Figs.~\ref{fig:sol_intensity_r15}--\ref{fig:sol_intensity_r25})~\cite{Teo1998,Bambi2013,Synge1966,FalckeMeliaAgol2000,Younsi2016,Cunningham1973,Luminet1979,MisnerThorneWheeler1973}.\\
For both throat radii, the static profiles show a single sharp peak at $b_{\rm ph}^{(0)}$ with the monotonic logarithmic-accumulation tail of a photon-sphere capture boundary~\cite{Luminet1979,Bambi2013,GrallaHolzWald2019}. For $r_{0}=1.5$~kpc the static peak sits at $b_{\rm ph}^{(0)}\approx1.75$~kpc; for $J>0$ the prograde peak moves \emph{inward} to $b_{-}\approx1.35$~kpc ($J=0.3$) and $\approx1.1$~kpc ($J=0.6$), while the retrograde peak moves \emph{outward} to $b_{+}\approx2.6$~kpc and $\approx3.65$~kpc, in agreement with $b_{\pm}=K(1\pm K\omega)$. For $r_{0}=2.5$~kpc the static peak is at $b_{\rm ph}^{(0)}\approx2.9$~kpc, with prograde shifts inward to $\approx2.6$ and $\approx2.4$~kpc and retrograde shifts outward to $\approx3.3$ and $\approx3.8$~kpc at $J=0.3$ and $0.6$ \cite{Teo1998,Hartle1967,HartleThorne1968}. The relative displacement $\sim K\omega(r_{\rm ph})\propto J/r_{\rm ph}^{2}$ is smaller for the larger throat, so the splitting in Fig.~\ref{fig:sol_intensity_r25} is more modest than in Fig.~\ref{fig:sol_intensity_r15}~\cite{NedkovaTinchevYazadjiev2013,Shaikh2018}. Because the soliton possesses a genuine photon sphere, both the prograde and retrograde shifts appear as resolved peaks, in contrast to the NFW case where the prograde branch was floored by the throat-grazing cutoff.\\
\begin{figure*}
    \centering
\subfigure[$r_0=1.5$ kpc]{
    \includegraphics[width=14cm,height=6cm]{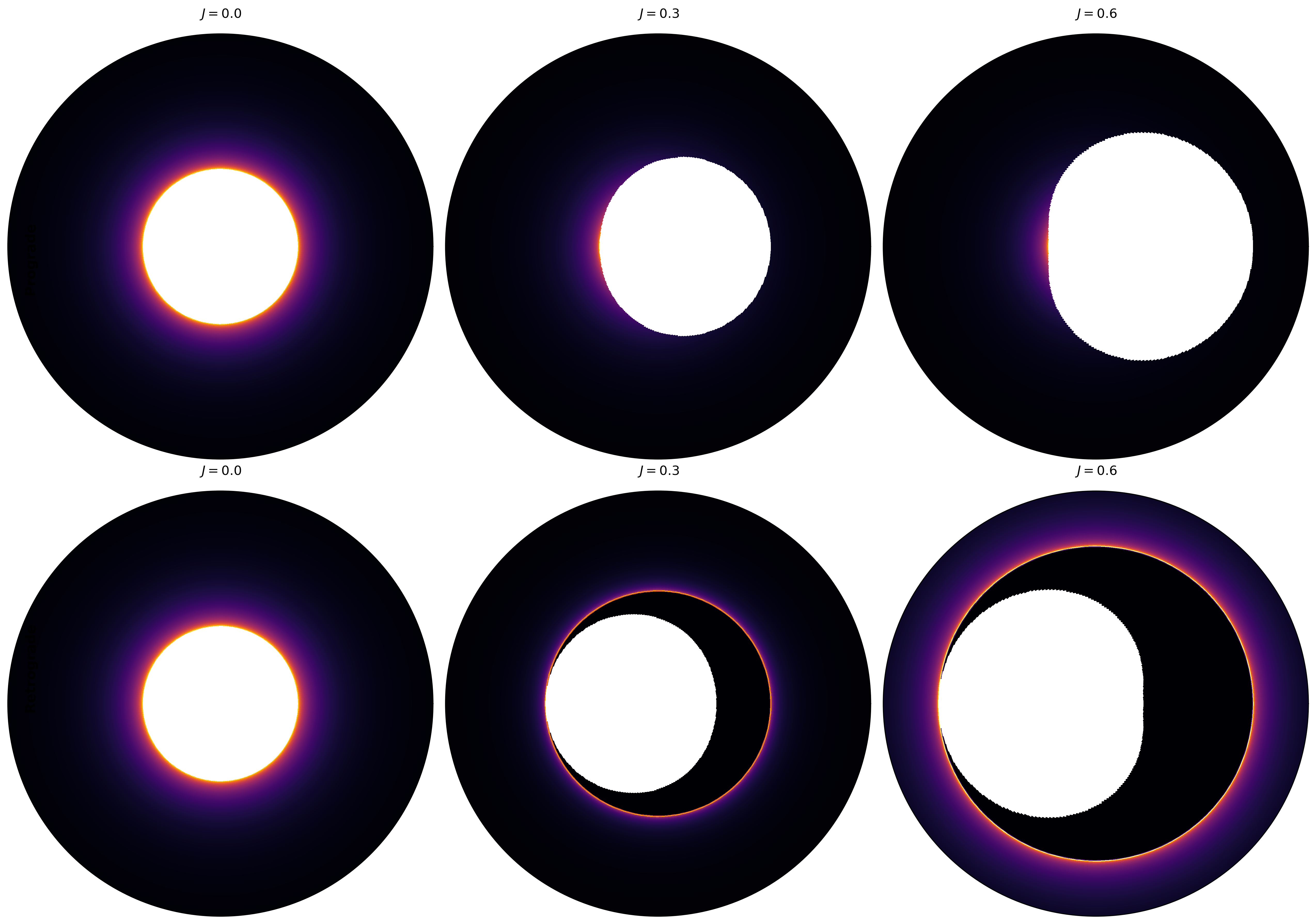}
    \label{fig:sol_shadow_r15}
}
\hfill
\subfigure[$r_0=2.5$ kpc]{
    \includegraphics[width=14cm,height=6cm]{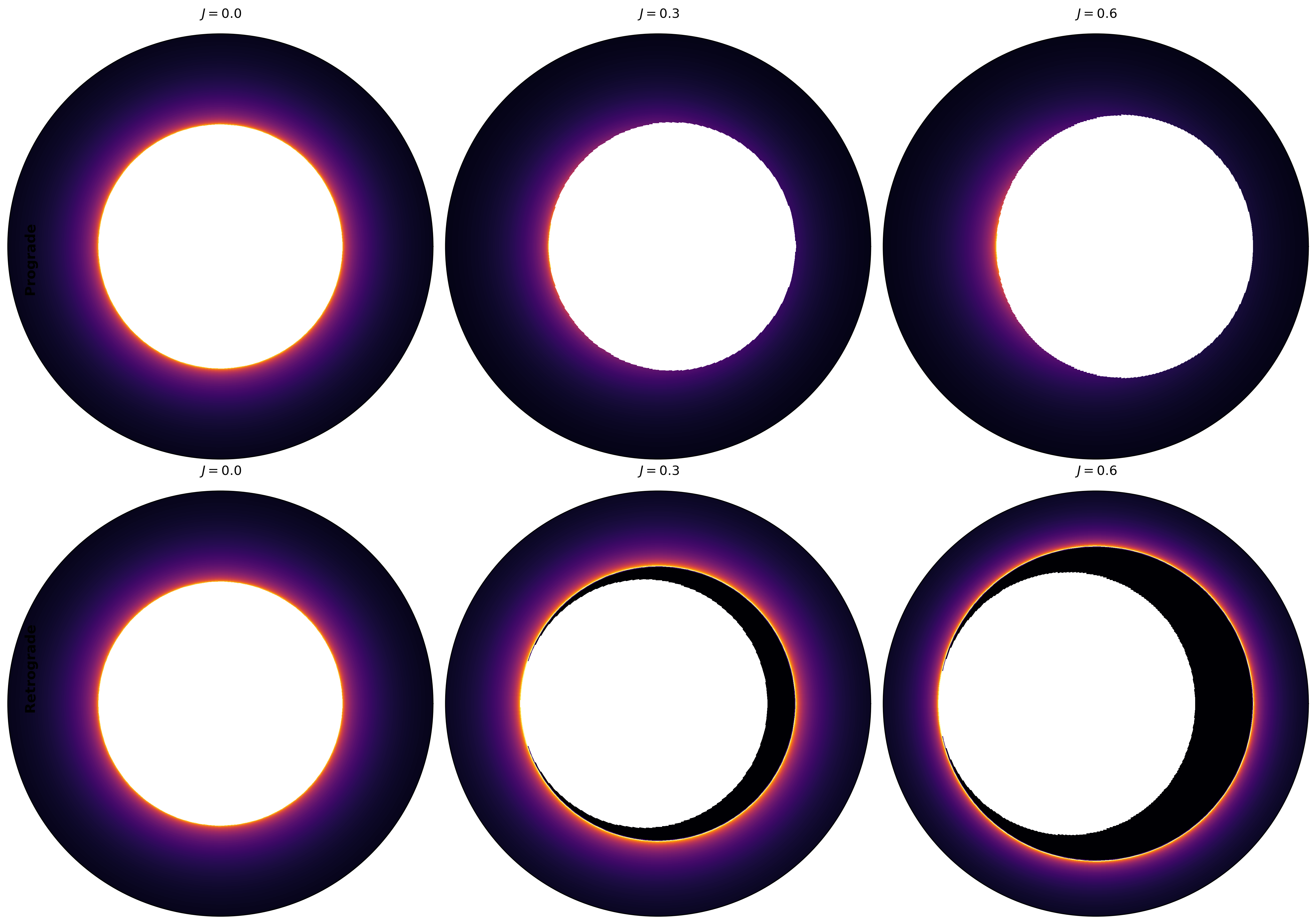}
    \label{fig:sol_shadow_r25}
}
\caption{Polar shadow maps of the slowly rotating soliton ($\psi$DM) wormhole for throat radii $r_0=1.5$~kpc and $r_0=2.5$~kpc, at angular momenta $J\in\{0,\,0.3,\,0.6\}$~kpc$^2$, using core radius $r_c=3.0$~kpc and boson mass $m_b=1.2\times10^{-17}$~eV. Top rows show prograde-rendered panels, bottom rows retrograde-rendered panels; the shadow boundary is ringed by a true photon ring sourced by the detached unstable circular orbit at $r_\text{ph}>r_0$, with maximum centroid offset $\tfrac{1}{2}\Delta b_\text{ph}\approx2J/r_\text{ph}$.}
\label{fig:sol_shadow_combined}
\end{figure*}
In the static case the shadow boundary in Figs.~\ref{fig:sol_shadow_r15}--\ref{fig:sol_shadow_r25} is a circle of radius $b_{\rm ph}^{(0)}$ centred at the origin, ringed by a thin brightening that is here a true photon ring sourced by the unstable circular orbit rather than the throat-grazing feature of the NFW case~\cite{Luminet1979,Bambi2013,GrallaHolzWald2019,Tsukamoto2017}. For $J>0$ the dark region in the prograde row displaces toward $\theta=0$, reaching $b_{+}$ on one side and $b_{-}$ on the other, with the retrograde panels mirrored toward $\theta=\pi$ and a maximum centroid offset $\tfrac12\Delta b_{\rm ph}\approx2J/r_{\rm ph}$~\cite{NedkovaTinchevYazadjiev2013,Jusufi2018,Konoplya2018,Tsupko2017}. The crescent asymmetry grows with $J$ and is stronger for the smaller throat where $K\omega(r_{\rm ph})$ is largest; at $J=0.6\,\mathrm{kpc}^{2}$ this parameter approaches the edge of the slow-rotation regime, so those panels are qualitative and an exact description requires an exact rotating soliton solution or a higher-order Hartle--Thorne expansion~\cite{Hartle1967,HartleThorne1968,Bronnikov2017}.\\
As in the NFW case, the centroid offset yields $J\approx r_{\rm ph}\,\Delta b_{\rm ph}/4$ independently of the soliton core parameters~\cite{Hioki2009,Konoplya2018,EHT2019}. The distinguishing feature is that the cored soliton profile supports a photon sphere detached from the throat, so the soliton shadow carries a genuine photon ring and a larger absolute $b_{\rm ph}^{(0)}$ at comparable throat radius; the differing radial structure of $\Phi(r)$ between the two dark-matter models is therefore in principle separable through the combination of shadow size, photon-ring sharpness, and rotational centroid offset~\cite{deBlok_2010,Marsh_2015,Ferreira_2021,Falcke2013,Bambi2013}.

\subsubsection{Accretion Disc Analysis of Rotating Soliton Wormhole}
The rotating accretion disc construction of Sec.~\ref{sec:rot_nfw_disc} is applied unchanged to the slowly rotating soliton ($\psi$DM) wormhole. The Teo metric, inverse-metric components Eq.~\eqref{eq:inv_metric_rot}, null super-Hamiltonian Eq.~\eqref{eq:H_rot}, geodesic system Eqs.~\eqref{eq:rdot_thetadot_rot}--\eqref{eq:ptheta_dot_rot}, frame-dragging $\omega(r)=2J/r^{3}$, circular-orbit frequency Eq.~\eqref{eq:omega_circular}, emitter four-velocity Eq.~\eqref{eq:ut_rot}, redshift factor Eq.~\eqref{eq:g_doppler_rot} clamped to $g\in[0.1,3.0]$, and the $g^{4}$ intensity accumulation Eq.~\eqref{eq:dI_rot} are all identical, now with the soliton redshift function $\Phi_{\rm sol}(r)$ and shape function $b_{\rm sol}(r)$ of Sec.~\ref{sec:soliton_theory} in place of their NFW counterparts~\cite{Teo1998,Hartle1967,HartleThorne1968,Errehymy_2024,Alshammari_2026}. As in the static soliton disc of Sec.~\ref{sec:soliton_accretion_disk}, the metric derivatives $\Phi'(r)$, $b'(r)$, and $\omega'(r)$ entering Eqs.~\eqref{eq:pr_dot_rot} and~\eqref{eq:omega_circular} are evaluated by second-order central finite differences with step $h=10^{-5}$, owing to the algebraic complexity of the closed-form soliton functions, and the radial coordinate is floored at $r_{0}+10^{-5}$ before each metric evaluation.\\
The observer is placed at $r_{\rm obs}=30$~kpc and polar inclination $\theta_{\rm obs}=60^{\circ}$, with a $300\times300$ pixel screen of half-width $12$~kpc and the same Bardeen--Cunningham camera relations Eq.~\eqref{eq:camera_rot}, fixed-step RK4 integration ($d\lambda=0.05$, $1500$ steps), equatorial-crossing detection by sign changes in $\theta(\lambda)-\pi/2$, emissivity Eq.~\eqref{eq:emissivity_rot} with $r_{\rm cut}=2.5$~kpc and validity window $1.1\,r_{0}<r_{\times}<20$~kpc, and termination at $r<1.02\,r_{0}$ or $r>35$~kpc. A single throat radius $r_{0}=1.0$~kpc is imaged for $J\in\{0,\,0.3,\,0.6\}\,\mathrm{kpc}^{2}$. The prograde (top) and retrograde (bottom) rows are generated by simultaneously reversing the sign of the wormhole angular momentum and of the photon angular momentum, $J\to-J$ and $L\to-L$, so that each row probes the corresponding co- or counter-rotating sense of disc material relative to the frame dragging~\cite{Cunningham1973,Luminet1979,Bambi2013}. Each panel is rendered with the square-root transfer function $\sqrt{I/I_{\max}}$.\\
\begin{figure*}
    \centering
    \includegraphics[width=15cm,height=8cm]{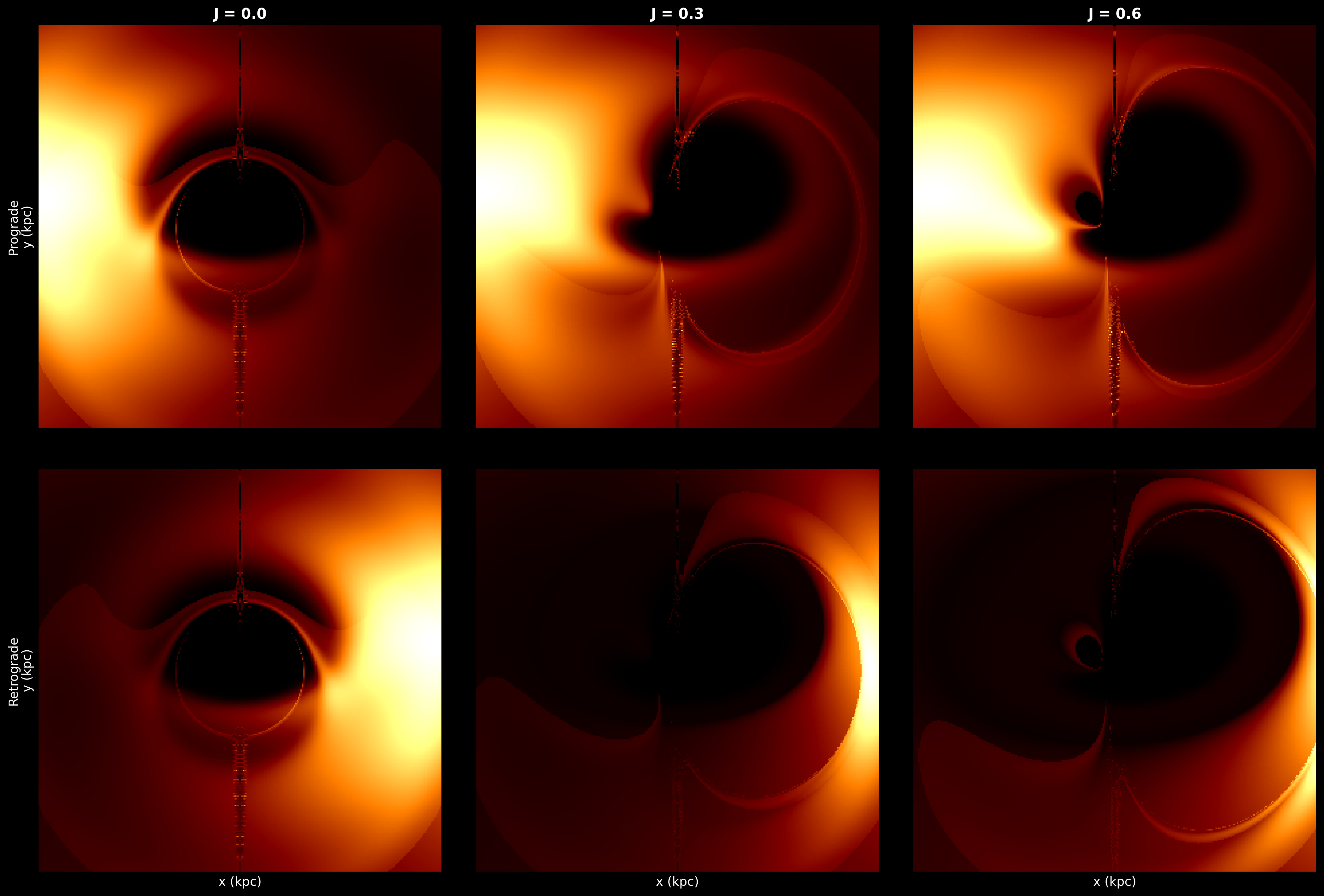}
    \caption{Accretion disc images of the slowly rotating soliton ($\psi$DM) wormhole with throat radius $r_0=1.0$~kpc, for angular momenta $J\in\{0,\,0.3,\,0.6\}$~kpc$^2$, using core radius $r_c=3.0$~kpc and boson mass $m_b=1.2\times10^{-17}$~eV. }
    \label{fig:rot_sol_disc}
\end{figure*}
Figure~\ref{fig:rot_sol_disc} shows the resulting images. The $J=0$ panels reproduce the static soliton Luminet-type disc of Sec.~\ref{sec:soliton_accretion_disk}: a bright lensed crescent with a strong left--right Doppler asymmetry from the prograde Keplerian motion of the disc material, a central dark region, and a thin photon-ring brightening encircling it, the latter sourced by the genuine soliton photon sphere standing off from the throat rather than by the throat itself~\cite{Luminet1979,Cunningham1973,Bambi2013,GrallaHolzWald2019,Tsukamoto2017}. With the row-flip convention $J\to-J$, $L\to-L$, the two $J=0$ panels are mirror images across the vertical axis, as required by the reflection symmetry of the static spacetime~\cite{Younsi2016,FalckeMeliaAgol2000}.\\
For $J>0$ frame dragging imposes a definite handedness. In the prograde row the central dark region elongates and shifts laterally, and the bright crescent intensifies on the approaching side, consistent with the photon-orbit splitting $b_{\pm}=K(1\pm K\omega)$ of Eq.~\eqref{eq:bpm}~\cite{NedkovaTinchevYazadjiev2013,Shaikh2018,Jusufi2018}; the retrograde row exhibits the mirrored configuration with the dark region displaced to the opposite side~\cite{Hioki2009,Konoplya2018,Tsupko2017}. Both the displacement and the contrast of the crescent increase monotonically with \(J\). This trend is consistent with the approximate relation
\[
\Delta b_{\rm ph}\approx \frac{4J}{r_{\rm ph}},
\]
given in Eq.~\eqref{eq:delta_bph}. At $J=0.6\,\mathrm{kpc}^{2}$, a compact secondary image of the lensed inner edge appears adjacent to the dark region, achieving the same handedness as the primary~\cite{GrallaHolzWald2019,Luminet1979,Bambi2013}. As established for the rotating NFW disc, the slow-rotation expansion parameter $K\omega(r_{\rm ph})$ exceeds unity at $J=0.6\,\mathrm{kpc}^{2}$ for these throat scales, so those panels are qualitative illustrations of the frame-dragging morphology rather than quantitatively exact images; a fully reliable description would require an exact rotating soliton wormhole solution or a higher-order Hartle--Thorne expansion~\cite{Hartle1967,HartleThorne1968,Bronnikov2017}, while the static and $J=0.3\,\mathrm{kpc}^{2}$ panels remain within the validity of the leading-order construction.\\
The observational implications mirror those of the rotating NFW disc: the centroid offset of the dark region scales linearly with $J$ at fixed $r_{0}$ in the perturbative regime, yielding $J\simeq r_{\rm ph}\,\Delta b_{\rm ph}/4$ independently of the soliton core parameters~\cite{Hioki2009,Konoplya2018}; the prograde--retrograde mirror symmetry is broken only by the sign of $J$ and so serves as a model-independent check on the rotation sense~\cite{Bardeen1973,FalckeMeliaAgol2000,EHT2019}; and the kiloparsec-scale shadow and asymmetry remain macroscopic (arcminute scale at the NGC~2366 distance $d=3.3$~Mpc), well above current VLBI resolution~\cite{EHT2019,EHT2022,Falcke2013,Bambi2013}. Relative to the cuspy NFW disc, the cored soliton produces a sharper, larger photon ring set by its detached photon sphere, so the combination of disc-image centroid offset, photon-ring radius, and crescent contrast distinguishes the two dark-matter profiles while jointly constraining the halo structure and the wormhole rotation~\cite{deBlok_2010,Marsh_2015,Ferreira_2021,NedkovaTinchevYazadjiev2013,Shaikh2018}.

\section{Detailed Comparative Analysis of the Models}\label{sec5}
The four wormhole configurations investigated in Sec.~IV, the static NFW, the static soliton, the slowly rotating NFW, and the slowly rotating soliton, share the same Morris--Thorne framework and the same astrophysical parameter set drawn from the dwarf galaxy NGC 2366, yet they produce distinct spacetime geometries and optical signatures. The contrast arises from two independent sources: the microphysical difference between a cuspy NFW halo and a cored solitonic wave-dark-matter distribution, and the dynamical difference between a static geometry and one carrying a first-order Lense--Thirring frame-dragging potential. We first compare the static models in terms of photon-orbit structure, shadow scaling, and disk morphology, then examine how rotation modifies each model, and close with the observational prospects this comparison opens up.

\subsection{Static Case} \label{sec:comparative_static}
The two static configurations investigated in the previous section, the NFW-supported and soliton-supported wormholes with the same NGC 2366 parameter set, yet they diverge sharply in their null-geodesic content and, consequently, in every observable derived from it. The origin of this divergence is not geometric in a generic sense but is traceable to a single quantity: the depth and shape of the redshift function $\Phi(r)$ sourced by each halo profile.\\
For the NFW halo, the redshift $\Phi_{\rm NFW}(r)$ remains shallow across the full range of throat radii considered, with $|\Phi_{\rm NFW}(r_0)| \lesssim 0.08$ even at $r_0 = 0.2$~kpc.
The condition for an unstable circular null orbit, \(r\,\Phi'(r)=1,\)
is not satisfied outside the throat. Therefore, no real photon sphere is formed, and hence the capture boundary becomes \[ b_{\rm ph}=r_0 e^{\Phi_{\rm NFW}(r_0)} \approx 1.05\,r_0.\]
Thus, it is determined almost entirely by the throat radius \cite{Kuhfittig_2014}.
In contrast, the soliton halo contains a larger fraction of its mass within the core radius \(r_c=3\,\mathrm{kpc}\) due to the steep density profile.
For a boson mass \(m_b=1.2\times10^{-17}\,\mathrm{eV}\) and central density \(\rho_c=2.06\times10^{-12}\,M_\odot\,\mathrm{pc}^{-3}\), the potential becomes sufficiently deep for the condition \(r\,\Phi'(r)=1\) to admit a physical root at \(r_{\rm ph}\) slightly outside \(r_0\). Hence, the critical impact parameter is separated from the throat and satisfies
\[
b_{\rm ph}^{\rm sol}
\approx 1.18\,r_0
\quad \text{to} \quad
1.26\,r_0,
\]
which is larger than the NFW value for the same throat radius \(r_0\). However, the difference follows directly from the mass distributions. The NFW potential increases logarithmically towards the center, whereas a sufficiently dense and compact soliton core can support a truly unstable circular orbit. Further, the relative potential depths also demonstrate this distinction. For the NFW profile,
\[
\rho_s r_s^3
\sim
10^{-3}\,M_\odot\,\mathrm{pc}^{-3}
\times
(1.447\,\mathrm{kpc})^3,
\]
which limits \(|\Phi_{\rm NFW}|_{\max}\) to approximately \(0.05\). In the soliton case, the quantity \(\rho_c r_c^4\) is sufficiently large for
\[
r\left|\Phi_{\rm sol}'\right|=1
\]
to be reached at intermediate radii when the boson mass is sufficiently small.
This structural difference determines the shadow properties of the two models. Since both spacetimes are static and spherically symmetric, hence their shadows are exactly circular and have no intrinsic shape distortion. Therefore, their comparison will be based on shadow size and boundary sharpness. For the NFW model, the shadow boundary is only slightly outside the throat,
\[
b_{\rm ph}-r_0\approx0.05\,r_0.
\]
Consequently, its angular size grows linearly with \(r_0\) and depends only on the halo parameters. The relation
\[
b_{\rm ph}\approx1.05\,r_0
\]
remains valid throughout the interval \(r_0\in[0.2,2.5]\,\mathrm{kpc}\).
For the soliton model, the shadow radius is determined by the photon sphere located outside the throat. Therefore, it depends on both the throat radius and the microscopic properties of the halo. At the distance of NGC 2366, \(d=3.3\,\mathrm{Mpc}\), the NFW model predicts that the angular diameter of the shadow is roughly \(2.2\) arcminutes, given a throat radius of \(r_0=1.0\,\mathrm{kpc}\). Interestingly, with the same throat radius, the shadow produced by the soliton is estimated to be approximately \(15\%\) and \(20\%\) larger.
Both values are far above the resolution of horizon-scale VLBI facilities, but the modest, arcsecond-to-arcminute-scale difference at fixed $r_0$ constitutes a discriminator between the two dark matter profiles, provided the throat radius can be independently bounded, for instance from rotation-curve fitting of the host galaxy.
\\
The lensing and intensity signatures follow the same logic. Both models produce an intensity profile that peaks sharply at $\tilde b = b_{\rm ph}$ and decays approximately logarithmically at larger impact parameter, falling below 20\% of peak within one to two kiloparsecs of the boundary; this is the generic signature of photon accumulation at a capture boundary in an optically thin medium and is expected on general grounds for any quasi-spherical compact geometry~\cite{Luminet1979,Bambi2013,GrallaHolzWald2019}. The two models differ, however, in the width and sharpness of that peak. In the NFW case the absence of a true photon sphere means deflection near the capture boundary is dominated by the single near-throat redshift factor, producing a comparatively narrow, soft peak that scales mildly with $r_0$. In the soliton scenario, photons that follow near critical trajectories around the detached photon sphere occupy a broader annular region. This produces a wider brightening feature. The accretion disk images in Figs.~\ref{fig:accretion_disk} and~\ref{fig:soliton_accretion_disk} demonstrate this behavior. In both cases, a bright lensed arc appears outside the central dark region, together with a pronounced Doppler asymmetry arising from the relativistic \(g^4\) factor. A faint secondary image is also visible inside the primary image. These features are standard signatures of strong deflection in optically thin accretion disks \cite{Luminet1979,Bambi2013,NedkovaTinchevYazadjiev2013,Shaikh2018}. The NFW dark region has a soft, diffuse edge set by throat-grazing deflection, while the soliton dark region is bounded by a somewhat sharper photon ring sourced by genuine photon-sphere winding, and its secondary images are correspondingly slightly more compact. The Doppler asymmetry itself is qualitatively similar in both cases, since the orbital frequency $\Omega(r_\times)=e^{\Phi}\sqrt{\Phi'/r}$ depends on a weak potential ($|\Phi|\ll1$) at these throat scales in both models, though the steeper fall-off of $\Phi_{\rm sol}$ relative to $\Phi_{\rm NFW}$ produces a subtly different degree of contrast between the approaching and receding sides of the disk, an effect that in principle offers an independent kinematic probe of the halo profile~\cite{deBlok_2010,Marsh_2015}, though isolating it from inclination and emissivity modeling is left for future work.
\\
A particularly consequential parameter dependence is the sensitivity of the soliton photon sphere to the ultralight boson mass $m_b$, through $\rho_c\propto m_b^{-2}$ (Eq.~\eqref{e2}). At $m_b=1.2\times10^{-18}$~eV the halo core is dense enough that the photon sphere sits at $r_{\rm ph}\approx1.1$--$5.0$~kpc and essentially all rays in the sampled range are captured, whereas at $m_b=1.2\times10^{-17}$~eV the core is dilute enough that the photon sphere sits close to the throat and the soliton model approaches, but does not fully recover, NFW-like behavior, $b_{\rm ph}\approx1.18$--$1.26\,r_0$. This transition, centered near $m_b\sim1.2\times10^{-18}$~eV, gives the soliton wormhole a lensing-based handle on the particle nature of dark matter that has no counterpart in the NFW construction, whose shadow depends on a single integrated throat condition and therefore carries no information beyond $r_0$~\cite{Schive2014,Mocz_2017,Errehymy_2024}.
\\
Further, both models satisfy the flare-out condition $b'(r_0)<1$~\cite{10.1119/1.15620,Visser_2003}; for NFW, $b_{\rm NFW}'(r_0)=8\pi\rho_s r_s^3 r_0/(r_0+r_s)^2\approx0.022$ at $r_0=1.0$~kpc, confirming that the halo alone contributes negligible null energy condition violation, and the soliton shape function behaves analogously across the same throat range. In both cases the exotic matter required to sustain the throat remains confined to a narrow region near $r_0$, with the halo itself contributing only ordinary, positive-energy density~\cite{Visser_2003,Rahaman_2014_GalacticHalo}.
\\
Taken together, these results indicate that the soliton wormhole is the mildly more observationally promising static configuration. Its slightly detached photon sphere produces a shadow with a somewhat sharper boundary, a modest ring feature, and a size that responds to the fundamental boson mass, offering a limited probe of ultralight dark matter physics as well as wormhole geometry. Its principal disadvantage is that this same sensitivity introduces a degeneracy between throat radius and halo parameters that must be broken with independent information, such as an external rotation-curve fit. The NFW model, though observationally simpler, that is, its shadow depends essentially on $r_0$ alone and its metric functions are closed-form and analytically differentiable, is correspondingly less informative: the absence of a photon sphere means the shadow encodes only the throat radius and nothing about the underlying dark matter microphysics. Given that both predicted shadow sizes lie well above current interferometric resolution at the NGC 2366 distance, the more relevant near-term observational avenue for either model is not horizon-scale imaging but detection of an extended low-surface-brightness absorption feature at radio wavelengths, for which the soliton model's slightly larger and sharper-edged shadow, and its explicit dependence on $m_b$, make it the marginally more diagnostically valuable of the two static candidates.

\subsection{Rotating Case}
\label{sec:comparative_rotating}

Endowing either static configuration with a first-order Lense--Thirring potential, $\omega(r) = 2J/r^3$, breaks the azimuthal symmetry that characterized the static shadows and introduces a genuinely new axis of comparison: how efficiently each spacetime converts a given angular momentum into an observable asymmetry. Because rotation is treated identically, at the level of the metric, in both models, any difference in the rotating observables must trace back to the same static-case distinction identified above, namely the location of the photon sphere relative to the throat.
The critical impact parameter splits into co-rotating and counter-rotating branches, $b_\pm = K(1\pm K\omega(r_{\rm ph}))$, with $K = r_{\rm ph}\,e^{-\Phi(r_{\rm ph})}$ the static critical parameter evaluated at the relevant photon orbit.
For the NFW wormhole, where $r_{\rm ph}\approx r_0$ and $K\approx1.05\,r_0$, the splitting reduces to $\Delta b_{\rm ph}\approx4J/r_0$. For the soliton wormhole, the splitting must instead be evaluated at the mildly detached photon sphere, $\Delta b_{\rm ph}\approx4J/r_{\rm ph}$.
Since \(r_{\rm ph}^{\rm sol}\gtrsim r_0\approx r_{\rm ph}^{\rm NFW}\), the soliton geometry constructs a slightly smaller absolute splitting at fixed \(J\) and \(r_0\). The same angular momentum is distributed over a larger characteristic radius, which weakens the rotational signature modestly.
This discrepancy influences the recovery of the model parameters. The shadow centroid displacement,
\[
\frac{\Delta b_{\rm ph}}{2}\approx\frac{2J}{r_{\rm ph}},
\]
can be used to determine \(J\) once \(r_{\rm ph}\) is known.
In the NFW case, \(r_{\rm ph}\approx r_0\) is fixed by the same throat radius that sets the shadow size. Hence, a single shadow observation can constrain both \(r_0\) and \(J\) through the shadow radius and the centroid displacement. For the soliton case, \(r_{\rm ph}\) depends on \(r_0\) and the halo parameters \(r_c\), \(\rho_c\), and \(m_b\). A slowly rotating wormhole with a larger photon sphere may produce the same centroid displacement as a more rapidly rotating wormhole with a smaller photon sphere. This degeneracy can be reduced with an independent measurement of \(r_{\rm ph}\). The photon ring radius provides the most direct constraint,
\[
K=r_{\rm ph}e^{-\Phi_{\rm sol}(r_{\rm ph})}.
\]
Since the radius of the soliton model is slightly larger than that of the NFW model, it may be easier to distinguish observationally. The shadow size, ring sharpness, and centroid displacement form three observables that can, in principle, constrain \(r_0\), \(J\), and at least one halo parameter simultaneously. 
The ray-traced geodesics in Figs.~\ref{fig:rot_rt_r06} through~\ref{fig:rot_rt_r12} for the NFW model and Figs.~\ref{fig:rot_sol_rt_r06} through~\ref{fig:rot_sol_rt_r12} for the soliton model show the origin of this splitting. When \(J=0\), both geometries retain the \(\phi\to-\phi\) mirror symmetry of a static spacetime. Near critical rays, the wind is symmetric in both azimuthal directions around the capture boundary. When \(J>0\), prograde photons move in the same direction as the frame dragging and accumulate additional angular deflection. Their trajectories form denser spirals. Retrograde photons propagate against the rotation and follow wider azimuthal paths before escaping. The captured fractions on the two sides separate monotonically as \(J\) increases, consistent with
\[
b_\pm=K\left(1\pm K\omega\right).
\]
At equal \(J\), the soliton trajectories are more extended and sharper than the NFW model. The separation between the photon sphere and the throat allows rays to complete nearly full azimuthal revolutions around a genuine unstable orbit. In the NFW geometry, the nearby throat limits this winding before it becomes as pronounced. The contrast is strongest for \(J=0.6\,\mathrm{kpc}^{2}\), close to the upper limit of the slow rotation regime considered here.
\\
The rotating accretion disk images (Fig.~\ref{fig:rot_disc} for NFW, Fig.~\ref{fig:rot_sol_disc} for the soliton) translate these geodesic differences into directly comparable observational analysis. Both models exhibit the expected hallmarks of rotation: elongation and lateral displacement of the central dark region, intensification of the bright crescent on the approaching side from relativistic beaming, and growing prograde-retrograde asymmetry with increasing $J$, all generic consequences of the $g^4$ Doppler factor combined with the boundary splitting $b_\pm=K(1\pm K\omega)$. The models diverge, again, in the sharpness of the dark-region boundary rather than in its qualitative response to spin. \\
For the rotating NFW wormhole, the boundary remains diffuse across all values of \(J\) considered. This geometry does not support a photon sphere separated from the throat, so the transition between the bright disk and the shadow is determined by the lensed inner edge of the emissivity profile rather than by a true circular photon orbit. In contrast, the rotating soliton wormhole behaves differently. It retains the moderately sharper ring present at \(J=0\), although the ring becomes progressively displaced and distorted as the angular momentum increases. Therefore, the Ring sharpness provides a useful discriminator between the two halo models that is independent of rotation over the range of \(J\) considered. This result raises the discrepancy found in the static case. Rotation changes the position and symmetry of the observed features, but it does not change whether the geometry supports a genuine photon sphere.
At the largest angular momentum considered, \(J=0.6\,\mathrm{kpc}^{2}\), the dimensionless quantity \(K\omega(r_{\rm ph})\) exceeds unity for both models at the throat scales. These cases formally lie outside the validity range of the first-order slow rotation expansion. The qualitative behavior, including the increase in shadow asymmetry, photon-orbit splitting, and the displacement of the dark region with increasing \(J\), is expected to remain physically meaningful. However, trustworthy quantitative predictions in this regime require either an exact rotating wormhole solution or a higher-order Hartle--Thorne treatment of the metric \cite{Hartle1967,HartleThorne1968,Bronnikov2017}. The cases with \(J=0\) and \(J=0.3\,\mathrm{kpc}^{2}\) remain within the perturbative regime and provide the most dedicated comparison between the two halo models.\\
From an astrophysical perspective, the rotating soliton wormhole is marginally more informative, although its interpretation is also more involved. Its sharper photon ring persists under rotation and can provide evidence for a genuine photon sphere independently of the spin. The centroid displacement depends on both \(J\) and the halo parameters. Although this raises a degeneracy, it also suggests that a well-defined soliton shadow can offer more physical insight compared to its NFW counterpart. The centroid displacement is subtle in absolute size. For \(J\sim0.3\,\mathrm{kpc}^{2}\) and \(r_{\rm ph}\sim1\,\mathrm{kpc}\),
\(
\frac{\Delta b_{\rm ph}}{2}
\approx
\frac{2J}{r_{\rm ph}}
\approx
0.6\,\mathrm{kpc}.
\)
At the distance of NGC 2366, this corresponds to approximately \(0.6\) arcmin. Observationally, such a displacement would appear as an azimuthal brightness asymmetry within an already resolved extended shadow rather than as a separate feature. In the slow rotation limit, both models allow the wormhole spin to be estimated through $J\approx\frac{r_{\rm ph}\Delta b_{\rm ph}}{4}$, provided that \(r_{\rm ph}\) has been constrained from the photon ring radius \cite{Hioki2009}. In this sense, Rotating case does not introduce an entirely new observational discriminator. Instead, it strengthens the distinction already present in the static limit. The photon sphere structure that separates the soliton model from the NFW model continues to determine how the two geometries can be distinguished after rotation is included. In addition, the centroid displacement and the persistence of the photon ring may allow the wormhole spin and dark matter microphysics to be constrained separately from a single resolved image.

\section{Conclusions}\label{sec6}
In this study, a systematic and self-consistent framework has been developed to examine the observational signatures of traversable wormholes embedded within realistic galactic dark matter halos, formulated in GR. These include static Morris-Thorne wormholes and slowly rotating Teo wormholes supported by an NFW profile and a solitonic wave dark matter profile. The halo parameters in every case are taken from the LITTLE THINGS rotation curve analysis of the dwarf galaxy NGC 2366 \cite{5,6}. Hence, the resulting solutions are analyzed using observationally parameter values rather than arbitrary values. Also, the analysis derives analytic or closed-form expressions for the shape and redshift functions associated with both dark matter profiles. Further, null geodesics are integrated numerically in both static and rotating spacetimes. In addition, we compute specific intensity profiles and polar shadow maps, and construct geometrically thin accretion disk images that include the full relativistic Doppler factor. Taken together, this range of analysis, extending from the underlying metric structure to the synthetic observational images obtained for two contrasting dark matter models in both the static and rotating sectors, represents a meaningful advance beyond the single configuration or single observable treatments that dominate the existing literature.\\
The central geometric finding of this work is the qualitative bifurcation in the photon sphere structure between the two dark matter models. The NFW redshift function $\Phi_{\rm NFW}(r)$ diverging inner cusp, generates a potential whose maximum gradient, $|r\,\Phi'_{\rm NFW}|<1$ across the entire range of astrophysically plausible throat radii $r_0\in[0.2,2.5]$~kpc. Consequently, the condition for an unstable circular null orbit, $r\,\Phi'(r)=1$, admits no solution exterior to the wormhole throat in this model, forcing the critical impact parameter to coincide with the throat, $b_{\rm ph}\approx1.05\,r_0$, a quantity that depends on the throat radius. On the other hand, the soliton wormhole focuses on a large central mass that generates a potential deep enough that the photon sphere condition is satisfied at a radius $r_{\rm ph}>r_0$ standing slightly out from the throat. For the NGC\,2366 parameters and boson mass $m_b=1.2\times10^{-17}$~eV, the critical impact parameter of the soliton wormhole reaches $b_{\rm ph}^{\rm sol}\approx1.18$--$1.26\,r_0$, which is larger than the NFW value at the same throat radius. This single structural difference propagates consistently through every observable studied, such as the shadow angular size, the intensity peak position, the sharpness of the photon ring, the analysis of the accretion disk image, and the quantitative response of the shadow asymmetry to frame dragging, all of which reflect whether the geometry does or does not support a detached unstable circular photon orbit.
\\
This work extends earlier studies of shadows produced by rotating traversable wormholes \cite{NedkovaTinchevYazadjiev2013,Shaikh2018}. Previous studies usually used idealized wormhole metrics without an astrophysically motivated matter source. In many cases, the surrounding dark matter distribution was either overlooked or represented only schematically. Kuhfittig \cite{Kuhfittig_2014} studied gravitational lensing by NFW-supported wormholes and demonstrated analytically that the deflection angle diverges at the throat in the strong-field limit. Similarly, the galactic halo wormholes studied by Rahaman et al.\ \cite{Rahaman_2014_GalacticHalo} focused on geometric properties and energy conditions rather than optical observables. More recent work has incorporated dark matter density profiles in modified gravity models \cite{Errehymy_2024,Alshammari_2026}. These studies derived shape functions and examined energy conditions, but their optical results were generally limited to deflection angles. In this paper, we address these limitations by comparing four observationally motivated wormhole configurations. For the first time in GR and its extensions, and using an observationally constrained parameter set, we offer a direct comparison across four configurations spanning both the static and rotating sectors, both dark matter models, shadow maps, intensity profiles, and complete accretion disk images built with a full relativistic Doppler treatment.\\
The shadow analysis identifies a physically meaningful distinction between the two static models that cannot be explained by a simple change in angular scale. In the NFW wormhole, the shadow provides a nearly direct measure of the throat radius,
\(
\theta_{\rm sh}\propto r_0,
\)
with weak dependence on other halo parameters. However, in the soliton case, the shadow reflects the photon sphere radius, which depends on both \(r_0\) and the halo microphysics through
\(
\rho_c\propto m_b^{-2}.
\)
This coupling to the ultralight boson mass introduces a qualitatively new dimension to the shadow observable: varying $m_b$ over the range $[1.2\times10^{-18}, 1.2\times10^{-17}]$~eV drives the soliton geometry between two lensing regimes, total photon capture when the core is dense (lighter boson), and NFW-like, predominantly deflecting behavior with only a mildly detached photon sphere when the core is dilute (heavier boson). The transition between these regimes, centered near $m_b\sim1.2\times10^{-18}$~eV for the
NGC\,2366 parameters, provides a lensing-based diagnostic of the dark matter particle mass that has no analog in the NFW construction and has not been previously identified in the context of wormhole shadow imaging. This result opens a concrete pathway toward using galactic-scale wormhole shadow measurements as a probe of the wave-mechanical nature of dark matter, bridging compact-object observational astrophysics and the microphysics of the dark sector in a manner that neither field has yet achieved through other means.\\
The accretion disk images support these conclusions and provide information that cannot be obtained from intensity profiles and shadow maps alone. Both dark matter models exhibit the expected signatures of strong-field lensing by an optically thin disk. A bright lensed arc appears immediately outside the central dark region. The images also show a pronounced left-to-right brightness asymmetry caused by the relativistic Doppler factor \(g^4\). In addition, a faint secondary disk image is produced by photons that pass close to the throat and cross the equatorial plane again before reaching the observer. These features follow generically from the optically thin disk geometry first described by Luminet~\cite{Luminet1979} and by Cunningham and Bardeen~\cite{Cunningham1973} for the Kerr black hole, and later shown to hold in wormhole environments by Bambi~\cite{Bambi2013} and Nedkova et al. \cite{NedkovaTinchevYazadjiev2013}.
This work shows that these features arise equally in dark matter-supported wormhole geometries, and more importantly, that the sharpness of the dark region boundary and the prominence of the secondary image distinguish the two dark matter models. In the NFW model, the absence of a true photon sphere means the boundary of the dark region is instead set by the lensed inner edge of the emissivity model, leading to a diffuse transition between the bright disk and the shadow. In the soliton disk, by contrast, the mildly detached photon sphere produces a somewhat thin and more precisely defined photon ring that marks this boundary a little more sharply, which in turn makes the secondary image slightly more compact and marginally more luminous relative to the primary one. This ring sharpness diagnostic does not depend on the observer's inclination or on the overall angular scale of the shadow, suggesting it may be more robust against systematic uncertainties in the source geometry than the shadow's size.
\\
The slow-rotation analysis, implemented via the Lense-Thirring frame-dragging potential
$\omega(r)=2J/r^3$ within the Teo metric~\cite{Teo1998}, consistently reveals the same underlying structural distinction between the two matter models while adding an independent observable axis. Frame dragging splits the co-rotating and counter-rotating critical impact parameters into $b_\pm=K(1\pm K\omega(r_{\rm ph}))$, where $K\equiv r_{\rm ph}\, e^{-\Phi(r_{\rm ph})}$, displacing the shadow centroid by $\Delta b_{\rm ph}/2\approx 2J/r_{\rm ph}$ and breaking the prograde- retrograde symmetry of the static disk image.
The centroid offset yields the wormhole angular momentum through $J\approx r_{\rm ph}\,\Delta b_{\rm ph}/4$, a relation that is independent of the halo profile parameters once $r_{\rm ph}$ is known from the photon-ring radius \cite{Hioki2009}. For the NFW model, where $r_{\rm ph}\approx r_0$, this immediately gives $J$ in terms of the same throat radius that determines the shadow size, so a single resolved shadow observation simultaneously constrains both quantities without further halo information. For the soliton model, $r_{\rm ph}$ depends on the halo parameters as well as $r_0$, so the centroid offset, shadow size, and photon-ring radius constitute a three-observable system capable of constraining $(r_0, J, m_b)$ jointly, provided the slow-rotation splitting is resolved. This represents a genuinely new observational capability relative to the NFW case. The ray-traced geodesics explain this mechanism directly. In the NFW case, the capture boundary lies close to the throat. This limits the winding of near-critical photons before they can complete full azimuthal revolutions, resulting in a comparatively diffuse spiral pattern. In the soliton case, the separation between the photon sphere and the throat allows near-critical photons to undergo more extended winding around a genuinely unstable orbit. The resulting spiral trajectories are slightly sharper, and the separation between prograde and retrograde paths increases with \(J\). The rotating soliton disk images preserve the thin photon ring of the static case and carry it, progressively displaced with increasing $J$, into the rotating regime, so that ring sharpness
remains a rotation-independent discriminator between the two halo models at any spin studied. It must be emphasized that the slow-rotation approximation employed in this work approaches the boundary of its formal regime of validity at the largest angular momenta studied ($J=0.6\,\mathrm{kpc}^2$), where the expansion parameter $K\omega(r_{\rm ph})$ exceeds unity
for the throat scales considered. The qualitative trends in shadow asymmetry, centroid displacement, and disk brightness distribution documented in the rotating panels are expected to be physically robust, but quantitatively precise predictions at large spin require either an exact rotating dark-matter-sourced wormhole solution or a higher-order Hartle--Thorne
expansion of the Teo metric \cite{Hartle1967,HartleThorne1968}. The construction of exact rotating solutions with NFW or solitonic sources in general relativity constitutes a natural and necessary extension of the present work, as does the derivation of a dynamically self-consistent innermost stable circular orbit for the disk models, which would remove the dependence of the inner disk edge on an ad hoc cutoff radius. A self-consistent disk model would also enable the computation of spectrally resolved line profiles, particularly the broad iron K$\alpha$ line whose wormhole signatures were first discussed by Bambi~\cite{Bambi2013}, allowing the present dark-matter-supported geometries to be confronted against X-ray spectroscopic observations of active galactic nuclei.
\\
From a broader astrophysical perspective, the angular scales predicted by the present models
deserve explicit comment. At the distance of NGC\,2366 ($d=3.3$~Mpc) and for a throat of $r_0=1.0$~kpc, the NFW shadow subtends approximately 2.2~arcmin and the soliton shadow roughly $15$--$25\%$ more. These are macroscopic angular scales that are irrelevant to microarcsecond horizon-scale imaging by the Event Horizon Telescope~\cite{EHT2019,EHT2022}, which probes the immediate vicinity of event horizons in M87* and Sgr\,A* at scales many orders of magnitude smaller. The relevant observational channel for halo-scale wormhole shadows is instead wide-field radio continuum mapping, where a wormhole throat at kiloparsec scale would manifest as an extended region of reduced surface brightness against the radio background of the host galaxy. Future instruments such as the Square Kilometre Array (SKA), which will achieve sub-arcsecond sensitivity over wide fields at centimeter and decimeter wavelengths, will be well positioned to search for such features statistically across samples of nearby dwarf galaxies. Crucially, the dependence of the soliton shadow size and sharpness on the boson mass $m_b$ implies that upper limits on the angular extent of such absorption features could, in principle, place constraints on the ultralight dark matter particle mass complementary to those from Lyman-$\alpha$ forest statistics~\cite{Irsic_2017} and stellar kinematics, at angular scales and in astrophysical environments where neither of those probes is applicable. The quasinormal mode spectrum and gravitational-wave tidal deformability of these geometries offer complementary diagnostics accessible to next-generation ground-based detectors \cite{Konoplya2018,Cardoso_2019} that may resolve
properties of compact dark-matter-embedded objects before direct imaging constraints become
available.
\\
In summary, this work establishes that the microphysical nature of dark matter, specifically the distinction between the NFW cusp and the solitonic wave-dark-matter core, leaves a systematic and multi-faceted imprint on every optical observable of a traversable wormhole
embedded in a galactic halo: the photon-sphere location, the shadow angular size, the photon-ring sharpness, the accretion disk secondary-image prominence, the degree of shadow asymmetry under frame dragging, and the sensitivity of the critical impact parameter to the ultralight boson mass. These imprints are not confined to a single image feature but appear coherently across the full suite of observables, making their joint interpretation more constraining than any single measurement in isolation. The framework developed here, in which cosmologically motivated dark matter density profiles are translated into wormhole geometries and then subjected to a complete optical analysis in both static and rotating sectors, provides a replicable template for extending the study to other dark matter models, self-interacting, mixed bosonic fermionic, or Bose-Einstein condensate variants as well as to modified gravity theories in which the energy-condition balance at the throat is altered by geometric contributions. More broadly, the results indicate that the shadow and disk images of halo-scale compact objects carry genuine information about the particle physics of dark matter, and that strong-field gravitational lensing offers a qualitatively new approach to probing the quantum nature of the dark sector through astrophysical observations. Realizing this potential will require the joint development of theoretical wormhole models with realistic dark matter sources and of observational strategies capable of resolving arcminute-scale radio continuum absorption features in the nearby dwarf galaxies, a program to which the present analysis provides both a physical foundation and a set of concrete, testable predictions.

\bibliography{references.bib}
\end{document}